\documentclass[10pt,natbib,times]{nature_fig}

\usepackage{sansmath}

\usepackage[T1]{fontenc}
\usepackage{times}
\usepackage{graphicx}
\usepackage{url}
\usepackage{epsfig}
\usepackage{upgreek}
\usepackage{xcolor}
\usepackage{lineno}
\usepackage{textcomp}
\usepackage{threeparttable}
\usepackage{threeparttablex}
\usepackage{booktabs}
\usepackage[caption = false]{subfig}
\usepackage{ulem}
\usepackage{amsmath}
\usepackage{mathrsfs}
\usepackage{tikz}
\usepackage{physics}
\usepackage{siunitx}
\usetikzlibrary{shapes.geometric, arrows}
\usepackage{lscape}
\RequirePackage{lineno}

\usepackage{txfonts}
\usepackage{float}
\usepackage{aas_macros}
\usepackage{pdflscape}
\usepackage{hyperref}
\usepackage{marvosym}
\usepackage{subeqnarray}

\usepackage{graphicx}
\usepackage{amssymb}
\usepackage{amsmath}
\usepackage{mathrsfs}
\usepackage{color}
\usepackage{rotating} 
\usepackage{empheq}
\usepackage{cases}
\usepackage{subeqnarray}

\usepackage{amssymb}
\usepackage{pdfpages}
\usepackage{caption}
\usepackage{mathrsfs}
\makeatletter
\usepackage[symbol]{footmisc}
\usepackage{longtable}
\usepackage{multirow}

\newcommand{\farcs}{\prime\prime}
\newcommand{\fcs}{$F_{\rm 5100}$}
\newcommand{\feii}{Fe {\sc ii}}

\newcommand{\hb}{H$\beta$}

\newcommand{\heii}{He {\sc ii}}
\newcommand{\mbh}{$M_\bullet$}
\newcommand{\oiii}{[O~{\sc iii}]}

\def\mbh{M_{\bullet}}

\def\ergs{\rm erg~s^{-1}}
\def\fblr{f_{_{\rm BLR}}}

\def\feii{Fe\,{\sc ii}}
\def\Hbeta{H$\beta$}

\def\kms{\rm km~s^{-1}}

\def\mathdotM{\dot{\mathscr{M}}}

\def\muc{\multicolumn}
\def\oiii{[O~{\sc iii}]$\lambda$\,5007\AA}
\def\pp{\prime\prime}
\def\Rhb{R_{\rm H\beta}}

\def\RFe{{\cal R}_{\rm Fe}}
\def\Rg{R_{\rm g}}

\def\sunm{M_{\odot}}
\def\tauhb{\tau_{\rm H\beta}}

\def\sSMBH{s\kern -1pt@\kern-1pt SMBH-disk}

\def\calR{{\cal R}_{\rm Fe}}
\def\calA{{\cal{A}}}

\def\MICA{{\ttfamily{MICA}}}
\def\JAVELIN{{\ttfamily{JAVELIN}}}

\def\ihep{Key Laboratory for Particle Astrophysics, Institute of High Energy Physics,
Chinese Academy of Sciences, 19B Yuquan Road, Beijing 100049, China}

\def\naoc{National Astronomical Observatories of China, Chinese Academy of Sciences,
 20A Datun Road, Beijing 100020, China}

\def\ucas{School of Astronomy and Space Science, School of Physical Sciences, University 
of Chinese Academy of Sciences, 19A Yuquan Road, Beijing 100049, China}

\def\wiro{Department of Physics and Astronomy, University of Wyoming, Laramie, WY 82071, USA}

\def\njuASS{School of Astronomy and Space Science, Nanjing University, Nanjing 210093, China}

\def\njuLab{Key Laboratory of Modern Astronomy and Astrophysics, Ministry of Education, China}

\def\wiro{Department of Physics and Astronomy, University of Wyoming, Laramie, WY 82071, USA}

\def\ynao{Yunnan Observatory, Chinese Academy of Sciences, Kunming 650011, China}

\def\CAHA{Centro Astronomico Hispano Alem\'an, Sierra de los filabres sn, E-04550 Gergal, Almer\'ia, Spain}

\def\CAHAIAA{Instituto de Astrofisica de Andaluc\'ia, Glorieta de la astronom\'ia sn, 18008 Granada, Spain}

\def\CSU{Technology and Engineering Center for Space Utilization, Chinese Academy of Sciences, Beijing 100094, P.\,R. China}

\def\aynu{School of Physics and Electrical Engineering, Anyang Normal University, Anyang, Henan 455000, China}

\title{Detection of unexpected leading delays in broad \Hbeta\ line reverberations in the quasar PHL\,1092}

\author{
Jian-Min Wang$^{1,2,3}$,
Chen Hu$^{1}$, 
Yong-Jie Chen$^{4,1,2}$,
Yu-Yang Songsheng$^1$,
Yi-Lin Wang$^{1,2}$,
Hao Zhang$^{1,2}$,
Pu Du$^{1}$,
Yan-Rong Li$^{1}$,
Bin Luo$^{5,6}$,
Michael S. Brotherton$^{7}$,
Jin-Ming Bai$^{8}$,
Wei-Jian Guo$^{3}$,
Seng Yang$^{9}$, 
Zhu-Heng Yao$^{3}$,
Jes\'us Aceituno$^{10,11}$
}

\begin{document}
\maketitle

\begin{affiliations}
{\small
\item{\ihep}
\item{\ucas}  
\item{\naoc}
\item{\CSU}
\item{\njuASS}
\item{\njuLab}
\item{\wiro}
\item{\ynao}
\item{\aynu}
\item{\CAHA}
\item{\CAHAIAA}
}
\end{affiliations}

\begin{abstract}
Delayed reverberations of broad emission lines in response to optical continuum variations\cite{Bahcall1972,Blandford1982} have been widely observed in active galactic nuclei (AGNs). 
They serve as a powerful tool for probing inner structures of AGNs\cite{Peterson1993,Kaspi2000} and estimating the masses of supermassive black holes (SMBHs)\cite{Peterson2014}.
The delays exhibit a strong correlation with approximately the square root of the optical luminosity -- a relationship known as the "standard structure" of AGN broad-line regions (BLRs)\cite{Peterson1993,Kaspi2000,Bentz2013,Du2019}.
Here, we report the discovery of leading delays in \Hbeta\ line reverberations (LDRs) in the quasar PHL\,1092 preceding variations of the 5100\,\AA\ continuum by 17–57 days, based on our eight-year continuous campaign of reverberation mapping of super-Eddington AGNs.
The LDRs suggest that the 5100\,\AA\ continuum regions are so extensive that they are larger than the BLRs.
This phenomenon not only fundamentally disrupts the well-established BLR size-luminosity relation but also violates the principle of causality.
This unprecedented LDRs challenge the conventional methods for estimating SMBH mass as well as the standard model of AGNs.
A preferred scenario to explain the LDRs is that the SMBH-disk contains a population of accreting stellar-mass black holes (sMBHs)\cite{Cheng1999,Gilbaum2022,Zhou2024} as extra heating sources of the disk. 
Consequently, continuum regions of the disk are efficiently stretched so that the 5100\,{\AA} regions exceed the BLRs, yielding the observed LDRs.
Generally, sMBH activities there could provide new physics of AGN phenomena, which can be tested by LIGO, LISA\cite{Amaro-Seoane2023}/Tianqin\cite{Luo2025} and ET\cite{DiGiovanni2025} detections of gravitational waves from sMBH mergers.
\end{abstract}

With great efforts of measuring the H$\beta$ lags ($\tauhb$) relative to the 5100\,\AA\ continuum from reverberation mapping (RM) campaigns, a robust empirical relation of $\Rhb\approx 33.6\,\ell_{44}^{\,0.53}$\,ltd has been established over more than four decades (see a brief summary in \S\,\ref{sec:overview} of Method), commonly referred to as the $R-L$ relation.
Here $\Rhb=c\tauhb$ denotes the size of the broad-line region (BLR), $c$ is the speed of light, and $\ell_{44}=L_{5100}/10^{44}\ergs$ represents the 5100\,\AA\ luminosity normalized to $10^{44}\ergs$.
This relation serves as a benchmark of AGN structure widely applied to all kinds of quasars including large samples\cite{Shen2012,Vestergaard2009} for cosmic black hole demography from $z\sim 7.5$ including the JWST quasars\cite{Yang2023} and little red dots (LRDs)\cite{Matthee2024}. 
However, the long-term SEAMBH (Super-Eddington Accreting Massive Black Holes) campaign\cite{Wang2013,Du2014,Wang2014a}, which is a program designed to investigate physics of super-Eddington AGNs, has found that $\Rhb$ significantly shrinks with increases of the relative strength of optical iron lines to H$\beta$ ($\RFe$)\cite{Du2019}. 
One of the primary goals of this campaign has been searching for the mostly extreme shortening of H$\beta$ lags in super-Eddington AGNs, where the intrinsic timescales $\Rhb/c$ approach ones comparable to the light-crossing time of the SMBH-disk itself.
If such an extreme object exists, what observational signatures would its reverberations exhibit? and what underlying physics would they reveal?

\begin{figure*}
\centering
\includegraphics[angle=0,width=1.0\textwidth]{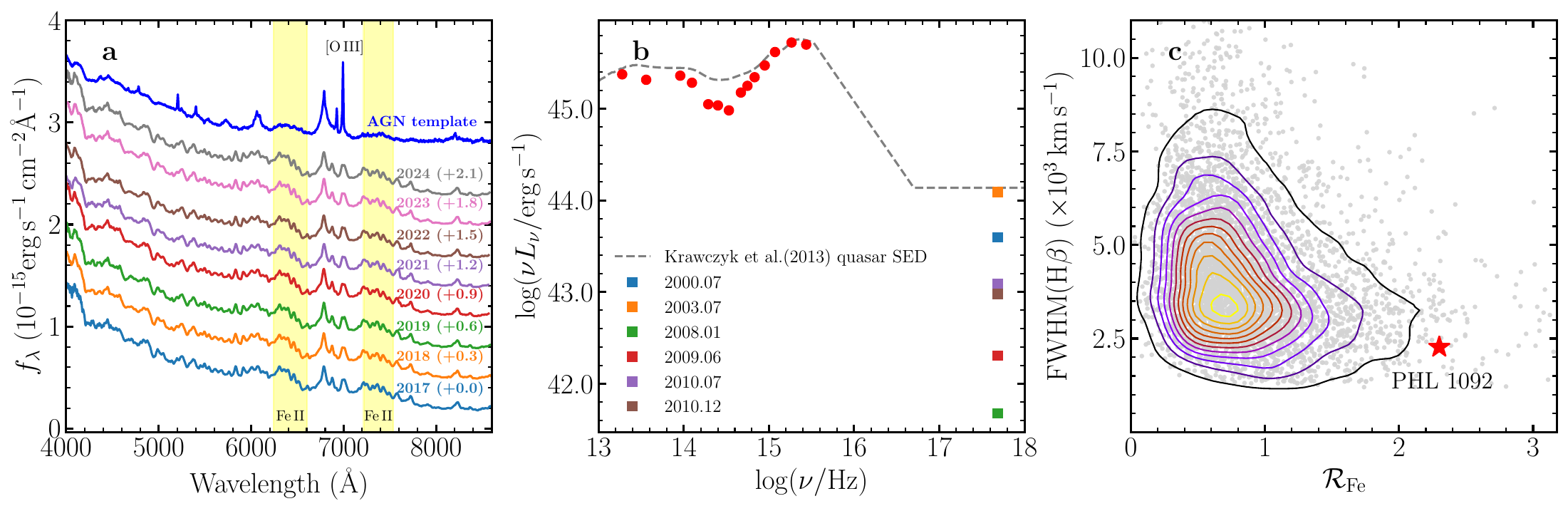}
\caption{\footnotesize 
Panel {\textbf{a}}: Annual mean spectra of PHL\,1092 (in the observer’s frame). Spectral fluxes have been vertically offset for clarity. 
Compared with the Sloan Digital Sky Survey (SDSS) quasar template (blue line)\cite{VandenBerk2001}, PHL\,1092 exhibits prominent \feii\ features and weak \oiii\ line emission—characteristics that distinguish it from normal quasars. 
The yellow shaded regions highlight the optical \feii\ lines.
Panel {\textbf{b}}: SED of PHL\,1092 (solid line) compared with the mean spectral energy distribution (SED) of luminous quasars (dotted line)\cite{Krawczyk2013}. 
Its SED is significantly bluer (i.e., has a steeper continuum) than the quasar mean SED. 
Its X-ray flux varies by more than two orders of magnitude\cite{Miniutti2009}.
SED data sources: ultraviolet (UV) from GALEX, optical from SDSS, near-infrared (NIR) from 2MASS, and mid-infrared (MIR) from WISE.
Panel {\textbf{c}}: Position of PHL\,1092 in the $\RFe$--FWHM(H$\beta$) diagram of the SDSS data\cite{Shen2014}, exhibiting PHL\,1092 as an extreme quasar.
}

\label{fig:spec}
\end{figure*}

PHL\,1092 is one of the long-term targets selected as an extremely high-$\calR$ AGN\cite{Marziani2001,Miniutti2009,Marinello2020} (see recent reviews on observational properties of SEAMBHs as a function of $\calR$\cite{Panda2023,Marziani2025}). 
Fig.\,\ref{fig:spec} presents its optical spectra, multi-wavelength spectral energy distributions (SEDs), and an extreme position in the $\calR-{\rm FWHM(H}\beta$) plane\cite{Marziani2001,Shen2014}.
Obviously, it is an AGN with a high Eddington ratio, and  its SED is notably bluer than the average SED of quasars.
Additionally, PHL\,1092 is known as a weak-line quasar\cite{Plotkin2010}, and its X-ray emissions exhibit strong variability\cite{Miniutti2009}.
Our campaign of PHL\,1092 has been conducted continuously since November 2017, spanning eight years.
Details of observations and data reduction are described in \S\,\ref{sec:observations}.

\begin{figure*}
\centering
\includegraphics[angle=0,width=0.49\textwidth]{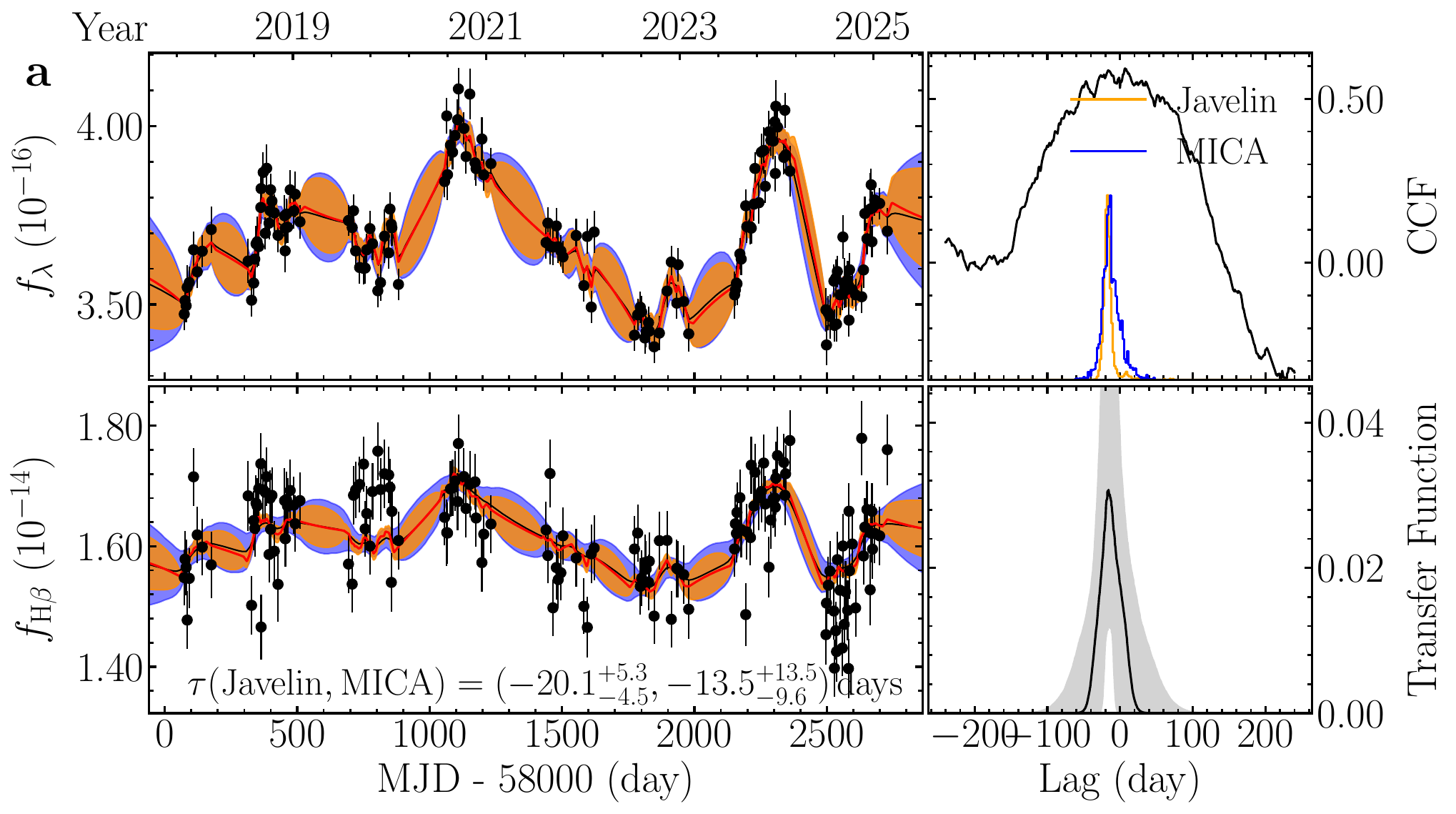}
\includegraphics[angle=0,width=0.49\textwidth]{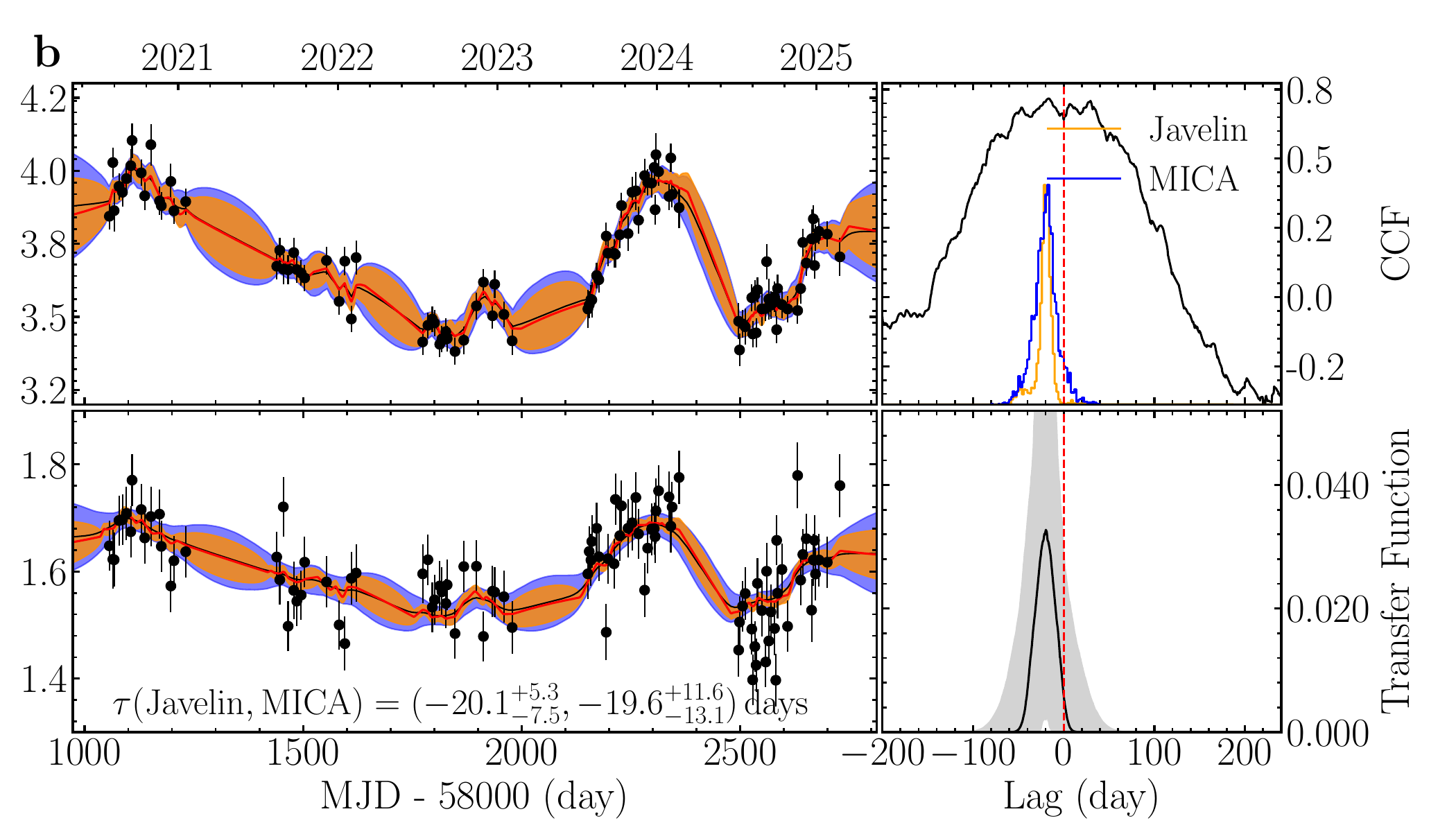}
\includegraphics[angle=0,width=0.49\textwidth]{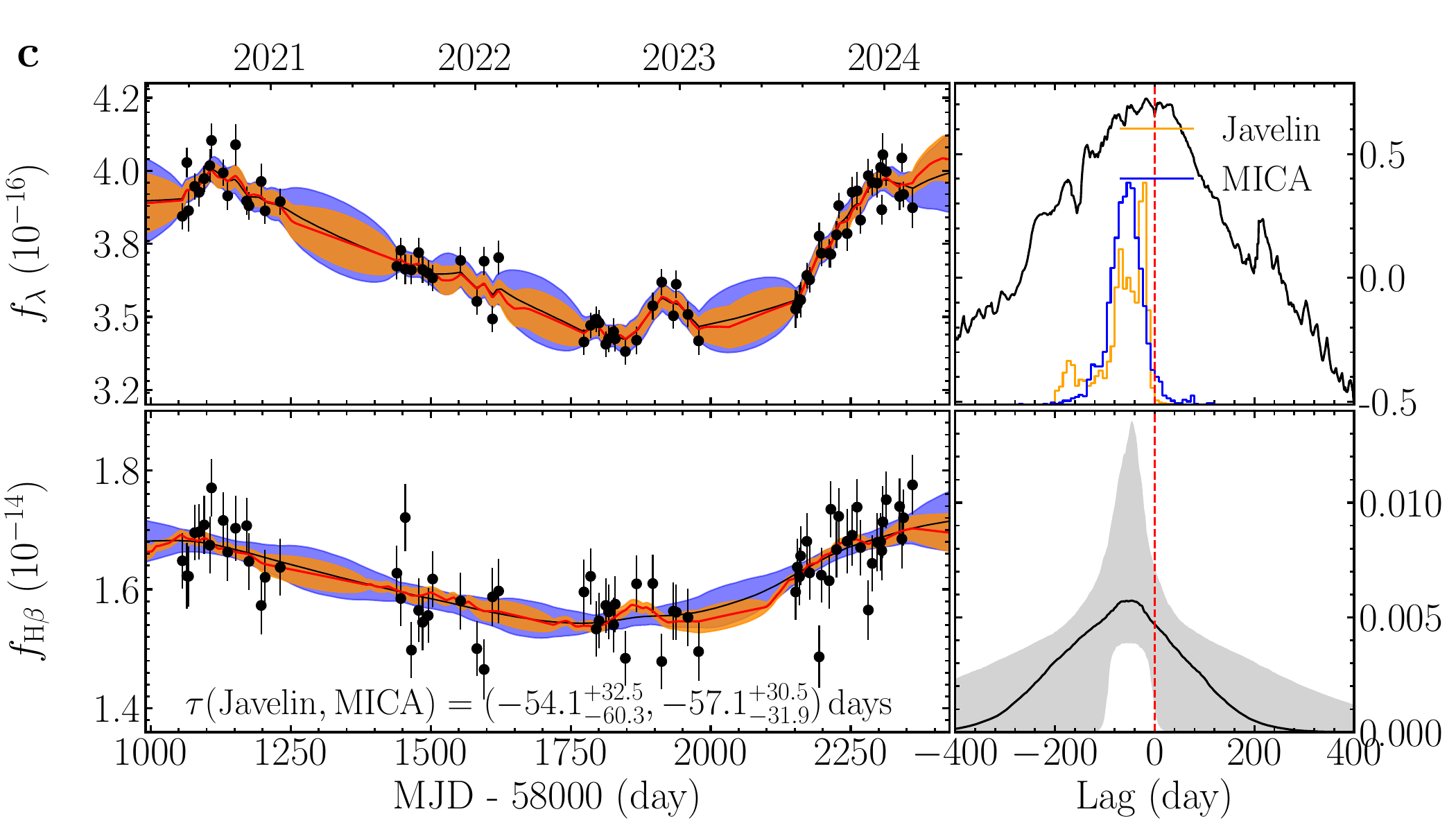}
\includegraphics[angle=0,width=0.49\textwidth]{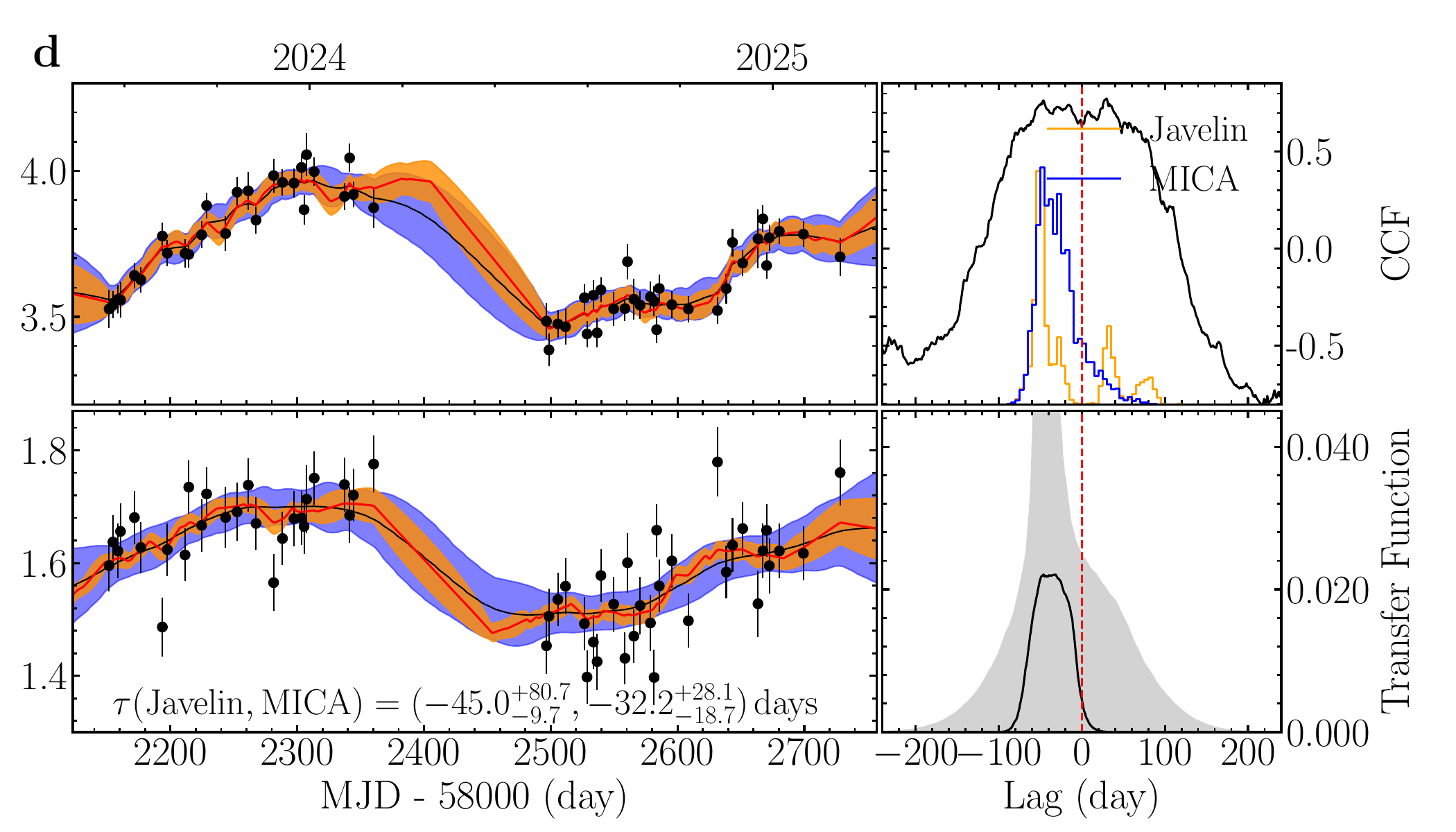}
\caption{\footnotesize 
Panels {\textbf{a}}–{\textbf{d}} display the LCs of the 5100\,\AA\ continuum and H$\beta$ line across different observational periods.
The cross-correlation function (CCF) reveals a strong correlation between the 5100\,\AA\ continuum and H$\beta$ line variations. 
Given the presence of multiple peaks in the CCF results, we used two independent methods —\JAVELIN\ and \MICA — to determine the reverberation delays.
The delay results derived from \JAVELIN\ (yellow) and \MICA\ (blue) are in good agreement with each other.
The red dotted lines mark $\tau_{\rm H\beta}=0$ (i.e., zero delay between H$\beta$ and the 5100\,\AA\ continuum).
The detected LDRs are consistent across all observational periods.
}
\label{fig:MICA-lags}
\end{figure*}

Tab.\,\ref{tab:PHL1092} provides the spectral features of PHL\,1092 and and additional relevant information.
Fig.\,\ref{fig:spec}a displays the annually averaged spectra, showing roughly constant profiles of the H$\beta$ line over the past eight years. 
This indicates that the BLRs are relatively stable but with only flux variations shown in the light curves (LCs).
Fig.\,\ref{fig:MICA-lags} shows the LCs of PHL\,1092 alongside the results of reverberation analysis.
Despite all the challenges of determining lags from a single annual campaign, we remained steadfast in our pursuit from our long term campaigns year after year.  
Given the BLR fluctuations observed in this object, we conducted RM analysis across different epochs using the \MICA\ and \JAVELIN\ codes (see \S\,\ref{sec:RManalysis}), intentionally excluding phases characterized by non-concordant reverberations discussed in \S\,\ref{sec:toymodel}.
Tab.\,\ref{tab:lags} provides flux variability ($F_{\rm var}$), and delays, which are in good agreement with one another.
We find that the H$\beta$ variations strongly correlate with the 5100\,\AA\ continuum, namely, \hb\ reverberations are indeed taking place, however,
the \hb\ reverberations are preceding the varying continuum with the lead of $\tauhb^{\,\ell}=-(17 \sim 57)\,$days given in Tab.\,\ref{tab:lags}. 
We note that the damped random walk (DRW) model doesn't fit well at epochs around MJD-58000\,+800 and +2500 (in timescales of one half years or so much shorter than the entire campaign). 
These could may result from non-concordant reverberations of the BLR composite responses with respect to spatially inhomogeneous variations of the extended ionizing sources. See \S\,\ref{sec:toymodel} for a toy model that discuss the non-concordant reverberations of the ideal cases.

Based on the spectral features, we derived H$\beta$ lags of $\tauhb^0=(128.7,\,15.9)\,$days using two relations: the $R-L$ relation\cite{Bentz2013} and the
$R_{\rm H\beta}-(L,\calR)$ relation of $R_{\rm H\beta}=44.7\,\ell_{44}^{0.45}10^{-0.35\calR}\,$ltd\cite{Du2019}, respectively.
While the modest shortening factors ($10^{-0.35\calR}\lesssim 5$) could be explained  by self-obscuration of the inner part of slim disks\cite{Wang2014b}, this effect fails to account for large shortening factors.
We note that the aliasing effects, which arise from seasonal gaps of long-term campaigns and observational errors, may randomly generate false-alarm LDRs. 
In \S\,\ref{sec:aliasing}, we investigate the effects through Monte Carlo simulations.
These simulations yield a false-alarm probability of $\lesssim 5\times 10^{-5}$ listed in Tab.\,\ref{tab:lags} and a corresponding significance level of $\gtrsim 4\,\sigma$ if we set the lags of $\tau_{\rm H\beta}^0=16.0\,$ltd from the $R_{\rm H\beta}-(L,\calR)$ relation.
Therefore, the LDR detections are robust in PHL\,1092.

\begin{figure*}
\centering
\includegraphics[angle=0,width=0.75\textwidth]{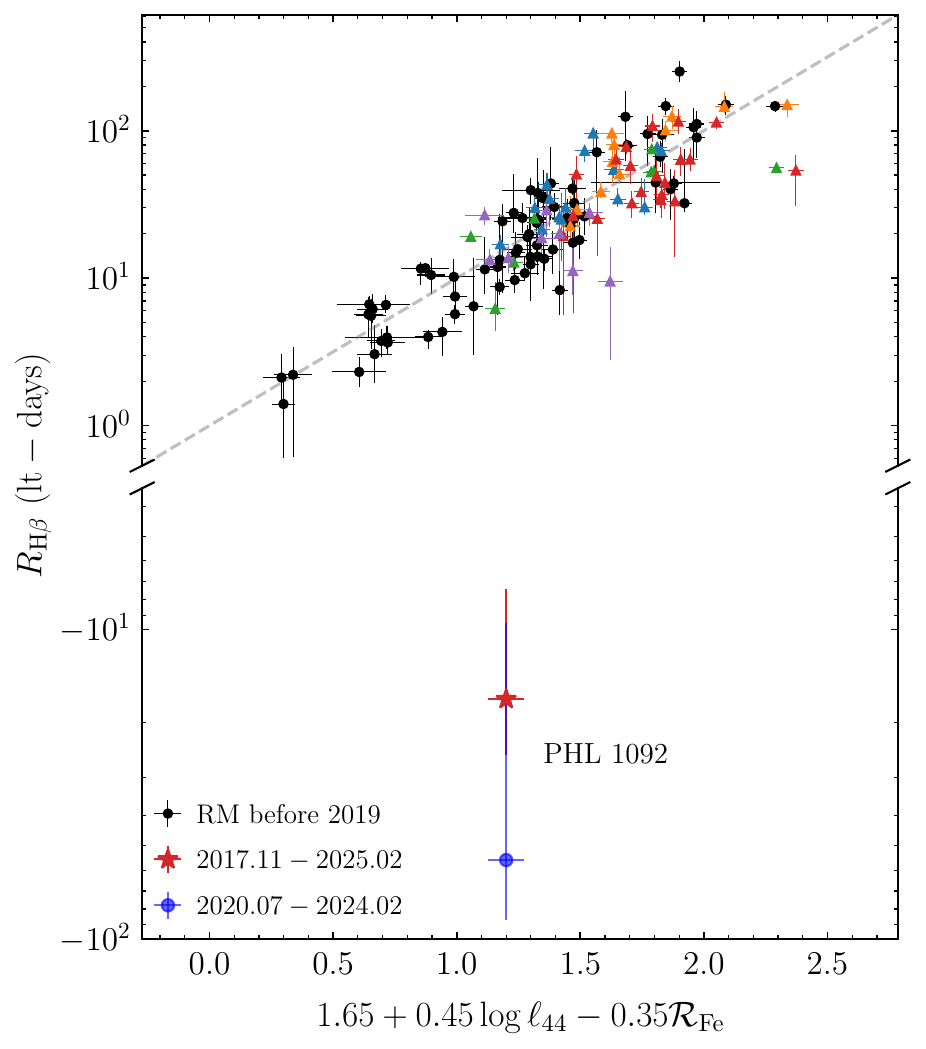}
\caption{\footnotesize 
The position of PHL\,1092 in the $R_{\rm H\beta}-(L,\calR)$ plane demonstrates that this object completely deviates from the known structures of AGNs.
Such a violation of reverberation causality requires additional energy sources (EESs) to extend the SMBH-disk.
RM data prior to 2019 (black points) are summarized in Ref.\cite{Du2019}, which includes measurements from\cite{Kaspi2000,Bentz2013,Du2014,Du2018} and other relevant literature (see the complete reference list in Ref.\cite{Du2019}).
Post-2019 RM points (color triangles) are from projects of LAMP\cite{U2022} (in purple), SEAMBH\cite{Hu2021,Hu2025} (in blue and orange), SAMP\cite{Woo2024} (in red), and MAHA (in green)\cite{Bao2022} with available $\calR$ measurements.
The dashed lines are adopted from Ref.\cite{Du2019}.
}
\label{fig:RL-Plane}
\end{figure*}

Fig.\,\ref{fig:RL-Plane} shows the position of PHL\,1092 in the $R_{\rm H\beta}-(L,\calR)$ plane, revealing a complete deviation from the well-established BLR scaling relations.
This points to a peculiar structure in this object different from the standard model of AGNs\cite{Netzer2013}.
Consequently, the conventional approach (Eq.\,\ref{eq:Mass} in \S\,\ref{sec:overview})\cite{Netzer2013,Peterson2014} for estimating SMBH masses becomes inapplicable to PHL\,1092.
Such an LDR phenomenon violates the causality of reverberations and has never been found in the past RM campaigns.

In order to explain the LDRs, we have to re-examine the validity of fundamental assumptions in reverberation physics.
The most important one is that the ionizing source is treated as a point, namely, the point source approximation (PSA)\cite{Blandford1982}, and the 5100\,\AA\ emission is conventionally used as a proxy for the ionizing emission. 
The PSA validity is examined in \S\,\ref{sec:overview} and \S\,\ref{sec:SSdisk}.
The observed LDRs in PHL\,1092 imply that the PSA is invalid, namely, the 5100\,\AA\ regions are extremely extensive, surpassing the BLR scale. 
This phenomenon triggers a series of significant challenges to the standard AGN model, as discussed in \S\,\ref{sec:question}. 
A fundamental issue is how to reconcile such extended structures with theoretical models of accretion disks\cite{Shakura1973}.
It is much beyond the scope of the present paper to establish a definitive model for the case of PHL\,1092, but we briefly discuss two possible mechanisms subsequently, which could jointly work for the LDRs.

First, we consider the extreme shortening of delays governed by accretion rates for the observed LDRs in PHL\,1092. 
It is known that the 5100\,\AA\ radius scales with the accretion rate as $R_{5100}\propto \left(\mathdotM/\mbh\right)^{1/3}$ in the regime of the Shakura-Sunyaev model (see Eq.\ref{eq:R5100SS} in \S\,\ref{sec:SSdisk}).
It increases until $R_{5100}$ reaches the photon trapping radius, given by $R_{\rm trap}\approx 10.4\,\mathdotM_3M_8\,$ltd (see Eq.\,22 in Ref.\cite{Wang1999}), where $\mathdotM_3=\mathdotM/10^3$ and $M_8=\mbh/10^8\,\sunm$.
In super-Eddington accretion disks, $R_{\rm 5100}$ can be obtained by the effective temperature of $T_{\rm eff}\approx T_c\tau_{\rm es}^{-1/4}$ (rather than Eq.\,\ref{eq:TeffSS}), where $T_c$ denotes the temperature of the mid-plane and $\tau_{\rm es}$ is the Thomson  scattering depth\cite{Wang1999}.
Thus, we have the maximum $R_{5100}$ radius of $R_{5100}^{\rm max}=15.8\,M_8^{1/2}\,$ltd that is independent of $\mathdotM$ when the rates are extremely high as $\mathdotM_3\ge 0.95\,M_8^{-1/2}$ (see Eq.\,18-20 in Ref.\cite{Du2016a}).
Conversely, for SEAMBHs, $R_{\rm H\beta}$ decreases with $\mathdotM$ (it correlates with ${\cal R}_{\rm Fe}$ as shown by Fig.\,4 in Ref.\cite{Marziani2025}). 
We expect a critical $\mathdotM$ (or ${\cal{R}}_{\rm Fe}$) at which $R_{5100}^{\rm max}\approx R_{\rm H\beta}$, namely, the observed delays of H$\beta$ reverberations relative to 5100\,\AA\ continuum would vanish. 
For PHL\,1092, this critical condition translates to ${\cal{R}}_{\rm Fe}=2.7-1.43\,\log M_8$, where $\ell_{44}\approx 12.6$ (see Tab.\,\ref{tab:PHL1092}).
Taking $\tauhb^0=15.9\,$days from the $R_{\rm H\beta}-(L,\calR)$ relation and $V_{\rm H\beta}
=2270\,\kms$,
we obtain $M_8=0.16$ from Eq.\,(\ref{eq:Mass})
(see Issue 1 in \S\,\ref{sec:question}, and \S\,\ref{sec:TF} about transfer functions for its order estimation). 
A lower limit of ${\calR}\gtrsim 3.8$ is necessary for the case of $R_{\rm H\beta}\sim R_{5100}$, which is much higher than the value 2.7 given in Tab.\,\ref{tab:PHL1092}. 
Furthermore, such extreme \feii\, emitters remain exceptionally rare in the SDSS database\cite{Shen2014,Du2016}. 
Therefore, the extremely super-Eddington accretion scenario alone is insufficient to explain the LDRs.

Second, extra energy sources (EESs: $Q_{\rm extra}$) heating the SMBH-disk are necessary to stretch the 5100\,\AA\ regions for the observed LDRs. 
Currently, continuum reverberations of accretion disks can be explained by reprocessing of X-rays as primary emissions in the lamp-post model\cite{Cackett2021,Kammoun2021}.
In the presence of EESs, $Q_{\rm extra}$ should be incorporated into the energy budget governing the effective temperature of the SMBH-disk ($T_{\rm eff}$) at a given radius $R$\cite{Gilbaum2022,Zhou2024,Cackett2007}
\begin{equation}\label{eq:extra}
\sigma_{\rm SB}T_{\rm eff}^{4}=Q_{\rm vis}+\frac{(1-\calA)L_{\rm X}}{4\pi R^2}\left(\frac{H}{R}\right)+Q_{\rm extra},
\end{equation}
where $\sigma_{\rm SB}$ is the Stefan-Boltzmann constant, $\calA$ is the disk albedo, $L_{\rm X}$ is the X-ray luminosity from the hot corona (in the lamp-post model\cite{Svensson1994,Collin2001}), $Q_{\rm vis}$ is the dissipation rates per area of gravitational energy of the SMBH-disk through viscosity in the standard model (see Eq.\,\ref{eq:Qvis})\cite{Shakura1973}, and $H$ is the half-height of the disk. 
Here only the modified energy equation (Eq.\,\ref{eq:extra}) is listed for discussions, and a complete set of \sSMBH\ equations is given by Ref.\cite{Gilbaum2022}.
$Q_{\rm extra}$ alters the radial distribution of the effective temperature.
Specifically, when $Q_{\rm extra}$ is sufficiently large relative to $Q_{\rm vis}$, $T_{\rm eff}$ distribution becomes notably flatter than the well-known power law of $T_{\rm eff}\propto R^{-3/4}$.
Recent studies have shown that irradiated photons are efficiently thermalized locally within the disk\cite{Secunda2024,Secunda2025} and are re-radiated instantaneously at characteristic energies of their local environment. 
With $Q_{\rm extra}$, both the ionizing photon field and the 5100\,\AA-emitting regions ($R_{5100}^{\bullet}$) are significantly extended, resulting in $R_{5100}^{\bullet}\gg R_{5100}$. In contrast, the broad-line region (BLR) is much less affected by $Q_{\rm extra}$.
This leads to a new structural configuration where $R_{5100}^{\,\bullet}>R_{\rm H\beta}$ (with $R_{\rm H\beta}$ being the BLR radius), corresponding to H$\beta$ reverberating with a leading delay of $\tauhb^{\,\ell}\approx \left(\Rhb-R_{5100}^{\,\bullet}\right)\sin\,\!i/c<0$, breaking down the PSA. 
Here $i$ is inclinations of the SMBH-disk.

\begin{figure*}
\centering
\includegraphics[trim=200 150 60 10, clip,width=1.1\textwidth]{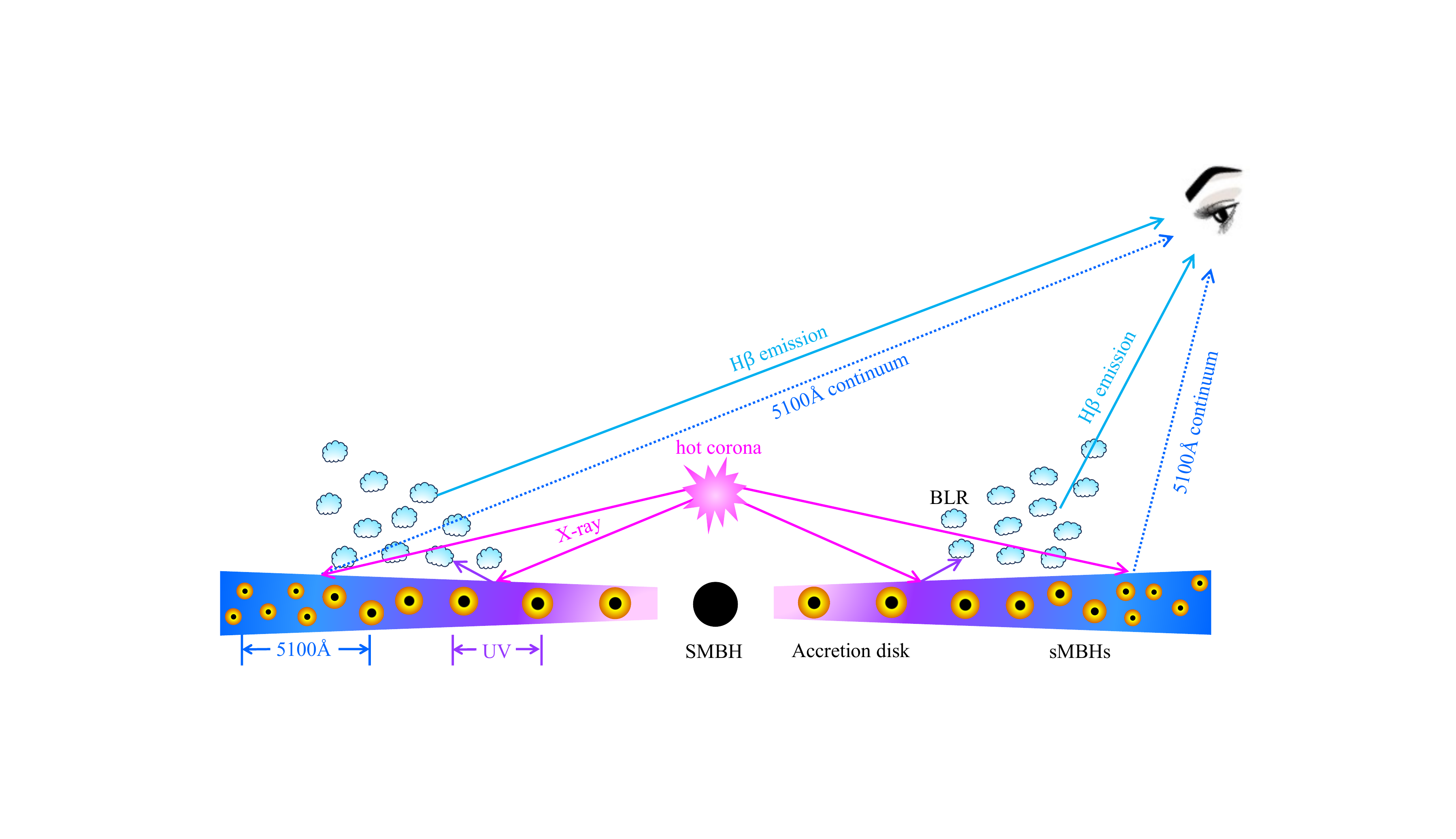}
\caption{\footnotesize 
This diagram illustrates the SMBH-disk with embedded stellar-mass black holes (sMBHs) — a system denoted as \sSMBH—proposed to explain the observed phenomena in PHL\,1092.
The sMBHs are represented as black dots surrounded by gradient yellow cocoons.
The total energy budget of the SMBH-disk must include energy contributions from the sMBH accretion altering the radial distribution of the disk’s effective temperature ($T_{\rm eff}$).
This gives rise to a large extension of the 5100\,\AA\ regions so that the BLR clouds (in shallow blue) are located within the former, appearing the observed leading delays of broad H$\beta$ line reverberations. 
The sMBHs are growing with inward migration but their number density decreases because of hierarchical mergers. 
}
\label{fig:model}
\end{figure*}

The $Q_{\rm extra}$ is suggested to originate  from accreting sMBHs (previously denoted as accretion-modified stars as AMS\cite{Wang2021a}) embedded in the SMBH-disk (hereafter refer to \sSMBH)\cite{Cheng1999,Tagawa2023,Wang2021a,Gilbaum2022,Zhou2024}.
Fig.\,\ref{fig:model} illustrates the new scenario for the \sSMBH\ and BLRs in PHL\,1092.
Actually, $Q_{\rm extra}$ depends on several factors\cite{Gilbaum2022}, including the total number of sMBHs, their spatial distribution across the disk, and their mass function.
The total numbers may be inferred from BLR metallicity according to stellar evolution\cite{Woosley2002}. 
Except for the alterations of $T_{\rm eff}$, $Q_{\rm extra}$ also modifies the albedo of the disk surface -- enhancing the reflection of X-rays from the corona.
Both the reflected and reprocessed components in the SEDs are regulated by $Q_{\rm extra}$.
Consequently, the $Q_{\rm extra}$-extended SMBH-disk is expected to be responsible for the observed LDRs in PHL\,1092.

There is growing evidence supporting the existence of \sSMBH, which is from optical variability in quasars\cite{Graham2020,Graham2023,He2025} and complemented by LIGO detections of gravitational waves (GWs) from events such as GW\,190521\cite{Abbott2020,Graham2020} and GW\,231123\cite{GW231123}.
These two GW events imply that their progenitors required a dense environment—specifically, AGN accretion disks—for formation. 
This is because such events cannot be explained by normal stellar evolution\cite{Woosley2002}; instead, they likely form via either rapid growth through accretion\cite{Graham2020,Samsing2022} or hierarchical mergers\cite{Yang2021,Mapelli2022,Vaccaro2025}.
Recent discoveries further reinforce this picture: the detection of millihertz (mHz) quasi-periodic oscillations (QPOs) in the actively accreting SMBH of 1ES\,1927+654 supports the presence of satellite black holes orbiting the central SMBH\cite{Masterson2025}, as do newly identified quasi-periodic eruptions (QPEs)\cite{Miniutti2019,Giustini2020,Arcodia2021,Chakraborty2021,Arcordia2024} associated with tidal disruption events\cite{Jiang2025}.
This body of evidence is consistent with our detection of leading delays of H$\beta$ reverberations in PHL\,1092.
Other potential models for explaining LDRs are briefly discussed in \S\,\ref{sec:other-models}, including 3D winds acting as the broad-line region (BLR), anisotropic radiation from the accretion disk, extended winds from super-Eddington accretion, and even complex turbulence differing from the standard Shakura-Sunyaev model. 
None of these models can fully explain the observed LDRs, and those that come close require highly specific configurations or conditions. 
We therefore favor AMS-driven heating as a nascent and parsimonious mechanism for generating LDRs -- one that is supported by observational hints\cite{Graham2020,Graham2023,Abbott2020,GW231123}.

Our monitoring campaign of PHL\,1092 reports the first detection of leading delays in the reverberations of the broad H$\beta$ line relative to the 5100\,\AA\ continuum. 
This finding indicates that the optical continuum-emitting regions extend beyond the BLRs.
This structural configuration deviates from the standard AGN model and fundamentally undermines the conventional BLR-based methodology for estimating SMBH masses.
Furthermore, there are a small number of PHL\,1092 analogs from our SEAMBH campaign, though these objects exhibit slightly more complex behaviors (see Issue 3 in \S\,\ref{sec:question}). 
This suggests that such systems are neither common nor rare.
The LDRs suggesting EESs powered by accretion of the sMBHs widely distributed over the SMBH-disk, provide compelling observational evidence for the existence of \sSMBH. 
If such accretion disks are ubiquitous in AGNs but vary in the number of embedded sMBHs, the shortening of $\tau_{\rm H\beta}$ should extend continuously to the regime of leading delays.
We thus expect the analogs to fill the gap between PHL\,1092 and the $R_{\rm H\beta}-(L,\calR)$ relation in Fig.\,\ref{fig:RL-Plane}.
Additionally, advancements in optical and near-infrared interferometry, such as the future Kilometers Baseline Interferometers\cite{Bourdarot2024}, are anticipated to directly image \sSMBH\ systems. 
Issue 6 in \S\,\ref{sec:question} briefly discusses the evolution of sMBHs inside the SMBH-disk for observational diverse properties in light of variations of broad emission lines and continuum.
Complementary to this, GW detections from nearby AGNs via missions such as LISA\cite{Amaro-Seoane2023}/Tianqin\cite{Luo2025}, the Einstein Telescope (ET)\cite{DiGiovanni2025}, or space-based detection networks\cite{Ruan2020} could further confirm the presence of these systems.

 
\begin{addendum}

\item[Correspondence] 
Correspondence and requests for materials
should be addressed to Jian-Min Wang (email: wangjm@ihep.ac.cn).

\item { The authors are grateful to three referees for their helpful reports clarifying ambiguous points and motivating us to think about the potential explanations of the LDRs.
This research is supported by grants NSFC-12333003, -92476203, -12521005, -12573016, the National Key R\&D Program of China (2021YFA1600404,  2023YFA1607903, and 2023YFA1607904), the National Science and Technology Major Project (2024ZD0300303), the China Manned Space Project (CMS-CSST-2025-A07), and the Strategic Priority Research Program of the Chinese Academy of Sciences (XDB1160202).}

\item[Author Contributions] 
JMW is the Principle Investigator of the SEAMBH project. 
He discovered the LDRs, made the theoretical model of the LDRs and wrote the manuscript. CH, YJC, YYSS, YLW, HZ, WJG, SY, ZHY and JA made observations, and CH and YJC reduced data. YJC, YRL and JMW made all LC fittings through {\tt{MICA}}. YYSS and JMW made Monte Carlo simulations on inhomogeneity of ionizing variations of extended sources, and found the non-concordant reverberations. PD and JMW estimated \feii\ contamination to H$\beta$ line.
All the authors discussed the contents of the paper.

\item[Competing Interests] The authors declare that they have no competing financial interests.

\end{addendum}

\clearpage

\clearpage

\clearpage

\begin{table}
\centering
\begin{threeparttable}
\caption{Information of PHL\,1092 and observations}\label{tab:PHL1092}
\vglue 0.15cm
{\rm\footnotesize
\begin{tabular}{cccccccccc}\hline\hline
$\alpha_{2000}$&$\delta_{2000}$&redshift&$M_{r}$&$L_{5100}$&$\RFe$ & FWHM &$N_{\rm sp}$&
\muc{2}{c}{Comparison stars}\\ \cline{9-10}
& & & & ($\ergs$)& & ($\kms$) &  &$R_*$ & P.A.          \\ \hline
01 39 55.76    &$+\,$06 19 22.55 &0.396   & 16.8&$1.26\times 10^{45}$ &2.7& 2270 &157 &
$129^{\pp}.9$&200.6$^{\circ}$\\ \hline
\end{tabular}
\begin{tablenotes}
\item {\footnotesize $M_{r}$ is the $r$-band magnitude.}
\item {\footnotesize $N_{\rm sp}$ is the numbers of spectroscopic observing epochs.}
\item {\footnotesize $R_*$ is the angular distance to the targets and P.A. is the position angle.}
\end{tablenotes}
}
\end{threeparttable}
\end{table}

\begin{table}
\centering
\begin{threeparttable}
\caption{H$\beta$ reverberation results of 8-year campaign of PHL\,1092}\label{tab:lags}
\vglue 0.15cm
{\rm\footnotesize
\begin{tabular}{lccccccccc}\hline\hline
Duration& \muc{2}{c}{Variability amplitudes*}&  &  \multicolumn{4}{c}{Time lags ($\tauhb^{\,\ell}$)}\\ \cline{2-3} \cline{5-9}
  & $F_{\rm var}$(5100\,\AA)&$F_{\rm var}({\rm H\beta})$  & & \JAVELIN &\MICA & Mean& $p$ & Significance\\
  & (\%) & (\%) & & (day) &(day) &(day) & & ($\sigma$)\\
\hline
2017.11$-$2025.02 & $4.36\pm1.00$ & $4.19\pm0.58$ & & $-20.1_{-7.5}^{+5.3}$ & $-13.5_{-9.6}^{+13.5}$ & $-16.8_{-8.6}^{+9.4}$& $5\times 10^{-5}$ & 4.1\\
2020.07$-$2024.02 & $5.35\pm1.77$ & $3.59\pm0.8$ & & $-54.1_{-60.3}^{+32.5}$ & $-57.1_{-31.9}^{+30.5}$ & $-55.6_{-46.1}^{+31.5}$& $2\times 10^{-4}$ & 3.7\\
2020.07$-$2025.02 & $5.13\pm1.42$ & $4.58\pm0.74$ &  & $-20.1_{-7.5}^{+5.3}$ & $-19.6_{-13.1}^{+11.6}$ &$-19.9_{-10.3}^{+8.5}$ & $ 2\times 10^{-4}$ & 3.7\\
2023.07$-$2025.02 & $4.82\pm1.70$ & $6.03\pm0.94$ &   & $-45.0_{-9.7}^{+80.7}$ &$-32.2_{-18.7}^{+28.1}$ & $-38.6^{+54.4}_{-14.2}$& $4\times 10^{-4}$& 3.4\\ \hline
\end{tabular}
\begin{tablenotes}
\item {\footnotesize $F_{\rm var}$: is calculated using Eq.\,(\ref{eq:Fvar}) and uncertainties according to Ref.\cite{Edelson2002}.}
\item {\footnotesize Mean: is the arithmetic mean values of \JAVELIN\ and \MICA\ measurements.}
\item {\footnotesize $p$: is the LDR probability due to aliasing effects and the according significance. We generated $2\times10^4$ pairs of LCs for 2017.11$-$2025.02, 5000 pairs for other periods.}
\end{tablenotes}
}
\end{threeparttable}
\end{table}

\clearpage
\begin{methods}
\section{Overview of RM AGNs and conventional RM physics}\label{sec:overview}
\subsection{1.1 Campaigns}
AGNs are prominently characterized by broad emission lines\cite{VandenBerk2001} (see extensive review in the textbook\cite{Netzer2013}) as well as in high$-z$ quasars\cite{Mortlock2011,Banados2018}. 
BLRs, which are photoionized by radiation from the SMBH-disk, are filling with numerous discrete clouds that emit emission lines.
The lines are broaden by orbiting motion around the central SMBH, but they are spatially unresolved so far except for a few AGNs observed by GRAVITY/VLTI along with spectroastronomy (note GRAVITY results are not spatially resolved images of BLRs)\cite{Amorim2024}.
Fortunately, the BLR clouds reverberate in response to continuum variability of the SMBH-disk.
The resulting echoes of broad emission lines convey invaluable information about the spatial distributions and kinematics of the BLRs\cite{Bahcall1972,Blandford1982,Peterson1993} enabling the measurement of SMBH masses through RM campaigns\cite{Peterson2014}. 
Since the 1980s, astronomers have initiated RM campaigns to probe the kinematics and structure of BLRs (see a classical review on the RM technique\cite{Peterson1993}). 
To date, approximately $\sim 300\,$AGNs have been studied, which are contributed from several major programs.
According to methods of spectral calibrations, we list them:
\begin{itemize}
\item RM campaigns with calibrations through comparison stars
\begin{itemize}
\item Steward PG program at Arizona (only 17 objects)\cite{Kaspi2000} with the Bok\,2.3\,m telescope,
\item Super-Eddington Accreting Massive Black Holes (SEAMBHs) at Yunnan Observatory and Centro Astronómico Hispano-Alemán (CAHA) Observatory (SEAMBHs: $\sim 60$)\cite{Du2015,Du2016a,Du2019,Hu2021,Hu2022,Hu2025}  with the Lijiang 2.4\,m and the CAHA\,2.2\,m telescopes,
\end{itemize}
\end{itemize}
\begin{itemize}
\item RM campaigns with calibrations through \oiii\
\begin{itemize}
\item AGN Watch (10 objects)\footnote{\url{https://www.asc.ohio-state.edu/astronomy/agnwatch/}}, Ohio State Project ($\sim 25$)\cite{Peterson1998,Denney2010,Grier2012,Fausnaugh2017} with several 1.5\,m and 2\,m telescopes, 
\item Lick AGN Monitoring Project (LAMP: $\sim 60$)\cite{Bentz2009,Barth2015,U2022} with the Shane 3.0\,m telescope,
\item Monitoring AGNs with H$\beta$ Asymmetry (MAHA) projects at Wyoming Infrared Observatory ($\sim 35$)\cite{Du2018,Bao2022,Zastrocky2024} with the WIRO 2.3\,m telescope,
\item the Seoul National University AGN Monitoring Project ($\sim 30$)\cite{Woo2024} with the Shane 3.0\,m telescope
\item and multiple fiber-fed spectrograph campaigns: Oz-DES Project ($\sim 8$)\cite{Malik2023} and SDSS-RM Project ($\sim 80$)\cite{Shen2024} with 4\,m and 2.5\,m telescopes, respectively. 
\end{itemize}
\end{itemize}
In Fig.\,\ref{fig:RL-Plane}, only objects with subtraction of host contamination are plotted.
These RM campaigns have greatly advanced the understanding of BLRs.

SEAMBH campaigns, focusing on AGNs with high Eddington ratios, have been conducted using the CAHA\,2.2\,m and Lijiang\,2.4\,m telescopes.
Target AGNs are selected primarily based on two key features: high values of $\calR$ and high accretion rates $\mathdotM$ (see Eq.\,\ref{eq:mdot}).
For discussions of PHL\,1092 analogs, refer to Issue 3 in \S\,\ref{sec:question}.
Phenomena of leading delays in H$\beta$ reverberations in AGNs are neither common nor extremely rare.
Despite of these objects we are not able to make a significant statistics of the phenomenon from this heterogeneous sample.

To date, over one million quasars have been identified\cite{Flesch2023}.
Although most of them have been photometrically monitored by ZTF\cite{Nakoneczny2025}, however, only about a few $10^{-4}$ of quasars have been mapped by campaigns, moreover no more than 20 AGNs and quasars have been spectroscopically monitored continually with high cadences longer than 7-8\,years (except for NGC\,5548\cite{Lu2022}).
Ref.\cite{Kaspi2000} performed the 7 year campaigns of 17 PG\,quasars, but cadences of most targets are quite poor.
As a result, the understanding of the inner structures of AGNs is extremely limited.
Long term campaigns of RM with high cadence through 2\,m telescopes are necessary to reveal anomalous reverberations of H$\beta$ line as we briefly discuss PHL\,1092 analogs in \S\,\ref{sec:question}.
So far, the SEAMBH campaigns are the longest RM campaigns (some of them are 8-9 years) with high and homogeneous cadence then.
In the meanwhile, large telescopes are also necessary to explore details of spatially inhomogeneous variations of ionizing sources in these analogs.
Although the campaigns are very costly in terms of observation time and human resources, it provides significant scientific returns for us revealing not only various versions of inner structures of AGNs beyond their standard model but also gravitational waves as fundamental physics.
It is just for 2\,m telescopes, we would like to emphasize here, that the significant goals can be achieved by discovering diverse behaviors of H$\beta$ line reverberations in AGNs.

The validation of significant reverberations typically relies on the maximum cross-correlation function (CCF) coefficient, denoted as $r_{\rm max}$.
In practice, $r_{\rm max}$ is a complex function influenced by multiple factors: the transfer function of the system, the observational cadence and its uniformity, seasonal gaps in data, and the total number of observational points.
While no strict universal criterion exists for validating reverberations, a common threshold adopted in the literature is $r_{\rm max}\gtrsim 0.5$\cite{Bentz2009}.
To contextualize this threshold, we checked $r_{\rm max}$ values in several popular studies\cite{Kaspi2000,Peterson1998,Bentz2009,U2022}. 
Notably, even datasets with suboptimal cadence have been shown to yield $r_{\rm max}\sim 0.8-0.9$ provided the total number of data points is not many.
For the present work, the measured $r_{\rm max}\sim 0.7$ is well above the typical acceptance threshold. 
This value supports a robust detection of reverberations, consistent with the standards established in the aforementioned references.

\subsection{1.2 Standard physics of reverberations}
The standard AGN model consists of three key components: the accretion disk, the BLR, and the dusty torus\cite{Netzer2013}.
In general, the radiation-dominated region of the accretion disk is significantly smaller than the BLR. 
Additionally, results from infrared RM campaigns indicate $R_{\rm H\beta}\approx (0.1\sim 0.2)\,R_{\rm torus}$, where $R_{\rm torus}$ denotes the inner radius of the dusty torus\cite{Minezaki2019}.
These three components may represent distinct regions (each governed by unique physical processes) of a continuous gaseous disk -- permeated with clumps -- that extends from the inner to outer scales of the AGNs\cite{Begelman1989}. 
Alternatively, they could be physically linked through the tidal capture of clumps\cite{Wang2017,Zhou2019}.

A robust relation between the BLR size and optical luminosity has been established for sub-Eddington accreting AGNs\cite{Kaspi2000,Bentz2013,U2022,Woo2024}, but super-Eddington AGNs have shortened H$\beta$ lags depending on accretion rates compared with the same luminosity AGNs\cite{Du2014,Du2015,Du2018,Du2019}.
We define the dimensionless accretion rate as $\mathdotM=\dot{M}_{\bullet}/\dot{M}_{\rm Edd}$, where $\dot{M}_{\bullet}$ is mass accretion rates of the central SMBH, $\dot{M}_{\rm Edd}=L_{\rm Edd}/c^2$ is the critical rates corresponding to the Eddington luminosity of $L_{\rm Edd}=4\pi G\mbh c/\kappa_{\rm es}$, and $\kappa_{\rm es}$ is the opacity of electron scatterings.
SEAMBHs are defined as $\mathdotM\gtrsim 3$ above which the inner structure of accretion disks can no longer be adequately described by the Shakura-Sunyaev model\cite{Laor1989}.
It is well understood that $\calR$ is a good proxy of accretion rates\cite{Boroson1992} (see also Fig.\,4 in Ref.\cite{Marziani2025}). 
Super-Eddington AGNs occupy about 1/3 of PG quasars known as narrow line Seyfert 1 galaxies, and similar fractions of SEAMBHs in SDSS quasars was found in light of $\calR$\cite{Du2016}.
The BLR scaling relation is expressed as 
\begin{subequations}
\begin{numcases}{\frac{\Rhb}{\Rg}= }
5.8\times 10^4\,\ell_{44}^{0.53}M_7^{-1}\quad {\rm (for~\mathdotM\le\mathdotM_{\rm C})},\label{Eq:R5100a}\\
7.7\times 10^4\,\ell_{44}^{0.45}M_7^{-1}10^{-0.35\RFe}
\quad {\rm (for~ \mathdotM>\mathdotM_{\rm C})},\label{Eq:RLdotMslim}
\end{numcases} 
\label{Eq:RLdotM}
\end{subequations}
where $\mathdotM_{\rm C}\approx11$ is the critical rates, $\Rg=G\mbh/c^2=1.5\times 10^{12}M_7\,$cm is the gravitational radius, $G$ is the gravitational constant and $M_7=\mbh/10^7\,\sunm$ is the SMBH mass. 
This indicates that the BLR shrinks with increases of $\mathdotM$, and sometimes shortened by factors up to $\sim 8$ (and $\mathdotM\sim 900$) so 
far\cite{Du2018} from the SEAMBH campaigns. 

Comparing with the disk radius (Eq.\,\ref{eq:R5100SS}), we find that $R_{5100}\ll \Rhb$, namely, the PSA is valid in sub-Eddington AGNs, underpinning the widely used methodology for estimating SMBH masses
\begin{equation}\label{eq:Mass}
\mbh=\fblr\, \frac{R_{\rm H\beta}V_{\rm H\beta}^2}{G}\approx 2.0\times 10^6\,\tau_{10}V_{3}^2\,\sunm,
\end{equation}
where $\fblr$ is the virial
factor ($\fblr=1$ is taken for SEAMBHs\cite{Du2014,Yang2024}), $V_{\rm H\beta}$ is the full-width-half-maximum (FWHM) of \Hbeta\ profiles, $\tau_{10}=\tauhb/10\,$days and $V_{3}=V_{\rm H\beta}/10^3\,\kms$.
The dimensionless accretion rates can be expressed by
\begin{equation}\label{eq:mdot}
\mathdotM\approx 20.1\left(\frac{\ell_{44}}{\cos i}\right)^{3/2}M_7^{-2},
\end{equation}
in light of the Shakura-Sunyaev model\cite{Du2015,Shakura1973}.
In addition to the assumption that motion of clouds is virialized, Eq.\,(\ref{eq:Mass}) also strongly depends on the PSA, specifically, the condition $R_{5100}\ll \Rhb$. However, this approximation is not universally valid, in particular, for extreme super-Eddington accreting massive black holes (SEAMBHs), $R_{5100}$ can become comparable to the size of the BLR, violating the PSA.
This caveat should be borne in mind.

\subsection{1.3 Non-point source effects}
When the PSA breaks down, what reverberation behaviors emerge? 
This question is addressed using toy models of extended ionizing sources in \S\,\ref{sec:toymodel}.
In this model, the 5100\,{\AA}/ionizing sources regions are so extensive that the reverberations of broad H$\beta$ line emissions arise from twice convolutions of independent transfer functions of the BLR and continuum regions [denoted by $\Psi_{\rm H\beta}$ and $\Psi_{\rm ion}$ in Eq.\,(\ref{eq:LHbeta}), respectively].

This double convolution causes the H$\beta$ line LC to exhibit non-concordant behaviors, reflecting composite responses from different regions of the BLR. 
In other words, the H$\beta$ line does not reverberate straightforwardly with respect to the continuum. 
Notably, spatial inhomogeneities of the ionizing source components further complicate these behaviors.
These processes not only cover the smooth effects of BLR geometries but also the effects of continuum geometry and spatial distributions.
For more realistic models, reverberation behaviors are expected to be even more complex.
This phenomena of non-concordant behaviors resulting from ionizing photon propagation of non-point source (NPS) to the extensive BLR regions are referred as the NPS effect in this paper.
Details are provided in \S\,\ref{sec:toymodel}.

\subsection{1.4 Anomalous behaviors of reverberations}
While most AGNs studied via RM exhibit regular reverberation behavior, anomalous reverberations have been detected in a small number of objects.
For instance, the AGN STORM campaign of NGC 5548 revealed that the broad H$\beta$ line failed to reverberate in response to 5100\,\AA\ continuum variations over a $\sim 50\,$day period\cite{Pei2017}.
A similar phenomenon was observed in PG\,1613+658 during the time interval JD\,2457000 + (1750–2000) as part of the MAHA campaign\cite{Bao2022}.
Such anomalous reverberations may arise from inhomogeneities in either the broad-line region (BLR) or the ionizing source.
Furthermore, NGC\,5548 follows a scaling relation of $R_{\rm H\beta}\propto L_{5100}^{0.8{-}0.9}$ —with substantially larger scatter and a significantly steeper slope compared to the canonical $R_{\rm H\beta}\propto L_{5100}^{0.5}$ relation established over the past three decades of RM campaigns\cite{Lu2022}.
Mrk\,817\cite{Homayouni2024} and NGC\,7469\cite{Peterson2014a} have also displayed anomalous reverberation behaviors (potentially attributable to the NPS effects discussed in \S\,\ref{sec:toymodel}). 
However, the conventional BLR-based method for SMBH mass estimation still yields valid results for these objects.
The reverberation behavior of PHL\,1092 differs fundamentally from all the aforementioned cases.

\section{Observations and data reduction}\label{sec:observations}
\subsection{2.1 Observations}
The 2.2\,m telescope at the CAHA Observatory was used for the monitoring campaigns.
Detailed procedures are provided in Ref.\cite{Hu2021,Hu2025}, however, a brief overview is presented here for the reader’s convenience.
Each night, broadband images of the target (PHL\,1092) were first acquired. 
A long slit—with a projected width of $3.^{\prime\prime}0$ — was then carefully aligned, and spectra of both the target and a nearby non-variable comparison star were obtained simultaneously.
To minimize slit-induced light loss (critical for high-precision flux calibration), the target and the comparison star were centered in the slit with typical positional accuracy: the offset from the slit center along the width direction was less than 0.5 pixels (corresponding to a projected offset of 0.$^{\farcs}$265).
Observational details for the target and the comparison star are given in Tab.\,\ref{tab:PHL1092}.

Three 120-second exposures of Johnson $V$-band images were typically acquired per night. 
%
%
Image reduction followed standard IRAF procedures. Fluxes of the target and the comparison star were measured via a circular aperture ($2.^{\farcs}65$), with differential magnitudes calculated relative to 7 other stars in the $9.^{\prime}7\times9.^{\prime}7$ field.

PHL\,1092 spectra were taken with the Calar Alto Faint Object Spectrograph (CAFOS) using Grism G-200 for its relatively high efficiency.
The resulting spectrum covers 4000\,$-$\,8500\,\AA, with a dispersion of 4.47\,\AA\,pixel$^{-1}$. 
Two 1800-second exposures were taken per night, and the typical signal-to-noise ratio (S/N) per pixel for a single exposure is $\sim 40$ at rest-frame 5100\,\AA. 
One or two spectrophotometric standards were observed each night during dusk and/or dawn, weather and time permitting.

\subsection{2.2 Data reduction} 
Spectra were reduced in IRAF via standard procedures: bias removal, flat-fielding, wavelength calibration, and 1D extraction. Extraction apertures were uniformly large (10.$^{\farcs}$6) to minimize light loss, particularly on nights with poor seeing. 
The target’s flux calibration used a sensitivity function derived from the comparison star, as detailed below.

First, on several clear nights, the comparison star was flux-calibrated via standard IRAF procedures using the night’s spectrophotometric standards. 
Second, these calibrated spectra were combined to produce a fiducial comparison star spectrum. 
Third, for each exposure, a Legendre polynomial was fitted by comparing the comparison star’s extracted count spectrum to the fiducial spectrum. %
This polynomial acts as the sensitivity function, accounting for atmospheric effects, slit loss, and instrument response. 
Telluric absorption was corrected using the comparison star as a telluric standard\cite{Kaspi2000,Lu2021}.
Finally, each spectrum of the target was flux-calibrated with the corresponding sensitivity function; the two nightly calibrated spectra were combined to form a single-night spectrum for subsequent measurements and analysis.

Flux calibration accuracy is better than 3\% in Ref.\cite{Hu2021}.
The comparison star’s $V$-band LC is shown in Fig.\,\ref{fig:seeing_cmp_lc}a.  
Its differential magnitude scatter is only 0.013 mag, far smaller than the target’s variations which ensures spectroscopic flux calibration accuracy.

\subsection{2.3 Seeing influence}
From $V$-band images, we derived FWHMs of surface brightness distributions for the target and the comparison star (Fig. \ref{fig:seeing_cmp_lc}{\textbf{b}} and \ref{fig:seeing_cmp_lc}{\textbf{c}}). 
Comparison star FWHMs serve as seeing measurements. 
Target FWHMs are nearly identical to that of the comparison star, indicating negligible host starlight contribution. 
Notably, continuum and \hb\ LC deviations from DRW fits (Fig. \ref{fig:Gauss}) show no correlation with seeing—confirming spectroscopic flux calibration is unaffected by seeing variations (due to our broad slit width and minimal host contribution). 
Deviations are driven by calibration accuracy.

\subsection{2.4 Light curve measurements}
Following the traditional method\cite{Kaspi2000,Peterson2004}, we obtain the broad H$\beta$ line and continuum LCs through integrating over a range of relevant wavelengths.
The window of 4810\,--\,4900\,\AA\ (in the rest frame) was set for the \hb\ line. 
A straight line was defined by the two continuum windows located on both sides of the \hb\ (4680\,--\,4710\,\AA\ and 5070\,--\,5100\,\AA, chosen to have minimum contribution of the \feii\ line emissions), and then the flux above this straight line was integrated in the emission-line window. 
The average flux in the redward continuum window was calculated as the integrated 5100\,\AA\ flux (\fcs). 
Fig.\,\ref{fig:MICA-lags} shows the LCs of different periods.
 
The \feii\ LC, shown in Fig.\,\ref{fig:LC_Fe}, is also measured by the integration method similar to Ref.\cite{Barth2015}, as done for the \hb\ line. 
The windows of 3795\,--\,3810\,\AA\ and 4690\,--\,4705\,\AA\ are used for defining the continuum, and the window of 4434\,--\,4584\,\AA\ is set for integrating the \feii\ flux (all three windows are in the rest frame). 
Note that this integration window of \feii\ is narrower than the window of 4434\,--\,4684\,\AA\ used for defining the \feii\ flux for calculating the $\RFe$, where the 4584\,--\,4684\,\AA\ window is masked here avoiding the possible influence of \heii\ emission line.

Spectral fitting method is employed to obtain LCs for campaigns with high spectral resolution spectrographs. 
It is necessary for the lines which blended with others (such as \heii), or in the cases that the host starlight is strong to contaminate the AGN continuum (the host is weak in PHL\,1092). 
However, it will introduce extra uncertainties by the imperfect modeling\cite{Hu2021}. 
We did this excise though the CAFOS resolution is only $\lambda/\Delta\lambda\sim 300$ since we took the slit width of $3^{\farcs}$ (see \S\,2.1).
The results yielded by the spectral fitting is consistent with that of the integrations but with larger uncertainties in some degrees.  

\subsection{2.5 Error bars}
Observed error bars consist of two parts $\delta_{\rm obs}=\sqrt{\delta_{\rm sys}^2+\delta_{\rm meas}^2}$, where $\delta_{\rm sys}$ and $\delta_{\rm meas}$ are the systematic and measured error bars, respectively.
Here we adopt two methods to estimate the systematic uncertainties, namely, the median filter and \MICA\ fit.
Following the median-filter approach\cite{Peterson1998,Bentz2009}, we smooth each LC with a five-point sliding median to capture the long-term trend. 
Subtracting this median-smoothed curve from the original, we compute the standard deviation of the residuals $\sigma_{\rm Med}$. 
The smoothed curve and residuals are shown in Fig.\,\ref{fig:Gauss}. 
We adopt $\sigma_{\rm Med}=\delta_{\rm sys}$ as an epoch-independent systematic uncertainty. 
Across our eight-year data set, the resulting systematic uncertainties for the continuum and H$\beta$ line are approximately 1.2\% and 2.7\% of the respective fluxes, consistent with the calibration accuracy being better than 3\%.

For the \MICA\ fitting, $\delta_{\rm sys}$ is treated as a free parameter during the LC modeling\footnote{More details of \MICA\ can be found in this URL: \url{https://mica2.readthedocs.io/en/latest/}.}. 
Applying to our data, we obtain the systematic uncertainties of about $1.3\%$ and $3.8\%$ for the continuum and H$\beta$, respectively. 
The {\MICA}-derived $\delta_{\rm sys}$ is slightly larger than the median-filter estimate because the long-term variability patterns of the continuum and H$\beta$ are not perfectly matched (eg, MJD 2250 in panel \textbf{a} of Fig.\,\ref{fig:MICA-lags}, which are caused by non-concordant reverberations of spatially inhomogeneous variations of extended ionizing sources). 
To achieve a statistically consistent fit (reduced $\chi^2 \approx 1$), \MICA\ necessarily inflates the systematic term. 
Considering these limitations, we adopt the median-filter estimate as our fiducial $\delta_{\rm sys}$ for LCs.

\subsection{2.6 Variations}
We use the popular methods\cite{Rodriguez1997} to calculate LC characteristics of the 
continuum and the H$\beta$ emission line. They are
\begin{equation}\label{eq:Fvar}
F_{\rm var}=\frac{\left(\sigma^2-\Delta^2\right)^{1/2}}{\langle F \rangle},
\end{equation}
where 
\begin{equation}
\sigma^2=\sum_{i=1}^{N} \frac{\left(F_i-\langle F \rangle\right)^2}{(N-1)},\quad
\Delta^2=\sum_{i=1}^{N} \frac{\Delta_i^2}{N},\quad
\langle F\rangle=\sum_{i=1}^N\frac{F_i}{N},
\end{equation}
and the error bar of $F_{\rm var}$ is given by\cite{Edelson2002}
\begin{equation}
\sigma_{\rm var}=\frac{1}{F_{\rm var}} \left(\frac{1}{2 \times N}\right)^{1/2}\frac{\sigma^2}{{\langle F \rangle}^2},
\end{equation}
where 
$N$ is the total number of data, $F_i$ is the flux of the $i$-th observation, and $\Delta_i$ is the 
uncertainty of $F_i$. We denoted $F_{\rm var}^{5100}$, $F_{\rm var}^{\rm combine}$ and 
$F_{\rm var}^{\rm H \beta}$ as variation amplitudes of the LCs for the 5100\,\AA\ continuum, 
the combined continuum and broad H$\beta$ line.

\subsection{2.7 Amplitudes of Fe\,{\sc ii} variations}
Considering difficulties that \feii\ variations are too small to measure, we estimate an intrinsic relation of \feii\ from H$\beta$ variations as followings. 
If the BLRs are virialized in the central SMBH potential, the \feii\ regions are $R_{\rm Fe}\approx \left(V_{\rm H\beta}/V_{\rm Fe}\right)^2R_{\rm H\beta}$,  where $V_{\rm H\beta}$ and $V_{\rm Fe}$ are the FWHMs of H$\beta$ and \feii\ lines. 
The relative variations of H$\beta$ line are given by $\delta F_{\rm H\beta}=\Delta F_{\rm H\beta}/\langle F_{\rm H\beta}\rangle$, where $F_{\rm H\beta}$ and $\Delta F_{\rm H\beta}$ are the steady/mean fluxes of the line and its variations, respectively.
Observationally, we take $\delta F_{\rm H\beta}=\left(F_{\rm max}-F_{\rm min}\right)/\langle F_{\rm H\beta}\rangle\approx 3\sigma$, $F_{\rm max}$ and $F_{\rm min}$ are the maximum and minimum  of H$\beta$ line variations (corresponding to $1.5\,\sigma\,$deviations from the mean values of fluxes), respectively.

According to photoionization of the BLR, we have $\Delta F_{\rm H\beta}=\Delta F_{\rm ion,H\beta}\left(\Delta t_{\rm ion}/\Delta t_{\rm H\beta}\right)$, where $\Delta F_{\rm ion,H\beta}$ is the variations of the H$\beta$ ionizing fluxes, and $(\Delta t_{\rm ion}, \Delta t_{\rm H\beta})$ are the variation timescales of the ionizing sources and H$\beta$ line, respectively. 
We have 
\begin{equation}
 \delta F_{\rm H\beta}=\left(\frac{\Delta F_{\rm ion, H\beta}}{F_{\rm H\beta}}\right) \left(\frac{\Delta t_{\rm ion}}{\Delta t_{\rm H\beta}}\right).      
\end{equation}
In light of the lamp-post model, the ionizing photons are the reprocessed emissions of X-ray photons by the disk surface. 
Given $\xi_{\rm X}$ is the reprocessing coefficient, we have $F_{\rm H\beta}=\xi_{\rm X}L_{\rm X}/4\pi R_{\rm H\beta}^2c$ and $\Delta F_{\rm ion,H\beta}=\xi_{\rm X}\Delta L_{\rm X}/4\pi R_{\rm H\beta}^2c$, yielding $\Delta F_{\rm ion,H\beta}/F_{\rm H\beta}=\Delta L_{\rm X}/L_{\rm X}$, where $L_{\rm X}$ and $\Delta L_{\rm X}$ are X-ray luminosity and its variations, respectively.
Similarly, we have the relative variations of the \feii,
\begin{equation}
 \delta F_{\rm Fe}=\left(\frac{\Delta F_{\rm ion, Fe}}{F_{\rm Fe}}\right) \left(\frac{\Delta t_{\rm ion}}{\Delta t_{\rm Fe}}\right)=\left(\frac{\Delta t_{\rm H\beta}}{\Delta t_{\rm Fe}}\right)\delta F_{\rm H\beta},
\end{equation}
where we use a similar relation of $\Delta F_{\rm ion,Fe}/F_{\rm Fe}=\Delta L_{\rm X}/L_{\rm X}$.
This shows the relations between H$\beta$ and \feii\ variations in the framework of the lamp-post model.
Considering that the $(\Delta t_{\rm H\beta},\Delta t_{\rm Fe})\approx (R_{\rm H\beta},R_{\rm Fe})/c$ and the virialization relations, we have
\begin{equation}\label{eq:deltaFe}
    \delta F_{\rm Fe}=\left(\frac{V_{\rm Fe}}{V_{\rm H\beta}}\right)^2\delta F_{\rm H\beta}\approx 0.2\,\delta F_{\rm H\beta},
\end{equation}
for PHL\,1092, where $(V_{\rm H\beta},V_{\rm Fe})\approx \left(2270,1054\right)\,\kms$ are from Fig.\,\ref{fig:Fe_fit}.
We thus expect the \feii's intrinsic variations of $\delta F_{\rm Fe}\approx 3\%$ only from the observed $\delta F_{\rm H\beta}\approx 15\%$. 
Fig.\,\ref{fig:LC_Fe} shows that the scatters of \feii\ variations are dominated by the system errors, which are consistent with the rough assessments of the \feii\ variations by Eq.\,(\ref{eq:deltaFe}).
Actually, the H$\beta$ LCs could be contaminated by the \feii\ line as the systematic errors, however, its contamination to H$\beta$ variations is small enough shown by Eq.\,(\ref{eq:deltaFe}) in PHL\,1092. 

It is worth of noting that our SEAMBH campaigns\cite{Hu2015} successfully measured \feii\ reverberations in 9 narrow-line Seyfert 1 galaxies (NLS1s). We find that these objects have $\delta F_{\rm Fe}\sim \delta F_{\rm H\beta}$ from Eq.\,(\ref{eq:deltaFe}) in light of $V_{\rm H\beta}\sim V_{\rm Fe}$ listed in Tab.\,2 of Ref.\cite{Hu2015}.
Independently, $\delta F_{\rm Fe}\sim \delta F_{\rm H\beta}$ indeed holds from the LCs in the nine NLS1s\cite{Hu2015}.
It is clear from $V_{\rm Fe}/V_{\rm H\beta}$ that PHL\,1092 has much larger structures of \feii\ line regions compared with other NLS1s.
Other two successful measurements of \feii\ reverberations\cite{Barth2013} have been done for NGC\,4593 and Mrk 1511, in which $\delta F_{\rm H\beta}\sim (60\%, 30\%)$, and $\delta F_{\rm Fe}\sim (40\%, 25\%)$ from their LCs, respectively. 
The FWHMs of NGC\,4593 and Mrk\,1511 were found that $V_{\rm H\beta}=(4395\pm362,4171\pm 137)\,\kms$ and $V_{\rm Fe}=(3330\pm153,3128\pm143)\,\kms$ from the Boroson \& Green template of \feii\ lines, or $V_{\rm Fe}=(5044\pm176,4459\pm116)\,\kms$ from the V\'eron-Cetty template.
Considering uncertainties of the templates, we take their mean values of $V_{\rm Fe}=(4187\pm164,3793\pm130)\,\kms$, and find $V_{\rm Fe}=(0.95,0.91)V_{\rm H\beta}$ for NGC\,4593 and Mrk\,1511, respectively. 
Interestingly, Eq.\,(\ref{eq:deltaFe}) roughly holds for the two objects.

Measurements of \feii\ line reverberations in AGNs are a well-known hard job, but the above brief discussion of Eq.\,(\ref{eq:deltaFe}) predicts promising AGN candidates. As an useful criterion,  relative broadness of \feii\ to the H$\beta$ line helps for future target campaigns of SEAMBHs to reveal \feii\ physics of super-Eddington accreting AGNs.

\subsection{2.8 Contamination of \feii\ to H$\beta$ line.}
We obtained the 5100\,\AA\ continuum and H$\beta$ LCs using a direct integration method. 
The Fe {\sc ii} emission within the continuum and H$\beta$ measurement windows, specifically the portions that cannot be removed by the integration method, may slightly influence the continuum and H$\beta$ LCs. 
To estimate the contribution from the residual Fe {\sc ii} emission in the measurement windows of the integration method, we performed multi-component spectral fitting on the mean spectrum of PHL\,1092. 
The components employed in the fitting include: (1) a power law to model the continuum, (2) two Gaussians to model the H$\beta$ emission line, (3) a template from Boroson \& Green (1992) to model the Fe {\sc ii} emission, and (4) a Gaussian for each of the coronal lines ([Fe {\sc vii}] $\lambda\lambda5158, 5176$ and [Ca {\sc v}] $\lambda$ 5309, as described by Ref.\cite{Hu2015}). 

From the fitting results (see Fig.\,\ref{fig:Fe_fit}), we found that the integration method may indeed retain some \feii\ contribution in the continuum and H$\beta$ windows, however, these fractions are small. The \feii\ residual contributes only 11\% in the continuum window and 6\% in the H$\beta$ window. 
Given that the Fe {\sc ii} variability during the campaign is estimated to be approximately 3-4\% (as seen in the Fe {\sc ii} LC), which is even smaller than the variation amplitude of the continuum and H$\beta$, the impact of the Fe {\sc ii} residuals on the continuum and H$\beta$ LCs can be considered negligible.

\subsection{2.9 Contamination of diffusive continuum.} There is growing evidence for continuum contamination contributed by the diffusive continuum risen by the BLRs, even up to 23\% on average in the sample\cite{Guo2022}.
We note that this number is based on several important assumptions, such as, the lags of diffusive continuum is half of the H$\beta$ and the BLR covering factor is 20\%, as in Ref.\cite{Netzer2022}, or even 50\% in Ref.\cite{Korista2019}. 
In the sample of Ref.\cite{Guo2022}, the averaged EW(H$\beta$) is about 70-80\,\AA\ (similar to normal quasars), while PHL\,1092 is a weak-line quasar and ${\rm EW(H\beta)}\approx 40\,$\AA.
Therefore the diffusive continuum in this object should be significantly smaller than the averaged values of AGN sample\cite{Guo2022} in light of the smaller covering factor.
Actually, the roles of the diffusive continuum are always forcing the lags toward zero. 
Therefore, the current leading delays cannot be resulted from the diffusive continuum, in particular, PHL\,1092 is a weak-line quasar.

\section{Lag measurements}\label{sec:RManalysis}
The interpolated cross-correlation function (ICCF)\cite{Gaskell1987} is widely employed to obtain H$\beta$ lags, but it is based on linear interpolations. 
The ICCF method often generates small multiple peaks due to noises of data, and it gets worse when there are season gaps in long term campaigns. 
It has been demonstrated that interpolations based on statistically and observationally motivated models are more reliable than the ICCF\cite{Jiang2017,Grier2017}, such as \JAVELIN\cite{Zu2011} and \MICA\footnote{\url{https://github.com/LiyrAstroph/MICA2}} (Multiple and Inhomogeneous Component Analysis\cite{Li2016}), and thus generate reasonable lags. 
Both \JAVELIN\ and \MICA\ employ a damped random walk (DRW) model to characterize the continuum LC but they make use of a top-hat transfer function and a transfer function composed of a sum of Gaussians, respectively. 
In this work, we adopt a single Gaussian in \MICA. Hence, the delays are determined by the center of the Gaussian function, while their corresponding uncertainties are estimated using the 15.87\% and 84.13\% quantiles derived from the posterior samples generated by the Markov Chain Monte Carlo (MCMC) method. 

As shown in Fig.\,\ref{fig:MICA-lags}, the H$\beta$ LCs are fit well by the DRW model except for two epochs (MJD-58000\,+800 and +2500).
These could be caused by the non-concordant  reverberations if composite responses to the spatially inhomogeneous variations of the extended continuum sources (see \S\,\ref{sec:toymodel}).
In order to avoid this complexity, we separate the LCs into three parts for determinations of reverberation delays as shown in panels {\textbf{b,\,c,\,d}} of Fig.\,\ref{fig:MICA-lags}.
Generally, the DRW models fit the LCs of different periods quite good.
All the separate parts and the entire LCs demonstrate that H$\beta$ line variations are leading the 5100\,\AA, namely, the leading delays of the H$\beta$ line reverberations.

Additionally, we also provide measurements of the H$\beta$ lags with respect to the broad-band photometry LCs in Fig.\,\ref{fig:lags_photometry}. 
The mean lags are $-12.0,\,-7.8,\,-31.2,\,-18.0\,$ days for the same periods with that of Fig.\,\ref{fig:MICA-lags}, respectively. 
We note some slight differences between the results from the photometry. 
These are caused by the fact that the photometry covers some contributions of emission lines. The consistent results also provide robustness of the LDRs in PHL\,1092.

\section{A toy model for anomalous reverberations}\label{sec:toymodel}
In order to understand the LDRs, we build up a toy model to illustrate the reverberations under spatially inhomogeneous variations of ionizing sources and the 5100\,\AA\ continuum.
Considering a central `lamp-post' generating X-ray variations $L_{\rm X}(t)$ as shown in Fig.\,\ref{fig:model}, we have the ionizing flux as responses to these variations as described by Eq.\,(\ref{eq:Lion}). 
Ionizing and the 5100\,\AA\ sources are simplified as two rings radii $R_{\rm ion}$ and $R_{5100}$, respectively, distributing over the rings.
Subsequently it drives the variation in the emission line as Eq.\,(\ref{eq:LHb}),
the variations at 5100\,\AA\ also align with the 'lamp-post' as indicated by Eq.\,(\ref{eq:L5100}).

In our simulations, the X-ray variation $L_{\rm X}(t)$ is modeled as a damped random walk with a timescale of $ \tau_{\rm d}=\tau_{\rm X} = \SI{150}{days} $ and an amplitude of $\sigma_{\rm d}= \sigma_{\rm X} = 0.25 $ in Eq.\,(\ref{eqn_sij}). 
As the simplest configuration, $N_*$ ionizing sources are distributed along a ring with a radius of $R_{\rm ion}$. 
The flux from each source can be expressed as follows:
\begin{equation} 
L_{{\rm ion},i}(t) = L_{\rm X}\left(t-t_{\rm ion}\right);\quad {\rm and\,\,} t_{\rm ion}=\left(1 - \cos\phi_i \sin i_{\rm o}\right)\left(\frac{R_{\rm ion}}{c}\right),
\end{equation}
where $\phi_i$ represents the azimuthal angle of the $i$-th source, and $i_{\rm o}$ denotes the inclination angle of the observer.
We take $i_{\rm o}=30^{\circ}$ for all the cases.
The clouds in the BLR will respond to the variations of each ionizing source independently, as given by:
\begin{equation}
L_{{\rm H\beta}, i}(t) = \int_{0}^{\infty} \dd{\tau} L_{{\rm ion},i}(t-\tau) \Psi_i(\tau),
\end{equation}
where $ L_{{\rm H\beta}, i} $ represents the variation in line emission caused by the $i$-th component of the ionizing source, and $\Psi_{i}$ is the corresponding transfer function.
Given the responsivity distribution $ g(\vb*{r}) $ of the BLR, the transfer function can be calculated as follows:
\begin{equation}
\Psi_i(\tau) = \int \dd[3]{r} \frac{g(\vb*{r})}{\abs{\vb*{r} - \vb*{r}_i}^2} \delta \left[\tau - \abs{\vb*{r} - \vb*{r}_i} + (\vb*{r} - \vb*{r}_i) \vdot \vu*{n}\right],
\end{equation}
where $ \vb*{r}_i $ denotes the position of the $i$-th source, and $ \vu*{n} = (\sin i_{\rm o}, 0, \cos i_{\rm o}) $ is the unit vector pointing from the source to the observer. 
For simplicity, we assume that the BLR forms a uniform ring with a radius of $ R_{\rm BLR} $. 
Consequently, we have
\begin{equation}
\Psi_i(\tau) = \int_{0}^{2\pi} \dd{\phi} \frac{\delta\left(\tau - \tau_i\right)}{\Delta_i^2} ,
\end{equation}
where
\begin{equation}
\Delta_i = \sqrt{(R_{\rm BLR}\cos\phi - R_{\rm ion}\cos\phi_i)^2 + (R_{\rm BLR}\sin\phi - R_{\rm ion}\sin\phi_i)^2}
\end{equation}
and
\begin{equation}
\tau_i = \left(\frac{1}{c}\right)\left[\Delta_i - \left(R_{\rm BLR}\cos\phi - R_{\rm ion}\cos\phi_i\right)\sin i_{\rm o}\right].
\end{equation}
The number and azimuthal angles of the 5100\,\AA\ sources are identical to those of the ionizing sources. 
Consequently, the continuum from the $i$-th source is
\begin{equation} 
L_{{\rm 5100},i}(t) = L_{\rm X}\left(t-t_{5100}\right); \quad {\rm and\,\,} t_{5100}= \left(1 - \cos\phi_i \sin i_{\rm o}\right)\left(\frac{R_{\rm 5100}}{c}\right). 
\end{equation}
The observed LCs of the 5100\,\AA\ continuum and H$\beta$ line are then
\begin{equation} 
L_{\rm 5100}(t) = \sum_{i=1}^{N_*} L_{{\rm 5100},i}(t) \quad 
\text{and} \quad L_{\rm H\beta}(t) = \sum_{i=1}^{N_*} L_{{\rm H\beta},i}(t), 
\end{equation}
respectively. Given the parameters of $R_{\rm ion}$, $R_{\rm 5100}$ and $R_{\rm H\beta}$ of simplified model, we can simulate the observed reverberations.

Fig.\,\ref{fig:lc0} -- \ref{fig:lc5-6} present results for various values of $R_{\rm ion}$, $R_{\rm 5100}$, and $R_{\rm H\beta}$, revealing diverse non-concordance reverberations of composite responses. 
In general, the BLRs exhibit leading delays of reverberations with respect to the 5100\,\AA\ provided $R_{\rm H\beta}<R_{5100}$, otherwise normal delayed reverberations occur.
Emission lines correlate strongly with the continuum’s large-scale structures over extended timescales but show poor responsiveness to small-scale structures—especially with multiple ionizing sources. 
Shaded areas mark periods of this poor responsiveness to illustrate the effect.
When BLR lies within the 5100\,\AA\ region, the line variations precede those in the 5100\,\AA\ continuum, and vice versa.
The time delay between the two LCs depends on the difference in their respective sizes.
Amid such complex situation of extended and inhomogeneous ionizing sources, composite-response reverberations appear non-concordant, with weak or over-responses in some periods. 
This non-concordant relation arises not from reprocessing physics but from differential lags caused by spatially inhomogeneous variations. 
Furthermore, measured lags reflect distances between the BLR and ionizing sources (not the central SMBH), invalidating the conventional SMBH mass formulation.

To better compare with observations, we resampled the mock LCs in the left panel of Fig.\,\ref{fig:gap}, with an average 6-day cadence and approximately 160-day season gaps.
Additionally, the observational errors for the continuum spectrum and emission lines are $1\%$ and $3\%$, respectively.
Fig.\,\ref{fig:gap} right panel shows negative lags, which are consistent with observed anomalous reverberations.
These simulations explain the aforementioned anomalies in NGC 5548, Mrk 817, and NGC 7469, as well as PHL\,1092’s over-responses around MJD 58000\,+800 and under-responses at +2500.
Future work will use a more realistic BLR model for inhomogeneous variations to explore anomalous reverberations of the broad H$\beta$ line, with detailed results to follow.

It should be noted that in these simulations $r_{\rm max}$ is significantly higher than the observational data.
This is because the simulated data are generated under ideal conditions — no random fluctuations are added, leading to $r_{\rm max}\sim 0.9$ (Figs. \ref{fig:lc0}–\ref{fig:lc5-6}). 
Additionally, the three components (ionizing sources, BLR, and 5100\,\AA\ regions) are simplified as rings with different radii. 
Recognizing this, more realistic models are needed to simulate PHL\,1092’s observations.

\section{Aliasing effects}\label{sec:aliasing}
To investigate whether the aliasing effect can result in a leading lag due to seasonal gaps, sampling, and scatters in the LCs from the current RM observations, we conducted a series of Monte Carlo simulations with the following steps. 
First, we fit the observed LCs of PHL\,1092 through the DRW model to obtain two parameters of timescales and amplitudes describing variations, namely,
\begin{equation}
S_{ij} = \sigma_{\rm d}^2\exp\left(-\frac{\left|t_i-t_j\right|}{\tau_{\rm d}}\right), 
\label{eqn_sij}
\end{equation} 
where $t_i$ and $t_j$ are the time of two points in the continuum LCs, $\sigma_{\rm d}$ is the variation amplitude at long timescale, $\tau_{\rm d}$ is the characteristic damped timescale.
We determine the two parameters by fitting the observed whole continuum LC using MCMC method. 
We obtain $\sigma_{\rm d}=1.42\times 10^{-3}$ (in arbitrary units) and $\tau_{\rm d}=267\,$days, where these values represent the median estimates derived from the posterior distribution of the MCMC samples. 
Using these parameter values, we generate mock LCs of the continuum with daily sampling.
Second, we make use of the Gaussian function as the BLR transfer function
\begin{equation}
    \Psi_{\rm BLR}(t)=\frac{1}{\sqrt{2\pi}\Delta \tau_0}\exp\left[\left(\frac{t-\tauhb}{\Delta \tau_0}\right)^2\right],
\end{equation}
to obtain \hb\ LCs, where $\Delta \tau_0$ is the width of the function.
We take $\tauhb=16\,$days (corresponding to the lag predicted by the new scaling relation\cite{Du2019}) and a standard deviation of 5\,days (equal to one-third of the lag), $\Delta \tau_0\approx \tauhb/3$, which means that $\tauhb-3\Delta \tau_0=0$ appears as zero-lag reverberation.
%
%
Actually, this is a conservative estimation, for example, $\Delta \tau_0=0.1\tauhb$ in Ref.\cite{Shen2015} and $\Delta \tau_0=0.2\tauhb$ in Ref.\cite{Grier2013} for mock LCs of reverberations.
Moreover, we also tested how the results of the simulations depend on the input width of the Gaussian functions. 
We find that results from $\Delta \tau_0=10\,$days are very similar to that of $\Delta \tau_0=5\,$days.
In practice, \MICA\ and \JAVELIN\ based on physical model for interpolations are a powerful tool to determine the lags, generating results very insensitive to the width of the transfer functions.
Subsequently, the two LCs are linearly interpolated onto the corresponding observed epochs to mimic real observations. 
To enhance the realism of the mock LCs, Gaussian noise is systematically introduced, ensuring that the relative errors are consistent with those observed in the actual data.
Finally, we measured the lags of the mock continuum and emission-line LCs using \MICA\, with the same settings applied to the measurements of the actual LCs of PHL\,1092.

We repeat the above procedure 5000-20000 times. 
Fig.\,\ref{fig:simlag_distribution} shows simulation results, yielding probabilities of random LDRs from aliasing effects and error bars listed in Tab.\,\ref{tab:lags}.
We found that the probability of getting a leading lag with $\tau_{\rm MICA} < -16$ days is $5\times 10^{-5}$ based on the lag distribution from the mock data for the whole length of the campaign.
The significance of other periods is also close to $4\,\sigma$ listed in Tab.\,\ref{tab:lags}.
We therefore conclude that the LDR detection is significant at a level of $4.0\,\sigma$ in PHL\,1092.

Finally, we plot residuals of the fittings in Fig.\,\ref{fig:Gauss}. 
The residual distributions are approximately Gaussian, indicating that the tests of aliasing effects are robust.
The observed LDRs in PHL\,1092 are intrinsic phenomena.

\section{Standard model of accretion disks}\label{sec:SSdisk} 
Considering potential configurations of the PSA invalidity, we have to examine sizes of accretion disks which play a key role in the reverberation physics, in particular, when the sizes are comparable to the BLRs. 
For the regime of $0.1\lesssim \mathdotM\lesssim 3$, the Shakura-Sunyaev model usually assumes a Keplerian rotation around the central SMBH with $\alpha$-viscosity description, and that the radial motion velocity of accreting gas is much smaller than the rotation velocity, enabling the local energy balance\cite{Shakura1973}. 
Additionally, the disk is cold (relative to local virial temperatures) because dissipated energy is radiated away locally and instantaneously. 
The energy dissipation rate per unit area of the disk is given by\cite{Lynden-Bell1969,Shakura1973},
\begin{equation}\label{eq:Qvis}
Q_{\rm vis}=\frac{3}{8\pi}\frac{G\mbh \dot{M}_{\bullet}}{R^3}
           =6.86\times 10^{10}\,\mathdotM M_7^{-1}r_3^{-3}\,{\rm erg\,cm^{-2}\,s^{-1}},
\end{equation}
where $r_3=R/10^3\,\Rg$, the boundary factor is neglected for $R/\Rg\gg1$. 
We would emphasize that this rate is fully independent of any viscosity.
The effective temperature of the disk surface approximately treated as a black body is given by $\sigma_{\rm SB}T_{\rm eff}^4=Q_{\rm vis}$, namely
\begin{equation}\label{eq:TeffSS}
T_{\rm eff}=5.89\times 10^3\,\mathdotM^{1/4}M_7^{-1/4}r_3^{-3/4}\,{\rm K}.
\end{equation}
Since the Planck function can be well approximated by the $\delta$-function of wavelength (see Eq.\,4.39 in Ref.\cite{Netzer1990}), we have the radius of emitting $\epsilon$ photons from the Wien's law,
\begin{equation}
\frac{R_{\epsilon}}{\Rg}=4.9\times 10^2\,\left(\epsilon/\epsilon_{5100}\right)^{-4/3}\mathdotM^{1/3}M_7^{-1/3},
\label{eq:R5100SS} 
\end{equation}
showing that $R_{\epsilon}$ increases with accretion rates, where $\epsilon_{5100}\approx 2.2\,$eV is photon energies at 5100\,\AA.
For $\epsilon=\epsilon_{5100}$ and a typical SMBH-disk with $\mathdotM=1$ and $M_7=1$, we have $R_{\rm 5100}\approx 500\,\Rg\ll R_{\rm BLR}$ (Eq.\,\ref{Eq:R5100a}), indicating validity of the PSA when accretion rates are not much super-Eddington ($\mathdotM\lesssim 10^{2}$).
Therefore Eq.\,(\ref{eq:Mass}) generally applies to most low-$\calR$ AGNs. 

Isotropy of emissions from slim disks has been examined\cite{Wang2014b} showing anisotropy of the ionizing energy source for $\mathdotM\gtrsim 10$ accretion disks. The anisotropy of Shakura-Sunyaev disk\cite{Shakura1973} can be approximated by $\cos i$, and hence as roughly isotropic sources if they are face-on.
Whatever accretion rates of slim disks, the self-obscuration of slim disks cannot lead to the LDRs.

\section{Challenging questions}\label{sec:question}
The LDRs of PHL\,1092 raise a series of fundamental questions listed below, which pertain to the inner structure of AGNs, in particular, the extra energy sources beyond the SMBH-disk could be involved. 
We discuss these questions with speculative insights, grounded in observational evidence.

First, with the PSA invalid, how to estimate SMBH masses for PHL\,1092 analogs remains open since Eq.\,(\ref{eq:Mass}) is no longer valid. 
In such a context, only order estimations of SMBH masses can be made. 
In \S\,\ref{sec:TF}, we provide some analytical discussions on transfer functions between the H$\beta$ and 5100\,{\AA} LCs for SMBH masses.
In order to simply illustrate, we consider the BLR simplified as a ring. 
If neglecting the spatial extension of the 5100\,{\AA} regions, the width of the transfer function is given by $2R_{\rm ring}\sin i/c$, where $R_{\rm ring}$ is the ring radius. See details in Fig.\,19 of page 114 in Ref.\cite{Netzer1990}.
Therefore, $R_{\rm ring}/c$ from the transfer function can be regarded as the rough width of the BLRs for SMBH masses.
In practice, using the tansfer function for $M_{\bullet}$ could be a quite reliable approximation when the reverberations are anomalous.
Transfer functions of the BLRs, which are obtained by \JAVELIN\ and \MICA\ measurements through fitting the continuum and H$\beta$ LCs, yield the full-width half maximum of the BLRs ($\Delta R_{\rm min}$).
The BLR size can be approximated by $\Delta R_{\rm min}$ to replace $\Rhb$ in Eq.\,(\ref{eq:Mass}) for SMBH masses.
For PHL\,1092, we have $\Rhb\approx \Delta R_{\rm min}\approx 25\,$ltd from \MICA\ analysis of the whole LC (see Fig.\,\ref{fig:MICA-lags} in \S\,\ref{sec:RManalysis} and related discussions in \S\,\ref{sec:TF}). 
We have $\mbh=2.5\times 10^7\sunm$ and accretion rates of $\mathdotM\approx 200$, suggesting that this is a super-Eddington accreting AGN.

Modeling the BLR reverberations through MCMC simulations is a powerful method of accurately measuring SMBH masses\cite{Pancoast2011,Li2018}, however, the method is not valid for the extremely anomalous reverberations (the PSA is invalid) because the 5100\,{\AA} region and BLR sizes are degenerate, and the 5100\,\AA\ sizes as a reference to the central SMBH distances is entirely unknown.
Fortunately, GRAVITY/VLTI measurement based on spectroastrometry\cite{Gravity2018} would be a key to measure SMBH masses since it models profiles of the broad emission line and its differential phase curve whatever the SMBH-disk sizes are.
If the SARM (SpectroAstrometry and Reverberation Mapping)\cite{Wang2020} analysis is employed, the SMBH masses and disk sizes are expected to be generated simultaneously.
This may allow us to determine spatial distributions of sMBHs over the \sSMBH.

If the spatial inhomogeneity scale is comparable to the radius of the {\sSMBH} ($\Delta R\sim R$, see Issue 6 in \S\,\ref{sec:question} below), the approximation of a monotonic temperature-radius ($T-R$) relation breaks down.
This will open a new research direction for high spatial resolution astronomy with future kilometer-baseline interferometers\cite{Bourdarot2024}. 
The measured masses will encompass all components within the spatially resolved radius (including gaseous mass, sMBHs, and the central SMBH). 
The pure mass of the central SMBH must be inferred from the detailed fine structure of differential phase curves though this remains uncertain in current GRAVITY observations\cite{Gravity2018}.

Second, what mechanism governs the shortening of \hb\ reverberation lags and the emergence of LDRs?
We define a parameter of $\delta\tau=\tau_{\rm H\beta}/\tau_{\rm H\beta}^0$, this parameter ranges approximately from $[-50,1]$ from PHL\,1092.
While delay shortening of SEAMBHs can be partially explained in some degrees by the self-obscuration of the inner part of slim disks\cite{Wang2014b}, this cannot account for the LDRs (i.e., $\delta\tau\le 0)$.
Fig.\,\ref{fig:RL-Plane} shows the large gap between the $R_{\rm H\beta}=-(17\sim57)\,$ltd and the $R_{\rm H\beta}-(L,\calR)$ relation, but there is growing evidence for analogs with anomalous reverberations (see a brief discussion in Issue 3 in \S\,\ref{sec:question}) filling the gap.
If the transition from shortened lags to LDRs is continuous, the shortening mechanism is likely not pure self-obscuration but other processes, e.g., EESs stretching the SMBH-disk system depending on the number of embedded stellar-mass black holes (sMBHs).
AGN disks are estimated to contain $\sim10^3-10^5$ massive stars\cite{Chen2024} over AGN episodic lifetimes, these stars can evolve into sMBHs (with approximately same number $\sim N_{\bullet}$) and Fe elements\cite{Wang2023}.
In the simplest version, a relation of $\calR\propto N_{\bullet}$ is expected (of course, it is still controversial\cite{Verner2004,Verner2009} if $\calR$ can be used for exact metallicity\cite{Netzer2007} since it is also sensitive to ionized gas density, temperatures and turbulence).
Therefore, the structures of high-$\calR$ AGNs controlled by the EESs can be tested by RM campaigns of broad emission lines of the BLR and continuum of accretion disks, or by micro-lensing effects for measurements of the SMBH-disk sizes\cite{Rauch1991,Morgan2010,Cornachione2020}.

Third, what are occurrence rates of the LDRs in AGNs analogous to PHL\,1092?
It is still too early to provide a definitive answer.
Our campaigns have monitored $\sim 20$ targets for over 7 years.
Among these, several show severely shortened H$\beta$ lags — $\tauhb$ approaches zero or becomes negative (shortening factors exceed 10), and some (e.g., SDSS\,J224028-10649) even oscillate between positive and negative lags across epochs. 
Examples include Mrk\,1239\footnote{see its complex spectra\cite{Pan2021}; the first 3-year of our RM campaign showed no anomalies\cite{Li2025}, but latest 3-year data exhibit anti-reverberation in our six-year campaign.}, PG\,0043+039, PG\,1244+026, PG\,1543+489, PHL\,1811 (a weak-line quasar) and SDSS\,J224028-10649.
We have $\sim 25\%$ of our total sample appearing anomalous reverberation but the sample is heterogeneous.
These objects violate the PSA and may be PHL\,1092 analogs, but their more complex delay behaviors suggest more intricate inner structures.
We will report these objects separately once their properties are definitive.
However, they lack distinct single-epoch spectral features to differentiate them from PHL\,1092, making it impossible to pre-select the analogs. 
To address this question, long-term, high-cadence RM campaigns of more high-$\calR$ AGNs are needed to systematically search for such LDRs.
Except for the LDRs in this paper, the long-standing unresolved issues of H$\beta$ reverberation trending in some AGNs\cite{Peterson2004} could be also driven by the sMBHs. 

Additionally, some targets have variation amplitudes too small to measure lags—even over several years. 
These objects typically have high-$\calR$ and may resemble the so-called "extremely stable quasars"\cite{Kang2024}, which is notable. 
Systematically investigating the temporal properties of AGNs as a function of $\calR$ is undoubtedly crucial.

Fourth, how many sMBHs are in an SMBH-disk if EESs are powered by them? 
A key clue to this question is why AGN BLRs are so metal-rich (i.e., high-$Z$). 
Previous work\cite{Artymowicz1993,Wang2011,Wang2012,Wang2023,Fan2023} suggests BLR metals form via star formation in nuclear regions, and sMBHs as remnants of massive stars their numbers should correlate with BLR metallicity.  
Thus, $\calR$ as a proxy of BLR metallicity\cite{Netzer2007} might also indicate numbers of sMBHs, which control the $\Rhb-(L,\calR)$ relation of the SEAMBHs\cite{Du2019}.
However, it is not a straightforward job to theoretically derive this relation, which involves evolution of massive stars formed from fast accretion\cite{Cantiello2021,Wang2023}.
About $10^3-10^5$ sMBHs are estimated in SMBH-disk\cite{Chen2024}.
Additionally, while AGN structures comprise accretion disks, BLRs, and dusty tori, their formation mechanism has remained unresolved since quasars were first discovered. 
A critical question arises: if sMBHs are sufficiently abundant, can they play important roles in AGN structure and evolution? 
Clearly, establishing relationships between $\calR$, metallicity (Z), and sMBH number ($N_{\bullet}$) — as well as linking these to the $R_{\rm H\beta}-(L,\calR)$ relation—will be a pivotal future task.

Fifth, what are the temporal properties of the \sSMBH? 
This question prompts us to recall that X-ray binaries exhibit quasi-periodic oscillations (QPOs) across various frequencies\cite{Kato2008,Ingram2019}. 
By contrast, AGN QPOs are extremely rare except for a few objects, such as X-ray QPOs in RE\,J1034+396\cite{Gierlinski2008}. 
This implies AGN SMBH-disk systems are not simply black hole mass-scaled versions of X-ray binary accretion disks.
A heuristic hypothesis is that the difference stems from the presence of \sSMBH\ in AGNs.
In this scenario, sMBHs suppress gaseous oscillations in the SMBH-disk. 
For instance, a kind of QPOs with a frequency of $\nu\sim c_{\rm s}/H$ will be destructed by the Bondi accretion of sMBHs with a timescale of $R_{\rm Bon}/c_{\rm s}$, where $c_{\rm s}$ is the sound speed, $R_{\rm Bon}$ is the Bondi radius and $R_{\rm Bon}\sim H$.
Nevertheless, sMBHs play a key role in AGN variability. 

Moreover, Eq.\,(\ref{eq:extra}) shows that if $Q_{\rm extra}\gtrsim Q_{\rm vis}$, the effective temperature distribution is drastically modified, producing warm outer regions of the SMBH-disk, namely, $R_{5100}^{\,\bullet}\gg R_{5100}$, where $R_{5100}^{\,\bullet}$ is the 5100\,{\AA} radius of the \sSMBH. 
This leads to that the reprocessed emissions appear far more temporally stable (longer variability timescales but much smaller amplitudes than normal AGNs).
This could explain why SEAMBH LCs are typically more stable than those of sub-Eddington AGNs\cite{Lu2019}, assuming the SMBH-disk of SEAMBHs contain more sMBHs than the sub-Eddington (which are usually low-$\calR$ objects. 
Emissions from sMBHs dominate the SMBH-disk so that they share high stability since the total variations are smeared by the widely spatial distribution of sMBHs across the disk). 
Notably, extremely stable quasars have recently been identified\cite{Kang2024} — these may be more extreme cases than PHL\,1092. 
A correlation between $F_{\rm var}$ and $\calR$ should be explored in a large sample, and investigating sMBH roles in AGN variability across multiwavelengths is highly prioritized.

Furthermore, growing evidence from microlensing indicates accretion disk sizes are significantly larger than standard model predictions\cite{Morgan2010,Cornachione2020}. 
Approximately ten AGNs have been jointly observed by {\textit{Swift}} and ground-based telescopes to constrain disk sizes.
The measured lags are $\sim 2-3$ times those of the standard accretion disk model\cite{Jiang2017,Cackett2021,Guo2022,Lewin2024}.
This substantial discrepancy could be explained by modified black-body radiation from the disk surface, driven by the disk’s detailed vertical structure\cite{Wang2025}. 
However, \sSMBH\ will exhibit distinct behavior in multi-color reverberations of the SMBH-disks, in particular, will show lags far longer than standard accretion disks. 
Indeed, Mrk 335 has a $\sim 8000\,${\AA} continuum size ten times larger than the standard model prediction\cite{Kara2023}. 
Future reverberation mapping of accretion disks in PHL\,1092 analogs will measure these stretched disks; the powerful Vera Rubin Telescope (LSST), in particular, will enable systematic accretion disk size measurements for a large AGN sample.

Additionally, sMBH roles in variability remain unclear, except for their involvement in outbursts via the Bondi explosions\cite{Wang2021a,Wang2021b}. 
Under the lamp-post model, reprocessing follows X-ray variations, but variability timescales are determined by light-crossing times of the \sSMBH.
However, the spread of UV and optical continuum variations reduces amplitudes, making it harder to measure delays of continuum reverberations via multi-wavelength campaigns.

Sixth, how do sMBHs migrate toward the central SMBH?
Studies of sMBH populations\cite{Secunda2019,Secunda2020,Wangmengye2023} show: (1) migration timescales depend on SMBH-disk accretion rates; (2) more massive sMBHs migrate inward faster than less massive ones; (3) hierarchical sMBH mergers\cite{YangY2019,Mapelli2022,Vaccaro2025} drive the growth of recently observed $\sim 100\,M_\odot$ sMBHs\cite{Abbott2020,GW231123}, even forming multi-body systems of several massive black holes in nuclear central regions; (4) multi-scattering of sMBHs with gaseous drag can form sMBH clusters inside/outside the SMBH-disk, causing spatial inhomogeneity in ionizing components (observationally referred to as hot spots).
This dynamical behavior, in principle, produces an interesting consequence of AGN temporal properties.
AGNs with higher accretion rates rapidly swallow sMBHs, gradually evolving into sub-Eddington AGNs with fewer, intermediate/massive black holes. 
In sub-Eddington AGNs, sMBHs play a minor role in SMBH-disk heating ($R_{5100}^{\,\bullet}\approx R_{5100}\ll \Rhb$) but can trigger strong AGN variability (e.g., outbursts\cite{Wang2021a,Wang2021b,Liu2024}).
By contrast, \sSMBH\ contains more sMBHs — enabling efficient heating and producing structures with $R_{5100}^{\,\bullet}< R_{5100}$, $R_{5100}^{\,\bullet}\sim \Rhb$ (zero-lag reverberations), or even $R_{5100}^{\,\bullet}>\Rhb$ (leading delay of responses, LDRs).
SEAMBH-focused campaigns are expected to detect all such anomalous reverberations, helping establish the potential sequence of dynamical evolution of sMBHs.

The existence of intermediate BHs around the central SMBH may drive a type I migration making the inner part of the SMBH-disk unstable.
This may yield the increasingly interesting phenomena of changing-look AGNs\cite{Komossa2024} through interaction of massive black holes with companion's disk\cite{WangBon2020}, which are showing increasingly more complicated behaviors\cite{Guo2025}.
It is also possible that successive mergers of sMBHs to form a few bodies of SMBHs, and even leave a binary-SMBH in galactic centers.
Except for galaxy mergers, this is a new way to form binary SMBHs in galaxies radiating nano-Hz GWs expected by PTA detections\cite{Reardon2023,EPTA2023,CPTA2023}.
Individual identifications of binary-SMBHs can be done by RM campiagns\cite{Wang2018,DuMAHAI2018,Songsheng2020,Fu2025} or spatially resolved by optical interferometers, such as GRAVITY/VLTI\cite{Songsheng2019,Songsheng2023}.

Seventh, non-concordant reverberations as the effects of the spatially inhomogeneous variations (incoherent in time and amplitudes) of an extended ionizing source have bee discovered from simple Monte Carlo simulations in \S\,\ref{sec:toymodel}.
This applies to PHL\,1092, which shows weak responses (anomalous behavior) at a few epochs.
When $R_{5100}\gtrsim R_{\rm H\beta}$ holds, the LDRs emerge, and non-concordant responses become more complex — this lies beyond the classical RM framework (see fundamental assumptions in Ref.\cite{Blandford1982}).
More realistic models are therefore needed to explain such anomalous reverberations.

Observationally, the LDRs in PHL\,1092 are the first extreme outliers from the known $R_{\rm H\beta}-(L,\calR)$ relation. 
Future campaigns require the highest possible cadence to probe ionizing source inhomogeneities in PHL\,1092, in particular, 6–8 m telescopes are necessary for high-fidelity RM improving spectral calibration and addressing annual scatter in spectroscopic data. 
While \S\,\ref{sec:toymodel} only explores ionizing source inhomogeneities and inhomogeneous BLRs cloud distributions are also plausible, both inhomogeneities could be detected by future high-fidelity campaigns. 
Non-concordant reverberation effects will then help unravel the complex structures and components around the central SMBH. 
However, annual lags may vary, as ionizing source components are random and varying somehow. 
These random lag values reflect inhomogeneity of the ionizing sources and their non-synchronized variabilities — potentially explaining the multiple H$\beta$ reverberation lags noted in Issue 3 of \S\,\ref{sec:question} for our sample.

In summary, LDRs pose significant challenges to the standard AGN model. 
As briefly discussed in \S\,\ref{sec:other-models}, sMBHs accreting within the SMBH-disk can provide additional energy sources and act as a new mechanism governing extremely anomalous H$\beta$ reverberations (e.g., PHL\,1092 as reported here). 
This may represent just the tip of the iceberg in AGNs hosting \sSMBH.

\section{Reverberations of the lamp-post model}\label{sec:TF}
\JAVELIN\ and \MICA\ don't make use of the PSA, but we illustrate the physical implications of the transfer functions. 
In the lamp-post model, the 5100\,\AA, ionizing photon and H$\beta$ fluxes are related to the X-ray flux as 
\begin{equation}\label{eq:L5100}
    L_{5100}(t)=\int \Psi_{5100}(\tau)L_{\rm X}(t-\tau)d\tau,
\end{equation}
\begin{equation}\label{eq:Lion}
    L_{\rm ion}(t)=\int \Psi_{\rm ion}(\tau)L_{\rm X}(t-\tau)d\tau,
\end{equation}
and 
\begin{equation}\label{eq:LHb}
    L_{\rm H\beta}(t)=\int \Psi_{\rm H\beta}(\tau)L_{\rm ion}(t-\tau)d\tau,
\end{equation}
where $\Psi$ are transfer functions of the 5100\,\AA, ionizing sources and H$\beta$ regions.
Inserting Eq.\,(\ref{eq:Lion}) into (\ref{eq:LHb}), we have
\begin{equation}\label{eq:LHbeta}
     L_{\rm H\beta}(t) =\int d\tau \Psi_{\rm H\beta}(\tau)
     \int d\tau^{\prime} \Psi_{\rm ion}(\tau^{\prime})L_{\rm X}(t-\tau-\tau^{\prime}).
\end{equation}
In the stretched SMBH-disk, $\Psi_{\rm ion}(\tau)\approx \delta(\tau-\tau_0)$ is still valid by $\delta$ function (ionizing source is a narrow ring) in the face-on case, we have
\begin{equation}\label{eq:LHb2}
     L_{\rm H\beta}(t) =\int \Psi_{\rm H\beta}(\tau-\tau_0)L_{\rm X}(t-\tau) d\tau.
\end{equation}
By performing a Fourier transformation to Eqs.(\ref{eq:L5100}) and (\ref{eq:LHb2}), we have 
\begin{equation}\label{eq:LHb3}
    \mathcal{F}[L_{\rm H\beta}(t)] = \frac{\mathcal{F}[\Psi_{\rm H\beta}(\tau-\tau_0)]}{\mathcal{F}[\Psi_{\rm 5100}(\tau)]}\times \mathcal{F}[L_{5100}(t)],
\end{equation}
where $\mathcal{F}$ denotes the Fourier transformation. Eq.(\ref{eq:LHb3}) corresponds to a convolution in the time domain as
\begin{equation}\label{eq:LHb4}
    L_{\rm H\beta}(t) = \int \Psi_{\rm R}(\tau) L_{5100}(t-\tau) d\tau,
\end{equation}
where 
\begin{equation}\label{eq:tfr}
    \Psi_{\rm R}(\tau) = \mathcal{F}^{-1}\left[\frac{\mathcal{F}[\Psi_{\rm H\beta}(\tau-\tau_0)]}{\mathcal{F}[\Psi_{\rm 5100}(\tau)]}\right].
\end{equation}
The current analysis of \JAVELIN\ and \MICA\ is equivalent to solving Eq.\,(\ref{eq:LHb4}). The obtained transfer functions therefore represent an approximation to $\Psi_{\rm R}(\tau)$. 

In cases of $R_{5100}\ll \Rhb$, it is safe to approximate $\Psi_{5100}(\tau)\approx\delta(\tau-\tau_{5100})$, where $\tau_1$ is the delay with respect to the X-ray source. 
Then Eq.\,(\ref{eq:tfr}) will be  simplified to $\Psi_{\rm R}(\tau) = \Psi_{\rm H\beta}\left(\tau-\tau_0+\tau_{5100}\right)$.
In cases of $R_{5100}\sim \Rhb$, $\Psi_{5100}(\tau)$ cannot be simply treated as a $\delta$ function. 
If we approximate $\Psi_{\rm H\beta}(\tau)$ and $\Psi_{5100}(\tau)$ as Gaussians with the respective  center $\tau_{\rm H\beta}$ and $\tau_{\rm 5100}$ and the respective standard deviation $\sigma_{\rm H\beta}$ and $\sigma_{5100}$, Eq.\,(\ref{eq:tfr}) illustrates that $\Psi_{\rm R}(\tau)$ is also a Gaussian. 
The Gaussian center is equal to $\tau_{\rm R}=\tau_{\rm H\beta}+\tau_0-\tau_{5100}$ and the Gaussian standard deviation is equal to $\sigma_{\rm R}=\left(\sigma_{\rm H\beta}^2-\sigma_{5100}^2\right)^{1/2}$. 
In practice, $\Psi_{\rm H\beta}(\tau)$ and $\Psi_{5100}(\tau)$ might exhibit more complicated shapes, rather than following a simple Gaussian profile. 
However, it remains valid that the time lags measured by \JAVELIN\ and \MICA\ analysis on the H$\beta$ and 5100\,\AA\ LCs can be approximately regarded as the difference between $\tau_{\rm H\beta}+\tau_0$ and $\tau_{5100}$, which typically represent the sizes of the H$\beta$ and 5100\,{\AA} regions, respectively. 
In other words, the measured time lags reflect the size difference and a negative H$\beta$ time lag with respect to 5100\,{\AA} means that the H$\beta$ region is located inside the 5100\,{\AA} regions.

We note that discussions presented here remain in an analytical form and our simulations in \S\,4 are also for characterized models. 
Realistic models of BLRs and accretion disks for the transfer functions along with a toy model of the anomalous reverberations will be considered for PHL\,1092 in an separate paper.

\section{Discussions on potential models}\label{sec:other-models}
The negative lags set strong constrains on the configurations of the BLR and SMBH-disk.
There are several potential models listed below and we briefly comment on them.
Based on the subsequent discussions, we would draw a conclusion that the leading delays are plausibly driven by stretching SMBH-disk through sMBHs heating.

\subsection{3D geometry of disk.} We made efforts to explain the shortening of \Hbeta\ lags in super-Eddington AGNs, where the inner funnel of slim disks makes anisotropic radiation toward BLR clouds and observers\cite{Wang2014b}.
This effect can account for a shortening factor of $\sim 10^{-0.35\calR}\lesssim 5$, however, cannot generate the negative lags observed in PHL\,1092 unless the BLR clouds are predominantly located in the pole regions. 
This model is therefore disfavored for explaining leading delays.

\subsection{An outflowing BLR over the disk.} 
Outflowing BLRs parallel to the disk surface have been extensively studied\cite{Chiang1996}, assuming the ionizing regions of accretion disks are point-like.
While lags in this scenario are complex, no leading delays have been found in their simulations in any case. 
The LDRs necessarily require an extended ionizing source. 
Additionally, pole-dominated winds can produce leading reverberation delays, but extended ionizing sources larger than the cross-sectional radius of the outflowing BLRs are still necessary. 

\subsection{Anomalous viscosity in accretion disk.} 
In fact, the energy dissipation rate per unit area is fully independent of viscosity/turbulence physics—unless the gas has non-Keplerian rotation.
This was elegantly demonstrated by Lynden-Bell\cite{Lynden-Bell1969} and is also covered in standard textbooks\cite{Frank2002}. 
By contrast, radiation from slim disks\cite{Abramowicz1988} depends on viscosity, as these disks are sub-Keplerian.
To generate LDRs, viscosity would need to be modified into a complex form producing non-monotonic effective temperature distributions in the disk’s outer regions (e.g., a double power-law with respect to disk radius). 
However, such an unusual viscosity profile requires specific physical mechanisms. 
Again, \sSMBH\ may fulfill this role in LDR formation.

\subsection{Extended winds from super-Eddington accretion.} 
Recently, extended winds from super-Eddington accretion have been proposed to play a key role in explaining "little red dots"\cite{Liu2025}, producing the observed Balmer break. 
Intriguingly, these winds might extend to the broad line region (BLR), potentially generating the LDRs.
For a simple disk wind estimate, its radius is given by $R_{\rm w}/\Rg=\tau_{\rm es}\beta_{\rm w}\dot{M}_{\rm w}/\dot{M}_{\rm Edd}$ from an approximate spherical wind model: $\dot{M}_{\rm w}=4\pi R_{\rm w}^2V_{\rm w}\rho_{\rm w}$, where $\beta_{\rm w}=V_{\rm w}/c$ is the wind velocity, $\tau_{\rm es}=\rho_{\rm w}\kappa_{\rm es}R_{\rm w}$ is the Thomson scattering depth of the winds, and $\rho_{\rm w}$ is wind mass density.
If 5100\,{\AA} photons are scattered by these winds, $R_{\rm w}\gtrsim R_{\rm H\beta}$ is required, implying a lower limit on the central super-massive black hole (SMBH) accretion rate $\mathdotM\gtrsim \dot{M}_{\rm w}/\dot{M}_{\rm Edd}$. 
This yields $\mathdotM\gtrsim 10^3\,\left(\tau_{\rm es}\beta_{\rm w}/0.1\right)\left(R_{\rm H\beta}/10^4\Rg\right)$, indicating extremely high slim disk accretion rates are needed to produce such strong winds.

Observationally, our campaigns have identified a few super-Eddington accreting massive black holes (SEAMBHs) with potentially high rates (e.g., IRAS 04416+1215: $\mathdotM=10^{2.63}$; SDSS J084533.28+474934.5: $\mathdotM=10^{2.76}$; SDSS J100402.61+285535.3: $\mathdotM=10^{2.89}$)\cite{Du2019}. 
However, these SEAMBHs do neither exhibit absorbing features of H$\beta$ nor the LDRs, and their velocity-resolved delays do not support strong outflows in SEAMBHs\cite{Du2016b}.
Though we are not able to rule out this possibility within ranges of the parameters, this simple estimate thus disfavors slim disk winds as the cause of LDRs. 
That said, the idea remains interesting and merits further qualitative exploration for both lag shortening and LDRs.

\subsection{Satellite sMBHs.} 
There is growing evidence for the presence of satellite black holes around the central SMBH.
For instance, mHz quasi-periodical oscillations (QPOs) of X-ray variations in AGN 1ES\,1927+654\cite{Masterson2025}, and  quasi-periodical ejections (QPEs) in galaxies of Seyfert 2 galaxy GSN\,069\cite{Miniutti2019} and others\cite{Giustini2020,Arcodia2021,Chakraborty2021,Arcordia2024}.
Moreover, high metallicity of AGN BLRs indicate formation of massive stars inside the accretion disks\cite{Cantiello2021,Wang2023,Chen2025}, and compact objects from evolution of massive stars remain there\cite{Cheng1999}.
The \sSMBH\ suggested for the LDRs is generally consistent with the above evidence and arguments.
sMBHs must play a role in variations of continuum and BLRs somehow\cite{Wang2021a,Wang2021b}, yielding the anomalous reverberations of broad H$\beta$ line.
In such a context, GWs are expected from detections of LIGO, LISA\cite{Amaro-Seoane2023}/Tianqin\cite{Luo2025} and ET\cite{DiGiovanni2025}. 
LIGO already has observational hints from joint analyses with long-term photometric monitoring campaigns\cite{Abbott2020,GW231123,Graham2020,Graham2023,McKernan2024,He2025,Yang2025}.

In summary, we prefer \sSMBH\ scenario over the aforementioned models to explain LDRs. 
Considering that the PHL\,1092 analogs are not so rare, we recognize that these alternative models as extreme configurations to explain the LDRs become slim. However, we emphasize that they require quantitative investigation in future.


\noindent {\bf References}

\bibliographystyle{naturemag}
\bibliography{reference}

@ARTICLE{WangBon2020,
       author = {{Wang}, Jian-Min and {Bon}, Edi},
        title = "{Changing-look active galactic nuclei: close binaries of supermassive black holes in action}",
      journal = {\aap},
         year = 2020,
        month = nov,
       volume = {643},
          eid = {L9},
        pages = {L9},
          doi = {10.1051/0004-6361/202039368},
archivePrefix = {arXiv},
       eprint = {2010.04417},
 primaryClass = {astro-ph.GA},
       adsurl = {https://ui.adsabs.harvard.edu/abs/2020A&A...643L...9W}
}

@ARTICLE{Guo2025,
       author = {{Guo}, Wei-Jian and {Zou}, Hu and {Greenwell}, Claire L. and {Alexander}, David M. and {Fawcett}, Victoria A. and {Pan}, Zhiwei and {Siudek}, Ma{\l}gorzata and {Aguilar}, Jessica Nicole and {Ahlen}, Steven and {Brooks}, David and {Claybaugh}, Todd and {Dawson}, Kyle and {de la Macorra}, Axel and {Doel}, Peter and {Font-Ribera}, Andreu and {Gazta{\~n}aga}, Enrique and {Gontcho A Gontcho}, Satya and {Gutierrez}, Gaston and {Kehoe}, Robert and {Kisner}, Theodore and {Landriau}, Martin and {Le Guillou}, Laurent and {Manera}, Marc and {Meisner}, Aaron and {Miquel}, Ramon and {Moustakas}, John and {Prada}, Francisco and {Rossi}, Graziano and {Sanchez}, Eusebio and {Schubnell}, Michael and {Sprayberry}, David and {Sui}, Jipeng and {Tarl{\'e}}, Gregory and {Weaver}, Benjamin Alan and {Xiao}, Yun-Ao and {Zou}, Siwei},
        title = "{Changing-look Active Galactic Nuclei from the Dark Energy Spectroscopic Instrument. II. Statistical Properties from the First Data Release}",
      journal = {\apjs},
         year = 2025,
        month = may,
       volume = {278},
       number = {1},
          eid = {28},
        pages = {28},
          doi = {10.3847/1538-4365/adc124},
archivePrefix = {arXiv},
       eprint = {2408.00402},
 primaryClass = {astro-ph.GA},
       adsurl = {https://ui.adsabs.harvard.edu/abs/2025ApJS..278...28G}
}

@ARTICLE{Komossa2024,
       author = {{Komossa}, S. and {Grupe}, D.},
        title = "{The Extremes of Continuum and Emission-Line Variability of AGN: Changing-Look Events and Binary SMBHS}",
      journal = {Serbian Astronomical Journal},
         year = 2024,
        month = dec,
       volume = {209},
        pages = {1-24},
          doi = {10.2298/SAJ2409001K},
       adsurl = {https://ui.adsabs.harvard.edu/abs/2024SerAJ.209....1K}
}

@ARTICLE{Artymowicz1993,
       author = {{Artymowicz}, Pawel and {Lin}, D.~N.~C. and {Wampler}, E.~J.},
        title = "{Star Trapping and Metallicity Enrichment in Quasars and Active Galactic Nuclei}",
      journal = {\apj},
         year = 1993,
        month = jun,
       volume = {409},
        pages = {592},
          doi = {10.1086/172690},
       adsurl = {https://ui.adsabs.harvard.edu/abs/1993ApJ...409..592A}
}

@ARTICLE{Amaro-Seoane2023,
       author = {{Amaro-Seoane}, Pau and {Andrews}, Jeff and {Arca Sedda}, Manuel and {Askar}, Abbas and {Baghi}, Quentin and {Balasov}, Razvan and {Bartos}, Imre and {Bavera}, Simone S. and {Bellovary}, Jillian and {Berry}, Christopher P.~L. and {Berti}, Emanuele and {Bianchi}, Stefano and {Blecha}, Laura and {Blondin}, St{\'e}phane and {Bogdanovi{\'c}}, Tamara and {Boissier}, Samuel and {Bonetti}, Matteo and {Bonoli}, Silvia and {Bortolas}, Elisa and {Breivik}, Katelyn and {Capelo}, Pedro R. and {Caramete}, Laurentiu and {Cattorini}, Federico and {Charisi}, Maria and {Chaty}, Sylvain and {Chen}, Xian and {Chru{\'s}li{\'n}ska}, Martyna and {Chua}, Alvin J.~K. and {Church}, Ross and {Colpi}, Monica and {D'Orazio}, Daniel and {Danielski}, Camilla and {Davies}, Melvyn B. and {Dayal}, Pratika and {De Rosa}, Alessandra and {Derdzinski}, Andrea and {Destounis}, Kyriakos and {Dotti}, Massimo and {Du{\c{t}}an}, Ioana and {Dvorkin}, Irina and {Fabj}, Gaia and {Foglizzo}, Thierry and {Ford}, Saavik and {Fouvry}, Jean-Baptiste and {Franchini}, Alessia and {Fragos}, Tassos and {Fryer}, Chris and {Gaspari}, Massimo and {Gerosa}, Davide and {Graziani}, Luca and {Groot}, Paul and {Habouzit}, Melanie and {Haggard}, Daryl and {Haiman}, Zoltan and {Han}, Wen-Biao and {Istrate}, Alina and {Johansson}, Peter H. and {Khan}, Fazeel Mahmood and {Kimpson}, Tomas and {Kokkotas}, Kostas and {Kong}, Albert and {Korol}, Valeriya and {Kremer}, Kyle and {Kupfer}, Thomas and {Lamberts}, Astrid and {Larson}, Shane and {Lau}, Mike and {Liu}, Dongliang and {Lloyd-Ronning}, Nicole and {Lodato}, Giuseppe and {Lupi}, Alessandro and {Ma}, Chung-Pei and {Maccarone}, Tomas and {Mandel}, Ilya and {Mangiagli}, Alberto and {Mapelli}, Michela and {Mathis}, St{\'e}phane and {Mayer}, Lucio and {McGee}, Sean and {McKernan}, Berry and {Miller}, M. Coleman and {Mota}, David F. and {Mumpower}, Matthew and {Nasim}, Syeda S. and {Nelemans}, Gijs and {Noble}, Scott and {Pacucci}, Fabio and {Panessa}, Francesca and {Paschalidis}, Vasileios and {Pfister}, Hugo and {Porquet}, Delphine and {Quenby}, John and {Ricarte}, Angelo and {R{\"o}pke}, Friedrich K. and {Regan}, John and {Rosswog}, Stephan and {Ruiter}, Ashley and {Ruiz}, Milton and {Runnoe}, Jessie and {Schneider}, Raffaella and {Schnittman}, Jeremy and {Secunda}, Amy and {Sesana}, Alberto and {Seto}, Naoki and {Shao}, Lijing and {Shapiro}, Stuart and {Sopuerta}, Carlos and {Stone}, Nicholas C. and {Suvorov}, Arthur and {Tamanini}, Nicola and {Tamfal}, Tomas and {Tauris}, Thomas and {Temmink}, Karel and {Tomsick}, John and {Toonen}, Silvia and {Torres-Orjuela}, Alejandro and {Toscani}, Martina and {Tsokaros}, Antonios and {Unal}, Caner and {V{\'a}zquez-Aceves}, Ver{\'o}nica and {Valiante}, Rosa and {van Putten}, Maurice and {van Roestel}, Jan and {Vignali}, Christian and {Volonteri}, Marta and {Wu}, Kinwah and {Younsi}, Ziri and {Yu}, Shenghua and {Zane}, Silvia and {Zwick}, Lorenz and {Antonini}, Fabio and {Baibhav}, Vishal and {Barausse}, Enrico and {Bonilla Rivera}, Alexander and {Branchesi}, Marica and {Branduardi-Raymont}, Graziella and {Burdge}, Kevin and {Chakraborty}, Srija and {Cuadra}, Jorge and {Dage}, Kristen and {Davis}, Benjamin and {de Mink}, Selma E. and {Decarli}, Roberto and {Doneva}, Daniela and {Escoffier}, Stephanie and {Gandhi}, Poshak and {Haardt}, Francesco and {Lousto}, Carlos O. and {Nissanke}, Samaya and {Nordhaus}, Jason and {O'Shaughnessy}, Richard and {Portegies Zwart}, Simon and {Pound}, Adam and {Schussler}, Fabian and {Sergijenko}, Olga and {Spallicci}, Alessandro and {Vernieri}, Daniele and {Vigna-G{\'o}mez}, Alejandro},
        title = "{Astrophysics with the Laser Interferometer Space Antenna}",
      journal = {Living Reviews in Relativity},
         year = 2023,
        month = dec,
       volume = {26},
       number = {1},
          eid = {2},
        pages = {2},
          doi = {10.1007/s41114-022-00041-y},
archivePrefix = {arXiv},
       eprint = {2203.06016},
 primaryClass = {gr-qc},
       adsurl = {https://ui.adsabs.harvard.edu/abs/2023LRR....26....2A}
}

@ARTICLE{Nakoneczny2025,
       author = {{Nakoneczny}, S.~J. and {Graham}, M.~J. and {Stern}, D. and {Helou}, G. and {Djorgovski}, S.~G. and {Bellm}, E.~C. and {Chen}, T.~X. and {Dekany}, R. and {Drake}, A. and {Mahabal}, A.~A. and {Prince}, T.~A. and {Riddle}, R. and {Rusholme}, B. and {Sravan}, N.},
        title = "{QZO: A Catalog of 5 Million Quasars from the Zwicky Transient Facility}",
      journal = {arXiv e-prints},
         year = 2025,
        month = feb,
          eid = {arXiv:2502.13054},
        pages = {arXiv:2502.13054},
          doi = {10.48550/arXiv.2502.13054},
archivePrefix = {arXiv},
       eprint = {2502.13054},
 primaryClass = {astro-ph.GA},
       adsurl = {https://ui.adsabs.harvard.edu/abs/2025arXiv250213054N}
}

@ARTICLE{Flesch2023,
       author = {{Flesch}, Eric Wim},
        title = "{The Million Quasars (Milliquas) Catalogue, v8}",
      journal = {The Open Journal of Astrophysics},
         year = 2023,
        month = dec,
       volume = {6},
          eid = {49},
        pages = {49},
          doi = {10.21105/astro.2308.01505},
archivePrefix = {arXiv},
       eprint = {2308.01505},
 primaryClass = {astro-ph.GA},
       adsurl = {https://ui.adsabs.harvard.edu/abs/2023OJAp....6E..49F}
}

@ARTICLE{Chen2025,
       author = {{Chen}, Yi-Xian and {Jiang}, Yan-Fei and {Goodman}, Jeremy},
        title = "{Accretion of Active Galactic Nucleus Stars Under the Influence of Disk Geometry}",
      journal = {\apj},
         year = 2025,
        month = jul,
       volume = {987},
       number = {2},
          eid = {188},
        pages = {188},
          doi = {10.3847/1538-4357/addd0a},
archivePrefix = {arXiv},
       eprint = {2505.13951},
 primaryClass = {astro-ph.HE},
       adsurl = {https://ui.adsabs.harvard.edu/abs/2025ApJ...987..188C}
}

@ARTICLE{Jiang2025,
       author = {{Jiang}, Ning and {Pan}, Zhen},
        title = "{Embers of Active Galactic Nuclei: Tidal Disruption Events and Quasiperiodic Eruptions}",
      journal = {\apjl},
         year = 2025,
        month = apr,
       volume = {983},
       number = {1},
          eid = {L18},
        pages = {L18},
          doi = {10.3847/2041-8213/adc456},
archivePrefix = {arXiv},
       eprint = {2503.17609},
 primaryClass = {astro-ph.HE},
       adsurl = {https://ui.adsabs.harvard.edu/abs/2025ApJ...983L..18J}
}

@ARTICLE{Arcordia2024,
       author = {{Arcodia}, R. and {Liu}, Z. and {Merloni}, A. and {Malyali}, A. and {Rau}, A. and {Chakraborty}, J. and {Goodwin}, A. and {Buckley}, D. and {Brink}, J. and {Gromadzki}, M. and {Arzoumanian}, Z. and {Buchner}, J. and {Kara}, E. and {Nandra}, K. and {Ponti}, G. and {Salvato}, M. and {Anderson}, G. and {Baldini}, P. and {Grotova}, I. and {Krumpe}, M. and {Maitra}, C. and {Miller-Jones}, J.~C.~A. and {Ramos-Ceja}, M.~E.},
        title = "{The more the merrier: SRG/eROSITA discovers two further galaxies showing X-ray quasi-periodic eruptions}",
      journal = {\aap},
         year = 2024,
        month = apr,
       volume = {684},
          eid = {A64},
        pages = {A64},
          doi = {10.1051/0004-6361/202348881},
archivePrefix = {arXiv},
       eprint = {2401.17275},
 primaryClass = {astro-ph.HE},
       adsurl = {https://ui.adsabs.harvard.edu/abs/2024A&A...684A..64A}
}

@ARTICLE{Chakraborty2021,
       author = {{Chakraborty}, Joheen and {Kara}, Erin and {Masterson}, Megan and {Giustini}, Margherita and {Miniutti}, Giovanni and {Saxton}, Richard},
        title = "{Possible X-Ray Quasi-periodic Eruptions in a Tidal Disruption Event Candidate}",
      journal = {\apjl},
         year = 2021,
        month = nov,
       volume = {921},
       number = {2},
          eid = {L40},
        pages = {L40},
          doi = {10.3847/2041-8213/ac313b},
archivePrefix = {arXiv},
       eprint = {2110.10786},
 primaryClass = {astro-ph.HE},
       adsurl = {https://ui.adsabs.harvard.edu/abs/2021ApJ...921L..40C}
}

@ARTICLE{Arcodia2021,
       author = {{Arcodia}, R. and {Merloni}, A. and {Nandra}, K. and {Buchner}, J. and {Salvato}, M. and {Pasham}, D. and {Remillard}, R. and {Comparat}, J. and {Lamer}, G. and {Ponti}, G. and {Malyali}, A. and {Wolf}, J. and {Arzoumanian}, Z. and {Bogensberger}, D. and {Buckley}, D.~A.~H. and {Gendreau}, K. and {Gromadzki}, M. and {Kara}, E. and {Krumpe}, M. and {Markwardt}, C. and {Ramos-Ceja}, M.~E. and {Rau}, A. and {Schramm}, M. and {Schwope}, A.},
        title = "{X-ray quasi-periodic eruptions from two previously quiescent galaxies}",
      journal = {\nat},
         year = 2021,
        month = apr,
       volume = {592},
       number = {7856},
        pages = {704-707},
          doi = {10.1038/s41586-021-03394-6},
archivePrefix = {arXiv},
       eprint = {2104.13388},
 primaryClass = {astro-ph.HE},
       adsurl = {https://ui.adsabs.harvard.edu/abs/2021Natur.592..704A}
}

@ARTICLE{Giustini2020,
       author = {{Giustini}, Margherita and {Miniutti}, Giovanni and {Saxton}, Richard D.},
        title = "{X-ray quasi-periodic eruptions from the galactic nucleus of RX J1301.9+2747}",
      journal = {\aap},
         year = 2020,
        month = apr,
       volume = {636},
          eid = {L2},
        pages = {L2},
          doi = {10.1051/0004-6361/202037610},
archivePrefix = {arXiv},
       eprint = {2002.08967},
 primaryClass = {astro-ph.HE},
       adsurl = {https://ui.adsabs.harvard.edu/abs/2020A&A...636L...2G}
}

@ARTICLE{Miniutti2019,
       author = {{Miniutti}, G. and {Saxton}, R.~D. and {Giustini}, M. and {Alexander}, K.~D. and {Fender}, R.~P. and {Heywood}, I. and {Monageng}, I. and {Coriat}, M. and {Tzioumis}, A.~K. and {Read}, A.~M. and {Knigge}, C. and {Gandhi}, P. and {Pretorius}, M.~L. and {Ag{\'\i}s-Gonz{\'a}lez}, B.},
        title = "{Nine-hour X-ray quasi-periodic eruptions from a low-mass black hole galactic nucleus}",
      journal = {\nat},
         year = 2019,
        month = sep,
       volume = {573},
       number = {7774},
        pages = {381-384},
          doi = {10.1038/s41586-019-1556-x},
archivePrefix = {arXiv},
       eprint = {1909.04693},
 primaryClass = {astro-ph.HE},
       adsurl = {https://ui.adsabs.harvard.edu/abs/2019Natur.573..381M}
}

@ARTICLE{Masterson2025,
       author = {{Masterson}, Megan and {Kara}, Erin and {Panagiotou}, Christos and {Alston}, William N. and {Chakraborty}, Joheen and {Burdge}, Kevin and {Ricci}, Claudio and {Laha}, Sibasish and {Arcavi}, Iair and {Arcodia}, Riccardo and {Cenko}, S. Bradley and {Fabian}, Andrew C. and {Garc{\'\i}a}, Javier A. and {Giustini}, Margherita and {Ingram}, Adam and {Kosec}, Peter and {Loewenstein}, Michael and {Meyer}, Eileen T. and {Miniutti}, Giovanni and {Pinto}, Ciro and {Remillard}, Ronald A. and {Sadaula}, Dev R. and {Shuvo}, Onic I. and {Trakhtenbrot}, Benny and {Wang}, Jingyi},
        title = "{Millihertz oscillations near the innermost orbit of a supermassive black hole}",
      journal = {\nat},
         year = 2025,
        month = feb,
       volume = {638},
       number = {8050},
        pages = {370-375},
          doi = {10.1038/s41586-024-08385-x},
archivePrefix = {arXiv},
       eprint = {2501.01581},
 primaryClass = {astro-ph.HE},
       adsurl = {https://ui.adsabs.harvard.edu/abs/2025Natur.638..370M}
}

@ARTICLE{Krawczyk2013,
       author = {{Krawczyk}, Coleman M. and {Richards}, Gordon T. and {Mehta}, Sajjan S. and {Vogeley}, Michael S. and {Gallagher}, S.~C. and {Leighly}, Karen M. and {Ross}, Nicholas P. and {Schneider}, Donald P.},
        title = "{Mean Spectral Energy Distributions and Bolometric Corrections for Luminous Quasars}",
      journal = {\apjs},
         year = 2013,
        month = may,
       volume = {206},
       number = {1},
          eid = {4},
        pages = {4},
          doi = {10.1088/0067-0049/206/1/4},
archivePrefix = {arXiv},
       eprint = {1304.5573},
 primaryClass = {astro-ph.CO},
       adsurl = {https://ui.adsabs.harvard.edu/abs/2013ApJS..206....4K}
}

@ARTICLE{Marziani2001,
       author = {{Marziani}, P. and {Sulentic}, J.~W. and {Zwitter}, T. and {Dultzin-Hacyan}, D. and {Calvani}, M.},
        title = "{Searching for the Physical Drivers of the Eigenvector 1 Correlation Space}",
      journal = {\apj},
         year = 2001,
        month = sep,
       volume = {558},
       number = {2},
        pages = {553-560},
          doi = {10.1086/322286},
archivePrefix = {arXiv},
       eprint = {astro-ph/0105343},
 primaryClass = {astro-ph},
       adsurl = {https://ui.adsabs.harvard.edu/abs/2001ApJ...558..553M}
}

@ARTICLE{Plotkin2010,
       author = {{Plotkin}, Richard M. and {Anderson}, Scott F. and {Brandt}, W.~N. and {Diamond-Stanic}, Aleksandar M. and {Fan}, Xiaohui and {MacLeod}, Chelsea L. and {Schneider}, Donald P. and {Shemmer}, Ohad},
        title = "{Multiwavelength Observations of Radio-quiet Quasars with Weak Emission Lines}",
      journal = {\apj},
         year = 2010,
        month = sep,
       volume = {721},
       number = {1},
        pages = {562-575},
          doi = {10.1088/0004-637X/721/1/562},
archivePrefix = {arXiv},
       eprint = {1007.5058},
 primaryClass = {astro-ph.CO},
       adsurl = {https://ui.adsabs.harvard.edu/abs/2010ApJ...721..562P}
}

@ARTICLE{Liu2025,
       author = {{Liu}, Hanpu and {Jiang}, Yan-Fei and {Quataert}, Eliot and {Greene}, Jenny E. and {Ma}, Yilun},
        title = "{The Balmer Break and Optical Continuum of Little Red Dots From Super-Eddington Accretion}",
      journal = {arXiv e-prints},
         year = 2025,
        month = jul,
          eid = {arXiv:2507.07190},
        pages = {arXiv:2507.07190},
          doi = {10.48550/arXiv.2507.07190},
archivePrefix = {arXiv},
       eprint = {2507.07190},
 primaryClass = {astro-ph.GA},
       adsurl = {https://ui.adsabs.harvard.edu/abs/2025arXiv250707190L}
}

@PROCEEDINGS{Netzer1990,
        title = "{Active Galactic Nuclei}",
    booktitle = {Active Galactic Nuclei},
         year = 1990,
       editor = {{Blandford}, R.~D. and {Netzer}, H. and {Woltjer}, L. and {Courvoisier}, T.~J. -L. and {Mayor}, M.},
        month = jan,
       adsurl = {https://ui.adsabs.harvard.edu/abs/1990agn..conf.....B}
}

@ARTICLE{Chiang1996,
       author = {{Chiang}, J. and {Murray}, N.},
        title = "{Reverberation Mapping and the Disk-Wind Model of the Broad-Line Region}",
      journal = {\apj},
         year = 1996,
        month = aug,
       volume = {466},
        pages = {704},
          doi = {10.1086/177543},
archivePrefix = {arXiv},
       eprint = {astro-ph/9511006},
 primaryClass = {astro-ph},
       adsurl = {https://ui.adsabs.harvard.edu/abs/1996ApJ...466..704C}
}

@ARTICLE{DiGiovanni2025,
       author = {{Di Giovanni}, Matteo},
        title = "{Einstein Telescope and Cosmic Explorer}",
      journal = {arXiv e-prints},
         year = 2025,
        month = may,
          eid = {arXiv:2505.11033},
        pages = {arXiv:2505.11033},
          doi = {10.48550/arXiv.2505.11033},
archivePrefix = {arXiv},
       eprint = {2505.11033},
 primaryClass = {gr-qc},
       adsurl = {https://ui.adsabs.harvard.edu/abs/2025arXiv250511033D}
}

@ARTICLE{Vaccaro2025,
       author = {{Vaccaro}, Maria Paola},
        title = "{Hierarchical Black Hole Mergers in AGN Disks: Tracing Massive Black Hole Growth Across Cosmic Time}",
      journal = {arXiv e-prints},
         year = 2025,
        month = aug,
          eid = {arXiv:2508.15337},
        pages = {arXiv:2508.15337},
          doi = {10.48550/arXiv.2508.15337},
archivePrefix = {arXiv},
       eprint = {2508.15337},
 primaryClass = {astro-ph.HE},
       adsurl = {https://ui.adsabs.harvard.edu/abs/2025arXiv250815337V}
}

@ARTICLE{Pan2021,
       author = {{Pan}, Xiang and {Zhou}, Hongyan and {Yang}, Chenwei and {Sun}, Luming and {Smith}, Paul S. and {Ji}, Tuo and {Jiang}, Ning and {Jiang}, Peng and {Liu}, Wenjuan and {Lu}, Honglin and {Shi}, Xiheng and {Dai}, Xuejie and {Zhang}, Shaohua},
        title = "{Mrk 1239: a Type-2 Counterpart of Narrow-line Seyfert-1?}",
      journal = {\apj},
         year = 2021,
        month = may,
       volume = {912},
       number = {2},
          eid = {118},
        pages = {118},
          doi = {10.3847/1538-4357/abf148},
archivePrefix = {arXiv},
       eprint = {2105.13672},
 primaryClass = {astro-ph.GA},
       adsurl = {https://ui.adsabs.harvard.edu/abs/2021ApJ...912..118P}
}

@ARTICLE{He2025,
       author = {{He}, Lei and {Liu}, Zheng-Yan and {Niu}, Rui and {Zhou}, Ming-Shen and {Zou}, Pu-Run and {Gao}, Bing-Zhou and {Liang}, Run-Duo and {Zhu}, Liang-Gui and {Wang}, Jian-Min and {Jiang}, Ning and {Cai}, Zhen-Yi and {Jiang}, Ji-an and {Dai}, Zi-Gao and {Yuan}, Ye-Fei and {Chen}, Yong-Jie and {Zhao}, Wen},
        title = "{A Systematic Search for AGN Flares in ZTF Data Release 23}",
      journal = {arXiv e-prints},
         year = 2025,
        month = jul,
          eid = {arXiv:2507.20232},
        pages = {arXiv:2507.20232},
          doi = {10.48550/arXiv.2507.20232},
archivePrefix = {arXiv},
       eprint = {2507.20232},
 primaryClass = {astro-ph.HE},
       adsurl = {https://ui.adsabs.harvard.edu/abs/2025arXiv250720232H}
}

@ARTICLE{GW231123,
       author = {{The LIGO Scientific Collaboration} and {the Virgo Collaboration} and {the KAGRA Collaboration}},
        title = "{GW231123: a Binary Black Hole Merger with Total Mass $190-265\, M_{\odot}$}",
      journal = {arXiv e-prints},
         year = 2025,
        month = jul,
          eid = {arXiv:2507.08219},
        pages = {arXiv:2507.08219},
          doi = {10.48550/arXiv.2507.08219},
archivePrefix = {arXiv},
       eprint = {2507.08219},
 primaryClass = {astro-ph.HE},
       adsurl = {https://ui.adsabs.harvard.edu/abs/2025arXiv250708219T}
}

@ARTICLE{Kara2023,
       author = {{Kara}, Erin and {Barth}, Aaron J. and {Cackett}, Edward M. and {Gelbord}, Jonathan and {Montano}, John and {Li}, Yan-Rong and {Santana}, Lisabeth and {Horne}, Keith and {Alston}, William N. and {Buisson}, Douglas and {Chelouche}, Doron and {Du}, Pu and {Fabian}, Andrew C. and {Fian}, Carina and {Gallo}, Luigi and {Goad}, Michael R. and {Grupe}, Dirk and {Gonz{\'a}lez Buitrago}, Diego H. and {Hern{\'a}ndez Santisteban}, Juan V. and {Kaspi}, Shai and {Hu}, Chen and {Komossa}, S. and {Kriss}, Gerard A. and {Lewin}, Collin and {Lewis}, Tiffany and {Loewenstein}, Michael and {Lohfink}, Anne and {Masterson}, Megan and {McHardy}, Ian M. and {Mehdipour}, Missagh and {Miller}, Jake and {Panagiotou}, Christos and {Parker}, Michael L. and {Pinto}, Ciro and {Remillard}, Ron and {Reynolds}, Christopher and {Rogantini}, Daniele and {Wang}, Jian-Min and {Wang}, Jingyi and {Wilkins}, Dan},
        title = "{UV-Optical Disk Reverberation Lags despite a Faint X-Ray Corona in the Active Galactic Nucleus Mrk 335}",
      journal = {\apj},
         year = 2023,
        month = apr,
       volume = {947},
       number = {2},
          eid = {62},
        pages = {62},
          doi = {10.3847/1538-4357/acbcd3},
archivePrefix = {arXiv},
       eprint = {2302.07342},
 primaryClass = {astro-ph.HE},
       adsurl = {https://ui.adsabs.harvard.edu/abs/2023ApJ...947...62K}
}

@BOOK{Frank2002,
       author = {{Frank}, Juhan and {King}, Andrew and {Raine}, Derek J.},
        title = "{Accretion Power in Astrophysics: Third Edition}",
         year = 2002,
       adsurl = {https://ui.adsabs.harvard.edu/abs/2002apa..book.....F}
}

@ARTICLE{Lynden-Bell1969,
       author = {{Lynden-Bell}, D.},
        title = "{Galactic Nuclei as Collapsed Old Quasars}",
      journal = {\nat},
         year = 1969,
        month = aug,
       volume = {223},
       number = {5207},
        pages = {690-694},
          doi = {10.1038/223690a0},
       adsurl = {https://ui.adsabs.harvard.edu/abs/1969Natur.223..690L}
}

@ARTICLE{Homayouni2024,
       author = {{Homayouni}, Y. and {Kriss}, Gerard A. and {De Rosa}, Gisella and {Plesha}, Rachel and {Cackett}, Edward M. and {Goad}, Michael R. and {Korista}, Kirk T. and {Horne}, Keith and {Fischer}, Travis and {Waters}, Tim and {Barth}, Aaron J. and {Kara}, Erin A. and {Landt}, Hermine and {Arav}, Nahum and {Boizelle}, Benjamin D. and {Bentz}, Misty C. and {Brotherton}, Michael S. and {Chelouche}, Doron and {Dalla Bont{\`a}}, Elena and {Dehghanian}, Maryam and {Du}, Pu and {Ferland}, Gary J. and {Fian}, Carina and {Gelbord}, Jonathan and {Grier}, Catherine J. and {Hall}, Patrick B. and {Hu}, Chen and {Ili{\'c}}, Dragana and {Joner}, Michael D. and {Kaastra}, Jelle and {Kaspi}, Shai and {Kova{\v{c}}evi{\'c}}, Andjelka B. and {Kynoch}, Daniel and {Li}, Yan-Rong and {Mehdipour}, Missagh and {Miller}, Jake A. and {Mitchell}, Jake and {Montano}, John and {Netzer}, Hagai and {Neustadt}, J.~M.~M. and {Partington}, Ethan and {Popovi{\'c}}, Luka {\v{C}}. and {Proga}, Daniel and {Storchi-Bergmann}, Thaisa and {Sanmartim}, David and {Siebert}, Matthew R. and {Treu}, Tommaso and {Vestergaard}, Marianne and {Wang}, Jian-Min and {Ward}, Martin J. and {Zaidouni}, Fatima and {Zu}, Ying},
        title = "{AGN STORM 2. V. Anomalous Behavior of the C IV Light Curve of Mrk 817}",
      journal = {\apj},
         year = 2024,
        month = mar,
       volume = {963},
       number = {2},
          eid = {123},
        pages = {123},
          doi = {10.3847/1538-4357/ad1be4},
archivePrefix = {arXiv},
       eprint = {2308.00742},
 primaryClass = {astro-ph.GA},
       adsurl = {https://ui.adsabs.harvard.edu/abs/2024ApJ...963..123H}
}

@ARTICLE{Peterson2014a,
       author = {{Peterson}, B.~M. and {Grier}, C.~J. and {Horne}, Keith and {Pogge}, R.~W. and {Bentz}, M.~C. and {De Rosa}, G. and {Denney}, K.~D. and {Martini}, Paul and {Sergeev}, S.~G. and {Kaspi}, S. and {Minezaki}, T. and {Zu}, Y. and {Kochanek}, C.~S. and {Siverd}, R.~J. and {Shappee}, B. and {Araya Salvo}, C. and {Beatty}, T.~G. and {Bird}, J.~C. and {Bord}, D.~J. and {Borman}, G.~A. and {Che}, X. and {Chen}, C. -T. and {Cohen}, S.~A. and {Dietrich}, M. and {Doroshenko}, V.~T. and {Drake}, T. and {Efimov}, Yu. S. and {Free}, N. and {Ginsburg}, I. and {Henderson}, C.~B. and {King}, A.~L. and {Koshida}, S. and {Mogren}, K. and {Molina}, M. and {Mosquera}, A.~M. and {Motohara}, K. and {Nazarov}, S.~V. and {Okhmat}, D.~N. and {Pejcha}, O. and {Rafter}, S. and {Shields}, J.~C. and {Skowron}, D.~M. and {Skowron}, J. and {Valluri}, M. and {van Saders}, J.~L. and {Yoshii}, Y.},
        title = "{Reverberation Mapping of the Seyfert 1 Galaxy NGC 7469}",
      journal = {\apj},
         year = 2014,
        month = nov,
       volume = {795},
       number = {2},
          eid = {149},
        pages = {149},
          doi = {10.1088/0004-637X/795/2/149},
archivePrefix = {arXiv},
       eprint = {1409.4448},
 primaryClass = {astro-ph.GA},
       adsurl = {https://ui.adsabs.harvard.edu/abs/2014ApJ...795..149P}
}

@ARTICLE{Pei2017,
       author = {{Pei}, L. and {Fausnaugh}, M.~M. and {Barth}, A.~J. and {Peterson}, B.~M. and {Bentz}, M.~C. and {De Rosa}, G. and {Denney}, K.~D. and {Goad}, M.~R. and {Kochanek}, C.~S. and {Korista}, K.~T. and {Kriss}, G.~A. and {Pogge}, R.~W. and {Bennert}, V.~N. and {Brotherton}, M. and {Clubb}, K.~I. and {Dalla Bont{\`a}}, E. and {Filippenko}, A.~V. and {Greene}, J.~E. and {Grier}, C.~J. and {Vestergaard}, M. and {Zheng}, W. and {Adams}, Scott M. and {Beatty}, Thomas G. and {Bigley}, A. and {Brown}, Jacob E. and {Brown}, Jonathan S. and {Canalizo}, G. and {Comerford}, J.~M. and {Coker}, Carl T. and {Corsini}, E.~M. and {Croft}, S. and {Croxall}, K.~V. and {Deason}, A.~J. and {Eracleous}, Michael and {Fox}, O.~D. and {Gates}, E.~L. and {Henderson}, C.~B. and {Holmbeck}, E. and {Holoien}, T.~W. -S. and {Jensen}, J.~J. and {Johnson}, C.~A. and {Kelly}, P.~L. and {Kim}, S. and {King}, A. and {Lau}, M.~W. and {Li}, Miao and {Lochhaas}, Cassandra and {Ma}, Zhiyuan and {Manne-Nicholas}, E.~R. and {Mauerhan}, J.~C. and {Malkan}, M.~A. and {McGurk}, R. and {Morelli}, L. and {Mosquera}, Ana and {Mudd}, Dale and {Muller Sanchez}, F. and {Nguyen}, M.~L. and {Ochner}, P. and {Ou-Yang}, B. and {Pancoast}, A. and {Penny}, Matthew T. and {Pizzella}, A. and {Poleski}, Rados{\l}aw and {Runnoe}, Jessie and {Scott}, B. and {Schimoia}, Jaderson S. and {Shappee}, B.~J. and {Shivvers}, I. and {Simonian}, Gregory V. and {Siviero}, A. and {Somers}, Garrett and {Stevens}, Daniel J. and {Strauss}, M.~A. and {Tayar}, Jamie and {Tejos}, N. and {Treu}, T. and {Van Saders}, J. and {Vican}, L. and {Villanueva}, Jr., S. and {Yuk}, H. and {Zakamska}, N.~L. and {Zhu}, W. and {Anderson}, M.~D. and {Ar{\'e}valo}, P. and {Bazhaw}, C. and {Bisogni}, S. and {Borman}, G.~A. and {Bottorff}, M.~C. and {Brandt}, W.~N. and {Breeveld}, A.~A. and {Cackett}, E.~M. and {Carini}, M.~T. and {Crenshaw}, D.~M. and {De Lorenzo-C{\'a}ceres}, A. and {Dietrich}, M. and {Edelson}, R. and {Efimova}, N.~V. and {Ely}, J. and {Evans}, P.~A. and {Ferland}, G.~J. and {Flatland}, K. and {Gehrels}, N. and {Geier}, S. and {Gelbord}, J.~M. and {Grupe}, D. and {Gupta}, A. and {Hall}, P.~B. and {Hicks}, S. and {Horenstein}, D. and {Horne}, Keith and {Hutchison}, T. and {Im}, M. and {Joner}, M.~D. and {Jones}, J. and {Kaastra}, J. and {Kaspi}, S. and {Kelly}, B.~C. and {Kennea}, J.~A. and {Kim}, M. and {Kim}, S.~C. and {Klimanov}, S.~A. and {Lee}, J.~C. and {Leonard}, D.~C. and {Lira}, P. and {MacInnis}, F. and {Mathur}, S. and {McHardy}, I.~M. and {Montouri}, C. and {Musso}, R. and {Nazarov}, S.~V. and {Netzer}, H. and {Norris}, R.~P. and {Nousek}, J.~A. and {Okhmat}, D.~N. and {Papadakis}, I. and {Parks}, J.~R. and {Pott}, J. -U. and {Rafter}, S.~E. and {Rix}, H. -W. and {Saylor}, D.~A. and {Schn{\"u}lle}, K. and {Sergeev}, S.~G. and {Siegel}, M. and {Skielboe}, A. and {Spencer}, M. and {Starkey}, D. and {Sung}, H. -I. and {Teems}, K.~G. and {Turner}, C.~S. and {Uttley}, P. and {Villforth}, C. and {Weiss}, Y. and {Woo}, J. -H. and {Yan}, H. and {Young}, S. and {Zu}, Y.},
        title = "{Space Telescope and Optical Reverberation Mapping Project. V. Optical Spectroscopic Campaign and Emission-line Analysis for NGC 5548}",
      journal = {\apj},
         year = 2017,
        month = mar,
       volume = {837},
       number = {2},
          eid = {131},
        pages = {131},
          doi = {10.3847/1538-4357/aa5eb1},
archivePrefix = {arXiv},
       eprint = {1702.01177},
 primaryClass = {astro-ph.GA},
       adsurl = {https://ui.adsabs.harvard.edu/abs/2017ApJ...837..131P}
}

@ARTICLE{Hu2015,
       author = {{Hu}, Chen and {Du}, Pu and {Lu}, Kai-Xing and {Li}, Yan-Rong and {Wang}, Fang and {Qiu}, Jie and {Bai}, Jin-Ming and {Kaspi}, Shai and {Ho}, Luis C. and {Netzer}, Hagai and {Wang}, Jian-Min and {SEAMBH Collaboration}},
        title = "{Supermassive Black Holes with High Accretion Rates in Active Galactic Nuclei. III. Detection of Fe II Reverberation in Nine Narrow-line Seyfert 1 Galaxies}",
      journal = {\apj},
         year = 2015,
        month = may,
       volume = {804},
       number = {2},
          eid = {138},
        pages = {138},
          doi = {10.1088/0004-637X/804/2/138},
archivePrefix = {arXiv},
       eprint = {1503.03611},
 primaryClass = {astro-ph.GA},
       adsurl = {https://ui.adsabs.harvard.edu/abs/2015ApJ...804..138H}
}

@INPROCEEDINGS{Begelman1989,
       author = {{Begelman}, M.~C. and {Frank}, J. and {Shlosman}, I.},
        title = "{Accretion Disks and the Link Between an AGN and its Host Galaxy}",
    booktitle = {Theory of Accretion Disks},
         year = 1989,
       editor = {{Meyer}, Friedrich},
       series = {NATO Advanced Study Institute (ASI) Series C},
       volume = {290},
        month = jan,
        pages = {373},
       adsurl = {https://ui.adsabs.harvard.edu/abs/1989ASIC..290..373B}
}

@ARTICLE{Minezaki2019,
       author = {{Minezaki}, Takeo and {Yoshii}, Yuzuru and {Kobayashi}, Yukiyasu and {Sugawara}, Shota and {Sakata}, Yu and {Enya}, Keigo and {Koshida}, Shintaro and {Tomita}, Hiroyuki and {Suganuma}, Masahiro and {Aoki}, Tsutomu and {Peterson}, Bruce A.},
        title = "{Reverberation Measurements of the Inner Radii of the Dust Tori in Quasars}",
      journal = {\apj},
         year = 2019,
        month = dec,
       volume = {886},
       number = {2},
          eid = {150},
        pages = {150},
          doi = {10.3847/1538-4357/ab4f7b},
archivePrefix = {arXiv},
       eprint = {1910.08722},
 primaryClass = {astro-ph.GA},
       adsurl = {https://ui.adsabs.harvard.edu/abs/2019ApJ...886..150M}
}

@ARTICLE{Miniutti2009,
       author = {{Miniutti}, G. and {Fabian}, A.~C. and {Brandt}, W.~N. and {Gallo}, L.~C. and {Boller}, Th.},
        title = "{PHL 1092 as a transient extreme X-ray weak quasar}",
      journal = {\mnras},
         year = 2009,
        month = jun,
       volume = {396},
       number = {1},
        pages = {L85-L89},
          doi = {10.1111/j.1745-3933.2009.00669.x},
archivePrefix = {arXiv},
       eprint = {0904.3194},
 primaryClass = {astro-ph.CO},
       adsurl = {https://ui.adsabs.harvard.edu/abs/2009MNRAS.396L..85M}
}

@ARTICLE{Marziani2025,
       author = {{Marziani}, Paola and {Garnica Luna}, Karla and {Floris}, Alberto and {del Olmo}, Ascensi{\'o}n and {Deconto-Machado}, Alice and {Buendia-Rios}, Tania M. and {Negrete}, C. Alenka and {Dultzin}, Deborah},
        title = "{Super-Eddington Accretion in Quasars}",
      journal = {Universe},
         year = 2025,
        month = feb,
       volume = {11},
       number = {2},
          eid = {69},
        pages = {69},
          doi = {10.3390/universe11020069},
archivePrefix = {arXiv},
       eprint = {2502.14713},
 primaryClass = {astro-ph.GA},
       adsurl = {https://ui.adsabs.harvard.edu/abs/2025Univ...11...69M}
}

@ARTICLE{Marinello2020,
       author = {{Marinello}, Murilo and {Rodr{\'\i}guez-Ardila}, Alberto and {Marziani}, Paola and {Sigut}, Aaron and {Pradhan}, Anil},
        title = "{Panchromatic properties of the extreme Fe II emitter PHL 1092}",
      journal = {\mnras},
         year = 2020,
        month = may,
       volume = {494},
       number = {3},
        pages = {4187-4202},
          doi = {10.1093/mnras/staa934},
archivePrefix = {arXiv},
       eprint = {2004.01811},
 primaryClass = {astro-ph.GA},
       adsurl = {https://ui.adsabs.harvard.edu/abs/2020MNRAS.494.4187M}
}

@ARTICLE{Grier2013,
       author = {{Grier}, C.~J. and {Peterson}, B.~M. and {Horne}, Keith and {Bentz}, M.~C. and {Pogge}, R.~W. and {Denney}, K.~D. and {De Rosa}, G. and {Martini}, Paul and {Kochanek}, C.~S. and {Zu}, Y. and {Shappee}, B. and {Siverd}, R. and {Beatty}, T.~G. and {Sergeev}, S.~G. and {Kaspi}, S. and {Araya Salvo}, C. and {Bird}, J.~C. and {Bord}, D.~J. and {Borman}, G.~A. and {Che}, X. and {Chen}, C. and {Cohen}, S.~A. and {Dietrich}, M. and {Doroshenko}, V.~T. and {Efimov}, Yu. S. and {Free}, N. and {Ginsburg}, I. and {Henderson}, C.~B. and {King}, A.~L. and {Mogren}, K. and {Molina}, M. and {Mosquera}, A.~M. and {Nazarov}, S.~V. and {Okhmat}, D.~N. and {Pejcha}, O. and {Rafter}, S. and {Shields}, J.~C. and {Skowron}, J. and {Szczygiel}, D.~M. and {Valluri}, M. and {van Saders}, J.~L.},
        title = "{The Structure of the Broad-line Region in Active Galactic Nuclei. I. Reconstructed Velocity-delay Maps}",
      journal = {\apj},
         year = 2013,
        month = feb,
       volume = {764},
       number = {1},
          eid = {47},
        pages = {47},
          doi = {10.1088/0004-637X/764/1/47},
archivePrefix = {arXiv},
       eprint = {1210.2397},
 primaryClass = {astro-ph.CO},
       adsurl = {https://ui.adsabs.harvard.edu/abs/2013ApJ...764...47G}
}

@ARTICLE{Shen2015,
       author = {{Shen}, Yue and {Brandt}, W.~N. and {Dawson}, Kyle S. and {Hall}, Patrick B. and {McGreer}, Ian D. and {Anderson}, Scott F. and {Chen}, Yuguang and {Denney}, Kelly D. and {Eftekharzadeh}, Sarah and {Fan}, Xiaohui and {Gao}, Yang and {Green}, Paul J. and {Greene}, Jenny E. and {Ho}, Luis C. and {Horne}, Keith and {Jiang}, Linhua and {Kelly}, Brandon C. and {Kinemuchi}, Karen and {Kochanek}, Christopher S. and {P{\^a}ris}, Isabelle and {Peters}, Christina M. and {Peterson}, Bradley M. and {Petitjean}, Patrick and {Ponder}, Kara and {Richards}, Gordon T. and {Schneider}, Donald P. and {Seth}, Anil and {Smith}, Robyn N. and {Strauss}, Michael A. and {Tao}, Charling and {Trump}, Jonathan R. and {Wood-Vasey}, W.~M. and {Zu}, Ying and {Eisenstein}, Daniel J. and {Pan}, Kaike and {Bizyaev}, Dmitry and {Malanushenko}, Viktor and {Malanushenko}, Elena and {Oravetz}, Daniel},
        title = "{The Sloan Digital Sky Survey Reverberation Mapping Project: Technical Overview}",
      journal = {\apjs},
         year = 2015,
        month = jan,
       volume = {216},
       number = {1},
          eid = {4},
        pages = {4},
          doi = {10.1088/0067-0049/216/1/4},
archivePrefix = {arXiv},
       eprint = {1408.5970},
 primaryClass = {astro-ph.IM},
       adsurl = {https://ui.adsabs.harvard.edu/abs/2015ApJS..216....4S}
}

@ARTICLE{Panda2023,
       author = {{Panda}, Swayamtrupta and {Marziani}, Paola},
        title = "{High Eddington quasars as discovery tools: current state and challenges}",
      journal = {Frontiers in Astronomy and Space Sciences},
         year = 2023,
        month = may,
       volume = {10},
          eid = {1130103},
        pages = {1130103},
          doi = {10.3389/fspas.2023.1130103},
archivePrefix = {arXiv},
       eprint = {2210.15041},
 primaryClass = {astro-ph.GA},
       adsurl = {https://ui.adsabs.harvard.edu/abs/2023FrASS..1030103P}
}

@ARTICLE{Hu2025,
       author = {{Hu}, Chen and {Yao}, Zhu-Heng and {Chen}, Yong-Jie and {Songsheng}, Yu-Yang and {Wang}, Yi-Lin and {Yang}, Sen and {Zhang}, Hao and {Guo}, Wei-Jian and {Du}, Pu and {Li}, Yan-Rong and {Xiao}, Ming and {Liu}, Jun-Rong and {Bai}, Hua-Rui and {Fang}, Feng-Na and {Fu}, Yi-Xin and {Peng}, Yue-Chang and {Zhai}, Shuo and {Bai}, Jin-Ming and {Ho}, Luis C. and {Brotherton}, Michael S. and {Aceituno}, Jes{\'u}s and {Winkler}, Hartmut and {Wang}, Jian-Min and {Seambh Collaboration}},
        title = "{Supermassive Black Holes with High Accretion Rates in Active Galactic Nuclei. XIV. Long-duration High-cadence Reverberation Mapping Results for 11 PG Quasars}",
      journal = {\apjs},
         year = 2025,
        month = jun,
       volume = {278},
       number = {2},
          eid = {61},
        pages = {61},
          doi = {10.3847/1538-4365/add40b},
archivePrefix = {arXiv},
       eprint = {2505.01993},
 primaryClass = {astro-ph.GA},
       adsurl = {https://ui.adsabs.harvard.edu/abs/2025ApJS..278...61H}
}

@ARTICLE{Reardon2023,
       author = {{Reardon}, Daniel J. and {Zic}, Andrew and {Shannon}, Ryan M. and {Hobbs}, George B. and {Bailes}, Matthew and {Di Marco}, Valentina and {Kapur}, Agastya and {Rogers}, Axl F. and {Thrane}, Eric and {Askew}, Jacob and {Bhat}, N.~D. Ramesh and {Cameron}, Andrew and {Cury{\l}o}, Ma{\l}gorzata and {Coles}, William A. and {Dai}, Shi and {Goncharov}, Boris and {Kerr}, Matthew and {Kulkarni}, Atharva and {Levin}, Yuri and {Lower}, Marcus E. and {Manchester}, Richard N. and {Mandow}, Rami and {Miles}, Matthew T. and {Nathan}, Rowina S. and {Os{\l}owski}, Stefan and {Russell}, Christopher J. and {Spiewak}, Ren{\'e}e and {Zhang}, Songbo and {Zhu}, Xing-Jiang},
        title = "{Search for an Isotropic Gravitational-wave Background with the Parkes Pulsar Timing Array}",
      journal = {\apjl},
         year = 2023,
        month = jul,
       volume = {951},
       number = {1},
          eid = {L6},
        pages = {L6},
          doi = {10.3847/2041-8213/acdd02},
archivePrefix = {arXiv},
       eprint = {2306.16215},
 primaryClass = {astro-ph.HE},
       adsurl = {https://ui.adsabs.harvard.edu/abs/2023ApJ...951L...6R}
}

@ARTICLE{EPTA2023,
       author = {{EPTA Collaboration} and {InPTA Collaboration} and {Antoniadis}, J. and {Arumugam}, P. and {Arumugam}, S. and {Babak}, S. and {Bagchi}, M. and {Bak Nielsen}, A. -S. and {Bassa}, C.~G. and {Bathula}, A. and {Berthereau}, A. and {Bonetti}, M. and {Bortolas}, E. and {Brook}, P.~R. and {Burgay}, M. and {Caballero}, R.~N. and {Chalumeau}, A. and {Champion}, D.~J. and {Chanlaridis}, S. and {Chen}, S. and {Cognard}, I. and {Dandapat}, S. and {Deb}, D. and {Desai}, S. and {Desvignes}, G. and {Dhanda-Batra}, N. and {Dwivedi}, C. and {Falxa}, M. and {Ferdman}, R.~D. and {Franchini}, A. and {Gair}, J.~R. and {Goncharov}, B. and {Gopakumar}, A. and {Graikou}, E. and {Grie{\ss}meier}, J. -M. and {Guillemot}, L. and {Guo}, Y.~J. and {Gupta}, Y. and {Hisano}, S. and {Hu}, H. and {Iraci}, F. and {Izquierdo-Villalba}, D. and {Jang}, J. and {Jawor}, J. and {Janssen}, G.~H. and {Jessner}, A. and {Joshi}, B.~C. and {Kareem}, F. and {Karuppusamy}, R. and {Keane}, E.~F. and {Keith}, M.~J. and {Kharbanda}, D. and {Kikunaga}, T. and {Kolhe}, N. and {Kramer}, M. and {Krishnakumar}, M.~A. and {Lackeos}, K. and {Lee}, K.~J. and {Liu}, K. and {Liu}, Y. and {Lyne}, A.~G. and {McKee}, J.~W. and {Maan}, Y. and {Main}, R.~A. and {Mickaliger}, M.~B. and {Ni{\c{t}}u}, I.~C. and {Nobleson}, K. and {Paladi}, A.~K. and {Parthasarathy}, A. and {Perera}, B.~B.~P. and {Perrodin}, D. and {Petiteau}, A. and {Porayko}, N.~K. and {Possenti}, A. and {Prabu}, T. and {Quelquejay Leclere}, H. and {Rana}, P. and {Samajdar}, A. and {Sanidas}, S.~A. and {Sesana}, A. and {Shaifullah}, G. and {Singha}, J. and {Speri}, L. and {Spiewak}, R. and {Srivastava}, A. and {Stappers}, B.~W. and {Surnis}, M. and {Susarla}, S.~C. and {Susobhanan}, A. and {Takahashi}, K. and {Tarafdar}, P. and {Theureau}, G. and {Tiburzi}, C. and {van der Wateren}, E. and {Vecchio}, A. and {Venkatraman Krishnan}, V. and {Verbiest}, J.~P.~W. and {Wang}, J. and {Wang}, L. and {Wu}, Z.},
        title = "{The second data release from the European Pulsar Timing Array. III. Search for gravitational wave signals}",
      journal = {\aap},
         year = 2023,
        month = oct,
       volume = {678},
          eid = {A50},
        pages = {A50},
          doi = {10.1051/0004-6361/202346844},
archivePrefix = {arXiv},
       eprint = {2306.16214},
 primaryClass = {astro-ph.HE},
       adsurl = {https://ui.adsabs.harvard.edu/abs/2023A&A...678A..50E}
}

@ARTICLE{Wang2018,
       author = {{Wang}, Jian-Min and {Songsheng}, Yu-Yang and {Li}, Yan-Rong and {Yu}, Zhe},
        title = "{Kinematic Signatures of Reverberation Mapping of Close Binaries of Supermassive Black Holes in Active Galactic Nuclei}",
      journal = {\apj},
         year = 2018,
        month = aug,
       volume = {862},
       number = {2},
          eid = {171},
        pages = {171},
          doi = {10.3847/1538-4357/aacdfa},
archivePrefix = {arXiv},
       eprint = {1806.06487},
 primaryClass = {astro-ph.GA},
       adsurl = {https://ui.adsabs.harvard.edu/abs/2018ApJ...862..171W}
}

@ARTICLE{Songsheng2019,
       author = {{Songsheng}, Yu-Yang and {Wang}, Jian-Min and {Li}, Yan-Rong},
        title = "{The VLT Interferometric Measurements of Active Galactic Nuclei: Effects of Angular Momentum Distributions of Clouds in the Broad-line Region}",
      journal = {\apj},
         year = 2019,
        month = oct,
       volume = {883},
       number = {2},
          eid = {184},
        pages = {184},
          doi = {10.3847/1538-4357/ab3c5e},
       adsurl = {https://ui.adsabs.harvard.edu/abs/2019ApJ...883..184S}
}

@ARTICLE{Songsheng2020,
       author = {{Songsheng}, Yu-Yang and {Xiao}, Ming and {Wang}, Jian-Min and {Ho}, Luis C.},
        title = "{Kinematic Signatures of Reverberation Mapping of Close Binaries of Supermassive Black Holes in Active Galactic Nuclei. II. Atlas of Two-dimensional Transfer Functions}",
      journal = {\apjs},
         year = 2020,
        month = mar,
       volume = {247},
       number = {1},
          eid = {3},
        pages = {3},
          doi = {10.3847/1538-4365/ab665a},
archivePrefix = {arXiv},
       eprint = {1912.12965},
 primaryClass = {astro-ph.GA},
       adsurl = {https://ui.adsabs.harvard.edu/abs/2020ApJS..247....3S}
}

@ARTICLE{Songsheng2023,
       author = {{Songsheng}, Yu-Yang and {Wang}, Jian-Min},
        title = "{Differential Interferometric Signatures of Close Binaries of Supermassive Black Holes in Active Galactic Nuclei. II. Merged Broad-line Regions}",
      journal = {\apj},
         year = 2023,
        month = mar,
       volume = {945},
       number = {2},
          eid = {89},
        pages = {89},
          doi = {10.3847/1538-4357/acbafd},
archivePrefix = {arXiv},
       eprint = {2302.08338},
 primaryClass = {astro-ph.GA},
       adsurl = {https://ui.adsabs.harvard.edu/abs/2023ApJ...945...89S}
}

@ARTICLE{Fu2025,
       author = {{Fu}, Yi-Xin and {Li}, Yan-Rong and {Wang}, Jian-Min and {Horne}, Keith and {Hern{\'a}ndez Santisteban}, Juan V. and {Vieliute}, Roberta and {Edelson}, Rick and {Liu}, Tingting and {Brotherton}, Michael S. and {Popovi{\'c}}, Luka {\v{C}}. and {Kova{\v{c}}evi{\'c}}, Andjelka B. and {Zhai}, Shuo},
        title = "{Continuum Reverberation Mapping of Accretion Disks Surrounding Supermassive Black Hole Binaries: Observational Signatures}",
      journal = {\mnras},
         year = 2025,
        month = sep,
          doi = {10.1093/mnras/staf1473},
archivePrefix = {arXiv},
       eprint = {2507.21671},
 primaryClass = {astro-ph.GA},
       adsurl = {https://ui.adsabs.harvard.edu/abs/2025MNRAS.tmp.1419F}
}

@ARTICLE{DuMAHAI2018,
       author = {{Du}, Pu and {Brotherton}, Michael S. and {Wang}, Kai and {Huang}, Zheng-Peng and {Hu}, Chen and {Kasper}, David H. and {Chick}, William T. and {Nguyen}, My L. and {Maithil}, Jaya and {Hand}, Derek and {Li}, Yan-Rong and {Ho}, Luis C. and {Bai}, Jin-Ming and {Bian}, Wei-Hao and {Wang}, Jian-Min and {MAHA Collaboration}},
        title = "{Monitoring AGNs with H{\ensuremath{\beta}} Asymmetry. I. First Results: Velocity-resolved Reverberation Mapping}",
      journal = {\apj},
         year = 2018,
        month = dec,
       volume = {869},
       number = {2},
          eid = {142},
        pages = {142},
          doi = {10.3847/1538-4357/aaed2c},
archivePrefix = {arXiv},
       eprint = {1810.11996},
 primaryClass = {astro-ph.GA},
       adsurl = {https://ui.adsabs.harvard.edu/abs/2018ApJ...869..142D}
}

@ARTICLE{CPTA2023,
       author = {{Xu}, Heng and {Chen}, Siyuan and {Guo}, Yanjun and {Jiang}, Jinchen and {Wang}, Bojun and {Xu}, Jiangwei and {Xue}, Zihan and {Caballero}, R. Nicolas and {Yuan}, Jianping and {Xu}, Yonghua and {Wang}, Jingbo and {Hao}, Longfei and {Luo}, Jingtao and {Lee}, Kejia and {Han}, Jinlin and {Jiang}, Peng and {Shen}, Zhiqiang and {Wang}, Min and {Wang}, Na and {Xu}, Renxin and {Wu}, Xiangping and {Manchester}, Richard and {Qian}, Lei and {Guan}, Xin and {Huang}, Menglin and {Sun}, Chun and {Zhu}, Yan},
        title = "{Searching for the Nano-Hertz Stochastic Gravitational Wave Background with the Chinese Pulsar Timing Array Data Release I}",
      journal = {Research in Astronomy and Astrophysics},
         year = 2023,
        month = jul,
       volume = {23},
       number = {7},
          eid = {075024},
        pages = {075024},
          doi = {10.1088/1674-4527/acdfa5},
archivePrefix = {arXiv},
       eprint = {2306.16216},
 primaryClass = {astro-ph.HE},
       adsurl = {https://ui.adsabs.harvard.edu/abs/2023RAA....23g5024X}
}

@ARTICLE{Wangmengye2023,
       author = {{Wang}, Mengye and {Ma}, Yiqiu and {Wu}, Qingwen},
        title = "{Accretion-modified stellar-mass black hole distribution and milli-Hz gravitational wave backgrounds from galaxy centre}",
      journal = {\mnras},
         year = 2023,
        month = apr,
       volume = {520},
       number = {3},
        pages = {4502-4516},
          doi = {10.1093/mnras/stad422},
archivePrefix = {arXiv},
       eprint = {2212.05724},
primaryClass = {astro-ph.HE},
       adsurl = {https://ui.adsabs.harvard.edu/abs/2023MNRAS.520.4502W}
}

@BOOK{Kato2008,
       author = {{Kato}, S. and {Fukue}, J. and {Mineshige}, S.},
        title = "{Black-Hole Accretion Disks -- Towards a New Paradigm --}",
         year = 2008,
       adsurl = {https://ui.adsabs.harvard.edu/abs/2008bhad.book.....K}
}

@ARTICLE{Yang2023,
       author = {{Yang}, Jinyi and {Wang}, Feige and {Fan}, Xiaohui and {Hennawi}, Joseph F. and {Barth}, Aaron J. and {Ba{\~n}ados}, Eduardo and {Sun}, Fengwu and {Liu}, Weizhe and {Cai}, Zheng and {Jiang}, Linhua and {Li}, Zihao and {Onoue}, Masafusa and {Schindler}, Jan-Torge and {Shen}, Yue and {Wu}, Yunjing and {Bhowmick}, Aklant K. and {Bieri}, Rebekka and {Blecha}, Laura and {Bosman}, Sarah and {Champagne}, Jaclyn B. and {Colina}, Luis and {Connor}, Thomas and {Costa}, Tiago and {Davies}, Frederick B. and {Decarli}, Roberto and {De Rosa}, Gisella and {Drake}, Alyssa B. and {Egami}, Eiichi and {Eilers}, Anna-Christina and {Evans}, Analis E. and {Farina}, Emanuele Paolo and {Habouzit}, Melanie and {Haiman}, Zoltan and {Jin}, Xiangyu and {Jun}, Hyunsung D. and {Kakiichi}, Koki and {Khusanova}, Yana and {Kulkarni}, Girish and {Loiacono}, Federica and {Lupi}, Alessandro and {Mazzucchelli}, Chiara and {Pan}, Zhiwei and {Rojas-Ruiz}, Sof{\'\i}a and {Strauss}, Michael A. and {Tee}, Wei Leong and {Trakhtenbrot}, Benny and {Trebitsch}, Maxime and {Venemans}, Bram and {Vestergaard}, Marianne and {Volonteri}, Marta and {Walter}, Fabian and {Xie}, Zhang-Liang and {Yue}, Minghao and {Zhang}, Haowen and {Zhang}, Huanian and {Zou}, Siwei},
        title = "{A SPectroscopic Survey of Biased Halos in the Reionization Era (ASPIRE): A First Look at the Rest-frame Optical Spectra of z > 6.5 Quasars Using JWST}",
      journal = {\apjl},
         year = 2023,
        month = jul,
       volume = {951},
       number = {1},
          eid = {L5},
        pages = {L5},
          doi = {10.3847/2041-8213/acc9c8},
archivePrefix = {arXiv},
       eprint = {2304.09888},
 primaryClass = {astro-ph.GA},
       adsurl = {https://ui.adsabs.harvard.edu/abs/2023ApJ...951L...5Y}
}

@ARTICLE{Gierlinski2008,
       author = {{Gierli{\'n}ski}, Marek and {Middleton}, Matthew and {Ward}, Martin and {Done}, Chris},
        title = "{A periodicity of \raisebox{-0.5ex}\textasciitilde1hour in X-ray emission from the active galaxy RE J1034+396}",
      journal = {\nat},
         year = 2008,
        month = sep,
       volume = {455},
       number = {7211},
        pages = {369-371},
          doi = {10.1038/nature07277},
       adsurl = {https://ui.adsabs.harvard.edu/abs/2008Natur.455..369G}
}

@ARTICLE{Ingram2019,
       author = {{Ingram}, Adam R. and {Motta}, Sara E.},
        title = "{A review of quasi-periodic oscillations from black hole X-ray binaries: Observation and theory}",
      journal = {\nar},
         year = 2019,
        month = sep,
       volume = {85},
          eid = {101524},
        pages = {101524},
          doi = {10.1016/j.newar.2020.101524},
archivePrefix = {arXiv},
       eprint = {2001.08758},
 primaryClass = {astro-ph.HE},
       adsurl = {https://ui.adsabs.harvard.edu/abs/2019NewAR..8501524I}
}

@ARTICLE{Rauch1991,
       author = {{Rauch}, Kevin P. and {Blandford}, Roger D.},
        title = "{Microlensing and the Structure of Active Galactic Nucleus Accretion Disks}",
      journal = {\apjl},
         year = 1991,
        month = nov,
       volume = {381},
        pages = {L39},
          doi = {10.1086/186191},
       adsurl = {https://ui.adsabs.harvard.edu/abs/1991ApJ...381L..39R}
}

@ARTICLE{Fan2023,
       author = {{Fan}, Xiao and {Wu}, Qingwen},
        title = "{In Situ Star Formation in Accretion Disks and Explanation of Correlation between the Black Hole Mass and Metallicity in Active Galactic Nuclei}",
      journal = {\apj},
         year = 2023,
        month = feb,
       volume = {944},
       number = {2},
          eid = {159},
        pages = {159},
          doi = {10.3847/1538-4357/acb532},
archivePrefix = {arXiv},
       eprint = {2212.06363},
 primaryClass = {astro-ph.GA},
       adsurl = {https://ui.adsabs.harvard.edu/abs/2023ApJ...944..159F}
}

@ARTICLE{Netzer2007,
       author = {{Netzer}, Hagai and {Trakhtenbrot}, Benny},
        title = "{Cosmic Evolution of Mass Accretion Rate and Metallicity in Active Galactic Nuclei}",
      journal = {\apj},
         year = 2007,
        month = jan,
       volume = {654},
       number = {2},
        pages = {754-763},
          doi = {10.1086/509650},
archivePrefix = {arXiv},
       eprint = {astro-ph/0607654},
 primaryClass = {astro-ph},
       adsurl = {https://ui.adsabs.harvard.edu/abs/2007ApJ...654..754N}
}

@ARTICLE{Wang2023,
       author = {{Wang}, Jian-Min and {Zhai}, Shuo and {Li}, Yan-Rong and {Songsheng}, Yu-Yang and {Ho}, Luis C. and {Chen}, Yong-Jie and {Liu}, Jun-Rong and {Du}, Pu and {Yuan}, Ye-Fei},
        title = "{Star Formation in Self-gravitating Disks in Active Galactic Nuclei. III. Efficient Production of Iron and Infrared Spectral Energy Distributions}",
      journal = {\apj},
         year = 2023,
        month = sep,
       volume = {954},
       number = {1},
          eid = {84},
        pages = {84},
          doi = {10.3847/1538-4357/acdf48},
archivePrefix = {arXiv},
       eprint = {2311.06782},
 primaryClass = {astro-ph.GA},
       adsurl = {https://ui.adsabs.harvard.edu/abs/2023ApJ...954...84W}
}

@ARTICLE{Wang2012,
       author = {{Wang}, Jian-Min and {Du}, Pu and {Baldwin}, Jack A. and {Ge}, Jun-Qiang and {Hu}, Chen and {Ferland}, Gary J.},
        title = "{Star Formation in Self-gravitating Disks in Active Galactic Nuclei. II. Episodic Formation of Broad-line Regions}",
      journal = {\apj},
         year = 2012,
        month = feb,
       volume = {746},
       number = {2},
          eid = {137},
        pages = {137},
          doi = {10.1088/0004-637X/746/2/137},
archivePrefix = {arXiv},
       eprint = {1202.0062},
 primaryClass = {astro-ph.CO},
       adsurl = {https://ui.adsabs.harvard.edu/abs/2012ApJ...746..137W}
}

@ARTICLE{Wang2011,
       author = {{Wang}, Jian-Min and {Ge}, Jun-Qiang and {Hu}, Chen and {Baldwin}, Jack A. and {Li}, Yan-Rong and {Ferland}, Gary J. and {Xiang}, Fei and {Yan}, Chang-Shuo and {Zhang}, Shu},
        title = "{Star Formation in Self-gravitating Disks in Active Galactic Nuclei. I. Metallicity Gradients in Broad-line Regions}",
      journal = {\apj},
         year = 2011,
        month = sep,
       volume = {739},
       number = {1},
          eid = {3},
        pages = {3},
          doi = {10.1088/0004-637X/739/1/3},
archivePrefix = {arXiv},
       eprint = {1107.3620},
 primaryClass = {astro-ph.GA},
       adsurl = {https://ui.adsabs.harvard.edu/abs/2011ApJ...739....3W}
}

@ARTICLE{Malik2023,
       author = {{Malik}, U. and {Sharp}, R. and {Penton}, A. and {Yu}, Z. and {Martini}, P. and {Lidman}, C. and {Tucker}, B.~E. and {Davis}, T.~M. and {Lewis}, G.~F. and {Aguena}, M. and {Allam}, S. and {Alves}, O. and {Andrade-Oliveira}, F. and {Asorey}, J. and {Bacon}, D. and {Bertin}, E. and {Bocquet}, S. and {Brooks}, D. and {Burke}, D.~L. and {Carnero Rosell}, A. and {Carollo}, D. and {Carrasco Kind}, M. and {Carretero}, J. and {Costanzi}, M. and {da Costa}, L.~N. and {Pereira}, M.~E.~S. and {De Vicente}, J. and {Desai}, S. and {Diehl}, H.~T. and {Doel}, P. and {Everett}, S. and {Ferrero}, I. and {Frieman}, J. and {Garc{\'\i}a-Bellido}, J. and {Gerdes}, D.~W. and {Gruen}, D. and {Gruendl}, R.~A. and {Gschwend}, J. and {Hinton}, S.~R. and {Hollowood}, D.~L. and {Honscheid}, K. and {James}, D.~J. and {Kuehn}, K. and {Marshall}, J.~L. and {Mena-Fern{\'a}ndez}, J. and {Menanteau}, F. and {Miquel}, R. and {Ogando}, R.~L.~C. and {Palmese}, A. and {Paz-Chinch{\'o}n}, F. and {Pieres}, A. and {Plazas Malag{\'o}n}, A.~A. and {Raveri}, M. and {Rodriguez-Monroy}, M. and {Romer}, A.~K. and {Sanchez}, E. and {Scarpine}, V. and {Sevilla-Noarbe}, I. and {Smith}, M. and {Soares-Santos}, M. and {Suchyta}, E. and {Swanson}, M.~E.~C. and {Tarle}, G. and {Taylor}, G. and {Tucker}, D.~L. and {Weaverdyck}, N. and {Wilkinson}, R.~D.},
        title = "{OzDES Reverberation Mapping Program: H{\ensuremath{\beta}} lags from the 6-yr survey}",
      journal = {\mnras},
         year = 2023,
        month = apr,
       volume = {520},
       number = {2},
        pages = {2009-2023},
          doi = {10.1093/mnras/stad145},
archivePrefix = {arXiv},
       eprint = {2210.03977},
 primaryClass = {astro-ph.GA},
       adsurl = {https://ui.adsabs.harvard.edu/abs/2023MNRAS.520.2009M}
}

@ARTICLE{Shen2024,
       author = {{Shen}, Yue and {Grier}, Catherine J. and {Horne}, Keith and {Stone}, Zachary and {Li}, Jennifer I. and {Yang}, Qian and {Homayouni}, Yasaman and {Trump}, Jonathan R. and {Anderson}, Scott F. and {Brandt}, W.~N. and {Hall}, Patrick B. and {Ho}, Luis C. and {Jiang}, Linhua and {Petitjean}, Patrick and {Schneider}, Donald P. and {Tao}, Charling and {Donnan}, Fergus. R. and {AlSayyad}, Yusra and {Bershady}, Matthew A. and {Blanton}, Michael R. and {Bizyaev}, Dmitry and {Bundy}, Kevin and {Chen}, Yuguang and {Davis}, Megan C. and {Dawson}, Kyle and {Fan}, Xiaohui and {Greene}, Jenny E. and {Gr{\"o}ller}, Hannes and {Guo}, Yucheng and {Ibarra-Medel}, H{\'e}ctor and {Jiang}, Yuanzhe and {Keenan}, Ryan P. and {Kollmeier}, Juna A. and {Lejoly}, Cassandra and {Li}, Zefeng and {de la Macorra}, Axel and {Moe}, Maxwell and {Nie}, Jundan and {Rossi}, Graziano and {Smith}, Paul S. and {Tee}, Wei Leong and {Weijmans}, Anne-Marie and {Xu}, Jiachuan and {Yue}, Minghao and {Zhou}, Xu and {Zhou}, Zhimin and {Zou}, Hu},
        title = "{The Sloan Digital Sky Survey Reverberation Mapping Project: Key Results}",
      journal = {\apjs},
         year = 2024,
        month = jun,
       volume = {272},
       number = {2},
          eid = {26},
        pages = {26},
          doi = {10.3847/1538-4365/ad3936},
archivePrefix = {arXiv},
       eprint = {2305.01014},
 primaryClass = {astro-ph.GA},
       adsurl = {https://ui.adsabs.harvard.edu/abs/2024ApJS..272...26S}
}

@ARTICLE{Fausnaugh2017,
       author = {{Fausnaugh}, M.~M. and {Grier}, C.~J. and {Bentz}, M.~C. and {Denney}, K.~D. and {De Rosa}, G. and {Peterson}, B.~M. and {Kochanek}, C.~S. and {Pogge}, R.~W. and {Adams}, S.~M. and {Barth}, A.~J. and {Beatty}, Thomas G. and {Bhattacharjee}, A. and {Borman}, G.~A. and {Boroson}, T.~A. and {Bottorff}, M.~C. and {Brown}, Jacob E. and {Brown}, Jonathan S. and {Brotherton}, M.~S. and {Coker}, C.~T. and {Crawford}, S.~M. and {Croxall}, K.~V. and {Eftekharzadeh}, Sarah and {Eracleous}, Michael and {Joner}, M.~D. and {Henderson}, C.~B. and {Holoien}, T.~W. -S. and {Horne}, Keith and {Hutchison}, T. and {Kaspi}, Shai and {Kim}, S. and {King}, Anthea L. and {Li}, Miao and {Lochhaas}, Cassandra and {Ma}, Zhiyuan and {MacInnis}, F. and {Manne-Nicholas}, E.~R. and {Mason}, M. and {Montuori}, Carmen and {Mosquera}, Ana and {Mudd}, Dale and {Musso}, R. and {Nazarov}, S.~V. and {Nguyen}, M.~L. and {Okhmat}, D.~N. and {Onken}, Christopher A. and {Ou-Yang}, B. and {Pancoast}, A. and {Pei}, L. and {Penny}, Matthew T. and {Poleski}, Rados{\l}aw and {Rafter}, Stephen and {Romero-Colmenero}, E. and {Runnoe}, Jessie and {Sand}, David J. and {Schimoia}, Jaderson S. and {Sergeev}, S.~G. and {Shappee}, B.~J. and {Simonian}, Gregory V. and {Somers}, Garrett and {Spencer}, M. and {Starkey}, D.~A. and {Stevens}, Daniel J. and {Tayar}, Jamie and {Treu}, T. and {Valenti}, Stefano and {Van Saders}, J. and {Villanueva}, Jr., S. and {Villforth}, C. and {Weiss}, Yaniv and {Winkler}, H. and {Zhu}, W.},
        title = "{Reverberation Mapping of Optical Emission Lines in Five Active Galaxies}",
      journal = {\apj},
         year = 2017,
        month = may,
       volume = {840},
       number = {2},
          eid = {97},
        pages = {97},
          doi = {10.3847/1538-4357/aa6d52},
archivePrefix = {arXiv},
       eprint = {1610.00008},
 primaryClass = {astro-ph.GA},
       adsurl = {https://ui.adsabs.harvard.edu/abs/2017ApJ...840...97F}
}

@ARTICLE{Grier2012,
       author = {{Grier}, C.~J. and {Peterson}, B.~M. and {Pogge}, R.~W. and {Denney}, K.~D. and {Bentz}, M.~C. and {Martini}, Paul and {Sergeev}, S.~G. and {Kaspi}, S. and {Minezaki}, T. and {Zu}, Y. and {Kochanek}, C.~S. and {Siverd}, R. and {Shappee}, B. and {Stanek}, K.~Z. and {Araya Salvo}, C. and {Beatty}, T.~G. and {Bird}, J.~C. and {Bord}, D.~J. and {Borman}, G.~A. and {Che}, X. and {Chen}, C. and {Cohen}, S.~A. and {Dietrich}, M. and {Doroshenko}, V.~T. and {Drake}, T. and {Efimov}, Yu. S. and {Free}, N. and {Ginsburg}, I. and {Henderson}, C.~B. and {King}, A.~L. and {Koshida}, S. and {Mogren}, K. and {Molina}, M. and {Mosquera}, A.~M. and {Nazarov}, S.~V. and {Okhmat}, D.~N. and {Pejcha}, O. and {Rafter}, S. and {Shields}, J.~C. and {Skowron}, J. and {Szczygiel}, D.~M. and {Valluri}, M. and {van Saders}, J.~L.},
        title = "{Reverberation Mapping Results for Five Seyfert 1 Galaxies}",
      journal = {\apj},
         year = 2012,
        month = aug,
       volume = {755},
       number = {1},
          eid = {60},
        pages = {60},
          doi = {10.1088/0004-637X/755/1/60},
archivePrefix = {arXiv},
       eprint = {1206.6523},
 primaryClass = {astro-ph.CO},
       adsurl = {https://ui.adsabs.harvard.edu/abs/2012ApJ...755...60G}
}

@ARTICLE{Denney2010,
       author = {{Denney}, K.~D. and {Peterson}, B.~M. and {Pogge}, R.~W. and {Adair}, A. and {Atlee}, D.~W. and {Au-Yong}, K. and {Bentz}, M.~C. and {Bird}, J.~C. and {Brokofsky}, D.~J. and {Chisholm}, E. and {Comins}, M.~L. and {Dietrich}, M. and {Doroshenko}, V.~T. and {Eastman}, J.~D. and {Efimov}, Y.~S. and {Ewald}, S. and {Ferbey}, S. and {Gaskell}, C.~M. and {Hedrick}, C.~H. and {Jackson}, K. and {Klimanov}, S.~A. and {Klimek}, E.~S. and {Kruse}, A.~K. and {Lad{\'e}route}, A. and {Lamb}, J.~B. and {Leighly}, K. and {Minezaki}, T. and {Nazarov}, S.~V. and {Onken}, C.~A. and {Petersen}, E.~A. and {Peterson}, P. and {Poindexter}, S. and {Sakata}, Y. and {Schlesinger}, K.~J. and {Sergeev}, S.~G. and {Skolski}, N. and {Stieglitz}, L. and {Tobin}, J.~J. and {Unterborn}, C. and {Vestergaard}, M. and {Watkins}, A.~E. and {Watson}, L.~C. and {Yoshii}, Y.},
        title = "{Reverberation Mapping Measurements of Black Hole Masses in Six Local Seyfert Galaxies}",
      journal = {\apj},
         year = 2010,
        month = sep,
       volume = {721},
       number = {1},
        pages = {715-737},
          doi = {10.1088/0004-637X/721/1/715},
archivePrefix = {arXiv},
       eprint = {1006.4160},
 primaryClass = {astro-ph.CO},
       adsurl = {https://ui.adsabs.harvard.edu/abs/2010ApJ...721..715D}
}

@ARTICLE{Zastrocky2024,
       author = {{Zastrocky}, T.~E. and {Brotherton}, Michael S. and {Du}, Pu and {McLane}, Jacob N. and {Olson}, Kianna A. and {Dale}, D.~A. and {Kobulnicky}, H.~A. and {Maithil}, Jaya and {Nguyen}, My L. and {Chick}, William T. and {Kasper}, David H. and {Hand}, Derek and {Adelman}, C. and {Carter}, Z. and {Murphree}, G. and {Oeur}, M. and {Roth}, T. and {Schonsberg}, S. and {Caradonna}, M.~J. and {Favro}, J. and {Ferguson}, A.~J. and {Gonzalez}, I.~M. and {Hadding}, L.~M. and {Hagler}, H.~D. and {Rogers}, C.~J. and {Stack}, T.~R. and {Chapman}, Franklin and {Bao}, Dong-Wei and {Fang}, Feng-Na and {Zhai}, Shuo and {Yang}, Sen and {Chen}, Yong-Jie and {Bai}, Hua-Rui and {Fu}, Yi-Xin and {Liu}, Jun-Rong and {Yao}, Zhu-Heng and {Peng}, Yue-Chang and {Songsheng}, Yu-Yang and {Li}, Yan-Rong and {Bai}, Jin-Ming and {Hu}, Chen and {Xiao}, Ming and {Ho}, Luis C. and {Wang}, Jian-Min},
        title = "{Monitoring AGNs with H{\ensuremath{\beta}} Asymmetry. IV. First Reverberation Mapping Results of 14 Active Galactic Nuclei}",
      journal = {\apjs},
         year = 2024,
        month = jun,
       volume = {272},
       number = {2},
          eid = {29},
        pages = {29},
          doi = {10.3847/1538-4365/ad3bad},
archivePrefix = {arXiv},
       eprint = {2404.07343},
 primaryClass = {astro-ph.GA},
       adsurl = {https://ui.adsabs.harvard.edu/abs/2024ApJS..272...29Z}
}

@ARTICLE{Bentz2009,
       author = {{Bentz}, Misty C. and {Walsh}, Jonelle L. and {Barth}, Aaron J. and {Baliber}, Nairn and {Bennert}, Vardha Nicola and {Canalizo}, Gabriela and {Filippenko}, Alexei V. and {Ganeshalingam}, Mohan and {Gates}, Elinor L. and {Greene}, Jenny E. and {Hidas}, Marton G. and {Hiner}, Kyle D. and {Lee}, Nicholas and {Li}, Weidong and {Malkan}, Matthew A. and {Minezaki}, Takeo and {Sakata}, Yu and {Serduke}, Frank J.~D. and {Silverman}, Jeffrey M. and {Steele}, Thea N. and {Stern}, Daniel and {Street}, Rachel A. and {Thornton}, Carol E. and {Treu}, Tommaso and {Wang}, Xiaofeng and {Woo}, Jong-Hak and {Yoshii}, Yuzuru},
        title = "{The Lick AGN Monitoring Project: Broad-line Region Radii and Black Hole Masses from Reverberation Mapping of H{\ensuremath{\beta}}}",
      journal = {\apj},
         year = 2009,
        month = nov,
       volume = {705},
       number = {1},
        pages = {199-217},
          doi = {10.1088/0004-637X/705/1/199},
archivePrefix = {arXiv},
       eprint = {0908.0003},
 primaryClass = {astro-ph.CO},
       adsurl = {https://ui.adsabs.harvard.edu/abs/2009ApJ...705..199B}
}

@ARTICLE{Du2016a,
       author = {{Du}, Pu and {Lu}, Kai-Xing and {Zhang}, Zhi-Xiang and {Huang}, Ying-Ke and {Wang}, Kai and {Hu}, Chen and {Qiu}, Jie and {Li}, Yan-Rong and {Fan}, Xu-Liang and {Fang}, Xiang-Er and {Bai}, Jin-Ming and {Bian}, Wei-Hao and {Yuan}, Ye-Fei and {Ho}, Luis C. and {Wang}, Jian-Min and {SEAMBH Collaboration}},
        title = "{Supermassive Black Holes with High Accretion Rates in Active Galactic Nuclei. V. A New Size-Luminosity Scaling Relation for the Broad-line Region}",
      journal = {\apj},
         year = 2016,
        month = jul,
       volume = {825},
       number = {2},
          eid = {126},
        pages = {126},
          doi = {10.3847/0004-637X/825/2/126},
archivePrefix = {arXiv},
       eprint = {1604.06218},
 primaryClass = {astro-ph.GA},
       adsurl = {https://ui.adsabs.harvard.edu/abs/2016ApJ...825..126D}
}

@ARTICLE{Du2016b,
       author = {{Du}, Pu and {Lu}, Kai-Xing and {Hu}, Chen and {Qiu}, Jie and {Li}, Yan-Rong and {Huang}, Ying-Ke and {Wang}, Fang and {Bai}, Jin-Ming and {Bian}, Wei-Hao and {Yuan}, Ye-Fei and {Ho}, Luis C. and {Wang}, Jian-Min and {SEAMBH Collaboration}},
        title = "{Supermassive Black Holes with High Accretion Rates in Active Galactic Nuclei. VI. Velocity-resolved Reverberation Mapping of the H{\ensuremath{\beta}} Line}",
      journal = {\apj},
         year = 2016,
        month = mar,
       volume = {820},
       number = {1},
          eid = {27},
        pages = {27},
          doi = {10.3847/0004-637X/820/1/27},
archivePrefix = {arXiv},
       eprint = {1602.01922},
 primaryClass = {astro-ph.GA},
       adsurl = {https://ui.adsabs.harvard.edu/abs/2016ApJ...820...27D}
}

@ARTICLE{Li2025,
       author = {{Li}, Yan-Rong and {Shangguan}, Jinyi and {Wang}, Jian-Min and {Davies}, Ric and {Santos}, Daryl J.~D. and {Eisenhauer}, Frank and {Songsheng}, Yu-Yang and {Winkler}, Hartmut and {Aceituno}, Jes{\'u}s and {Bai}, Hua-Rui and {Bai}, Jin-Ming and {Brotherton}, Michael S. and {Cao}, Yixian and {Chen}, Yong-Jie and {Du}, Pu and {Fang}, Feng-Na and {Feng}, Jia-Qi and {Feuchtgruber}, Helmut and {F{\"o}rster Schreiber}, Natascha M. and {Fu}, Yi-Xin and {Genzel}, Reinhard and {Gillessen}, Stefan and {Ho}, Luis C. and {Hu}, Chen and {Liu}, Jun-Rong and {Lutz}, Dieter and {Ott}, Thomas and {Petrov}, Romain G. and {Rabien}, Sebastian and {Shimizu}, Taro and {Sturm}, Eckhard and {Tacconi}, Linda J. and {Wang}, Yi-Lin and {Yao}, Zhu-Heng and {Zhai}, Shuo and {Zhang}, Hao and {Zhao}, Yi-Peng and {Zhao}, Yu and {SARM Collaboration}},
        title = "{Spectroastrometry and Reverberation Mapping of Active Galactic Nuclei. II. Measuring Geometric Distances and Black Hole Masses of Four Nearby Quasars}",
      journal = {\apj},
         year = 2025,
        month = jul,
       volume = {988},
       number = {1},
          eid = {42},
        pages = {42},
          doi = {10.3847/1538-4357/addf40},
archivePrefix = {arXiv},
       eprint = {2502.18856},
 primaryClass = {astro-ph.GA},
       adsurl = {https://ui.adsabs.harvard.edu/abs/2025ApJ...988...42L}
}

@ARTICLE{Mapelli2022,
       author = {{Mapelli}, Michela and {Bouffanais}, Yann and {Santoliquido}, Filippo and {Arca Sedda}, Manuel and {Artale}, M. Celeste},
        title = "{The cosmic evolution of binary black holes in young, globular, and nuclear star clusters: rates, masses, spins, and mixing fractions}",
      journal = {\mnras},
         year = 2022,
        month = apr,
       volume = {511},
       number = {4},
        pages = {5797-5816},
          doi = {10.1093/mnras/stac422},
archivePrefix = {arXiv},
       eprint = {2109.06222},
 primaryClass = {astro-ph.HE},
       adsurl = {https://ui.adsabs.harvard.edu/abs/2022MNRAS.511.5797M}
}

@ARTICLE{VandenBerk2001,
       author = {{Vanden Berk}, Daniel E. and {Richards}, Gordon T. and {Bauer}, Amanda and {Strauss}, Michael A. and {Schneider}, Donald P. and {Heckman}, Timothy M. and {York}, Donald G. and {Hall}, Patrick B. and {Fan}, Xiaohui and {Knapp}, G.~R. and {Anderson}, Scott F. and {Annis}, James and {Bahcall}, Neta A. and {Bernardi}, Mariangela and {Briggs}, John W. and {Brinkmann}, J. and {Brunner}, Robert and {Burles}, Scott and {Carey}, Larry and {Castander}, Francisco J. and {Connolly}, A.~J. and {Crocker}, J.~H. and {Csabai}, Istv{\'a}n and {Doi}, Mamoru and {Finkbeiner}, Douglas and {Friedman}, Scott and {Frieman}, Joshua A. and {Fukugita}, Masataka and {Gunn}, James E. and {Hennessy}, G.~S. and {Ivezi{\'c}}, {\v{Z}}eljko and {Kent}, Stephen and {Kunszt}, Peter Z. and {Lamb}, D.~Q. and {Leger}, R. French and {Long}, Daniel C. and {Loveday}, Jon and {Lupton}, Robert H. and {Meiksin}, Avery and {Merelli}, Aronne and {Munn}, Jeffrey A. and {Newberg}, Heidi Jo and {Newcomb}, Matt and {Nichol}, R.~C. and {Owen}, Russell and {Pier}, Jeffrey R. and {Pope}, Adrian and {Rockosi}, Constance M. and {Schlegel}, David J. and {Siegmund}, Walter A. and {Smee}, Stephen and {Snir}, Yehuda and {Stoughton}, Chris and {Stubbs}, Christopher and {SubbaRao}, Mark and {Szalay}, Alexander S. and {Szokoly}, Gyula P. and {Tremonti}, Christy and {Uomoto}, Alan and {Waddell}, Patrick and {Yanny}, Brian and {Zheng}, Wei},
        title = "{Composite Quasar Spectra from the Sloan Digital Sky Survey}",
      journal = {\aj},
         year = 2001,
        month = aug,
       volume = {122},
       number = {2},
        pages = {549-564},
          doi = {10.1086/321167},
archivePrefix = {arXiv},
       eprint = {astro-ph/0105231},
 primaryClass = {astro-ph},
       adsurl = {https://ui.adsabs.harvard.edu/abs/2001AJ....122..549V}
}

@ARTICLE{Chen2024,
       author = {{Chen}, Yi-Xian and {Lin}, Douglas N.~C.},
        title = "{The Population of Massive Stars in Active Galactic Nuclei Disks}",
      journal = {\apj},
         year = 2024,
        month = jun,
       volume = {967},
       number = {2},
          eid = {88},
        pages = {88},
          doi = {10.3847/1538-4357/ad3c3a},
archivePrefix = {arXiv},
       eprint = {2404.08780},
 primaryClass = {astro-ph.GA},
       adsurl = {https://ui.adsabs.harvard.edu/abs/2024ApJ...967...88C}
}

@ARTICLE{Morgan2010,
       author = {{Morgan}, Christopher W. and {Kochanek}, C.~S. and {Morgan}, Nicholas D. and {Falco}, Emilio E.},
        title = "{The Quasar Accretion Disk Size-Black Hole Mass Relation}",
      journal = {\apj},
         year = 2010,
        month = apr,
       volume = {712},
       number = {2},
        pages = {1129-1136},
          doi = {10.1088/0004-637X/712/2/1129},
archivePrefix = {arXiv},
       eprint = {1002.4160},
 primaryClass = {astro-ph.CO},
       adsurl = {https://ui.adsabs.harvard.edu/abs/2010ApJ...712.1129M}
}

@ARTICLE{Jiang2017,
       author = {{Jiang}, Yan-Fei and {Green}, Paul J. and {Greene}, Jenny E. and {Morganson}, Eric and {Shen}, Yue and {Pancoast}, Anna and {MacLeod}, Chelsea L. and {Anderson}, Scott F. and {Brandt}, W.~N. and {Grier}, C.~J. and {Rix}, H. -W. and {Ruan}, John J. and {Protopapas}, Pavlos and {Scott}, Caroline and {Burgett}, W.~S. and {Hodapp}, K.~W. and {Huber}, M.~E. and {Kaiser}, N. and {Kudritzki}, R.~P. and {Magnier}, E.~A. and {Metcalfe}, N. and {Tonry}, J.~T. and {Wainscoat}, R.~J. and {Waters}, C.},
        title = "{Detection of Time Lags between Quasar Continuum Emission Bands Based On Pan-STARRS Light Curves}",
      journal = {\apj},
         year = 2017,
        month = feb,
       volume = {836},
       number = {2},
          eid = {186},
        pages = {186},
          doi = {10.3847/1538-4357/aa5b91},
archivePrefix = {arXiv},
       eprint = {1612.08747},
 primaryClass = {astro-ph.HE},
       adsurl = {https://ui.adsabs.harvard.edu/abs/2017ApJ...836..186J}
}

@ARTICLE{Cackett2007,
       author = {{Cackett}, Edward M. and {Horne}, Keith and {Winkler}, Hartmut},
        title = "{Testing thermal reprocessing in active galactic nuclei accretion discs}",
      journal = {\mnras},
         year = 2007,
        month = sep,
       volume = {380},
       number = {2},
        pages = {669-682},
          doi = {10.1111/j.1365-2966.2007.12098.x},
archivePrefix = {arXiv},
       eprint = {0706.1464},
 primaryClass = {astro-ph},
       adsurl = {https://ui.adsabs.harvard.edu/abs/2007MNRAS.380..669C}
}

@ARTICLE{Ruan2020,
       author = {{Ruan}, Wen-Hong and {Liu}, Chang and {Guo}, Zong-Kuan and {Wu}, Yue-Liang and {Cai}, Rong-Gen},
        title = "{The LISA-Taiji network}",
      journal = {Nature Astronomy},
         year = 2020,
        month = feb,
       volume = {4},
        pages = {108-109},
          doi = {10.1038/s41550-019-1008-4},
archivePrefix = {arXiv},
       eprint = {2002.03603},
 primaryClass = {gr-qc},
       adsurl = {https://ui.adsabs.harvard.edu/abs/2020NatAs...4..108R}
}

@ARTICLE{Woosley2002,
       author = {{Woosley}, S.~E. and {Heger}, A. and {Weaver}, T.~A.},
        title = "{The evolution and explosion of massive stars}",
      journal = {Reviews of Modern Physics},
         year = 2002,
        month = nov,
       volume = {74},
       number = {4},
        pages = {1015-1071},
          doi = {10.1103/RevModPhys.74.1015},
       adsurl = {https://ui.adsabs.harvard.edu/abs/2002RvMP...74.1015W}
}

@ARTICLE{Matthee2024,
       author = {{Matthee}, Jorryt and {Naidu}, Rohan P. and {Brammer}, Gabriel and {Chisholm}, John and {Eilers}, Anna-Christina and {Goulding}, Andy and {Greene}, Jenny and {Kashino}, Daichi and {Labbe}, Ivo and {Lilly}, Simon J. and {Mackenzie}, Ruari and {Oesch}, Pascal A. and {Weibel}, Andrea and {Wuyts}, Stijn and {Xiao}, Mengyuan and {Bordoloi}, Rongmon and {Bouwens}, Rychard and {van Dokkum}, Pieter and {Illingworth}, Garth and {Kramarenko}, Ivan and {Maseda}, Michael V. and {Mason}, Charlotte and {Meyer}, Romain A. and {Nelson}, Erica J. and {Reddy}, Naveen A. and {Shivaei}, Irene and {Simcoe}, Robert A. and {Yue}, Minghao},
        title = "{Little Red Dots: An Abundant Population of Faint Active Galactic Nuclei at z {\ensuremath{\sim}} 5 Revealed by the EIGER and FRESCO JWST Surveys}",
      journal = {\apj},
         year = 2024,
        month = mar,
       volume = {963},
       number = {2},
          eid = {129},
        pages = {129},
          doi = {10.3847/1538-4357/ad2345},
archivePrefix = {arXiv},
       eprint = {2306.05448},
 primaryClass = {astro-ph.GA},
       adsurl = {https://ui.adsabs.harvard.edu/abs/2024ApJ...963..129M}
}

@ARTICLE{Woo2024,
       author = {{Woo}, Jong-Hak and {Wang}, Shu and {Rakshit}, Suvendu and {Cho}, Hojin and {Son}, Donghoon and {Bennert}, Vardha N. and {Gallo}, Elena and {Hodges-Kluck}, Edmund and {Treu}, Tommaso and {Barth}, Aaron J. and {Cho}, Wanjin and {Foord}, Adi and {Geum}, Jaehyuk and {Guo}, Hengxiao and {Jadhav}, Yashashree and {Jeon}, Yiseul and {Kabasares}, Kyle M. and {Kang}, Won-Suk and {Kim}, Changseok and {Kim}, Minjin and {Kim}, Tae-Woo and {Le}, Huynh Anh N. and {Malkan}, Matthew A. and {Mandal}, Amit Kumar and {Park}, Daeseong and {Spencer}, Chance and {Shin}, Jaejin and {Sung}, Hyun-il and {U}, Vivian and {Williams}, Peter R. and {Yee}, Nick},
        title = "{The Seoul National University AGN Monitoring Project. III. H{\ensuremath{\beta}} Lag Measurements of 32 Luminous Active Galactic Nuclei and the High-luminosity End of the Size{\textendash}Luminosity Relation}",
      journal = {\apj},
         year = 2024,
        month = feb,
       volume = {962},
       number = {1},
          eid = {67},
        pages = {67},
          doi = {10.3847/1538-4357/ad132f},
archivePrefix = {arXiv},
       eprint = {2311.15518},
 primaryClass = {astro-ph.GA},
       adsurl = {https://ui.adsabs.harvard.edu/abs/2024ApJ...962...67W}
}

@ARTICLE{Amorim2024,
       author = {{GRAVITY Collaboration} and {Amorim}, A. and {Bourdarot}, G. and {Brandner}, W. and {Cao}, Y. and {Cl{\'e}net}, Y. and {Davies}, R. and {de Zeeuw}, P.~T. and {Dexter}, J. and {Drescher}, A. and {Eckart}, A. and {Eisenhauer}, F. and {Fabricius}, M. and {Feuchtgruber}, H. and {F{\"o}rster Schreiber}, N.~M. and {Garcia}, P.~J.~V. and {Genzel}, R. and {Gillessen}, S. and {Gratadour}, D. and {H{\"o}nig}, S. and {Kishimoto}, M. and {Lacour}, S. and {Lutz}, D. and {Millour}, F. and {Netzer}, H. and {Ott}, T. and {Paumard}, T. and {Perraut}, K. and {Perrin}, G. and {Peterson}, B.~M. and {Petrucci}, P.~O. and {Pfuhl}, O. and {Prieto}, M.~A. and {Rabien}, S. and {Rouan}, D. and {Santos}, D.~J.~D. and {Shangguan}, J. and {Shimizu}, T. and {Sternberg}, A. and {Straubmeier}, C. and {Sturm}, E. and {Tacconi}, L.~J. and {Tristram}, K.~R.~W. and {Widmann}, F. and {Woillez}, J.},
        title = "{The size-luminosity relation of local active galactic nuclei from interferometric observations of the broad-line region}",
      journal = {\aap},
         year = 2024,
        month = apr,
       volume = {684},
          eid = {A167},
        pages = {A167},
          doi = {10.1051/0004-6361/202348167},
archivePrefix = {arXiv},
       eprint = {2401.07676},
 primaryClass = {astro-ph.GA},
       adsurl = {https://ui.adsabs.harvard.edu/abs/2024A&A...684A.167G}
}

@ARTICLE{Secunda2025,
       author = {{Secunda}, Amy and {Jiang}, Yan-Fei and {Greene}, Jenny E.},
        title = "{Continuum Reverberation in Active Galactic Nuclei Disks Only with Sufficient X-Ray Luminosity and Low Albedo}",
      journal = {\apj},
         year = 2025,
        month = may,
       volume = {984},
       number = {1},
          eid = {19},
        pages = {19},
          doi = {10.3847/1538-4357/adc25b},
archivePrefix = {arXiv},
       eprint = {2501.06304},
 primaryClass = {astro-ph.HE},
       adsurl = {https://ui.adsabs.harvard.edu/abs/2025ApJ...984...19S}
}

@ARTICLE{Bourdarot2024,
       author = {{Bourdarot}, G. and {Eisenhauer}, F.},
        title = "{Kilometer-baseline interferometry: science drivers for the next generation instrument}",
      journal = {arXiv e-prints},
         year = 2024,
        month = oct,
          eid = {arXiv:2410.22063},
        pages = {arXiv:2410.22063},
          doi = {10.48550/arXiv.2410.22063},
archivePrefix = {arXiv},
       eprint = {2410.22063},
 primaryClass = {astro-ph.IM},
       adsurl = {https://ui.adsabs.harvard.edu/abs/2024arXiv241022063B}
}

@ARTICLE{Verner2009,
       author = {{Verner}, Ekaterina and {Bruhweiler}, Frederick and {Johansson}, Sveneric and {Peterson}, Bruce},
        title = "{Fe ii emission spectra in AGN: observations and theoretical interpretation}",
      journal = {Physica Scripta Volume T},
         year = 2009,
        month = may,
       volume = {134},
          eid = {014006},
        pages = {014006},
          doi = {10.1088/0031-8949/2009/T134/014006},
       adsurl = {https://ui.adsabs.harvard.edu/abs/2009PhST..134a4006V}
}

@ARTICLE{Verner2004,
       author = {{Verner}, E. and {Bruhweiler}, F. and {Verner}, D. and {Johansson}, S. and {Kallman}, T. and {Gull}, T.},
        title = "{Fe II Diagnostic Tools for Quasars}",
      journal = {\apj},
         year = 2004,
        month = aug,
       volume = {611},
       number = {2},
        pages = {780-785},
          doi = {10.1086/422303},
archivePrefix = {arXiv},
       eprint = {astro-ph/0404593},
 primaryClass = {astro-ph},
       adsurl = {https://ui.adsabs.harvard.edu/abs/2004ApJ...611..780V}
}

@ARTICLE{Kang2024,
       author = {{Kang}, Wen-Yong and {Wang}, Jun-Xian and {Cai}, Zhen-Yi and {Wang}, Hao-Chen and {Ren}, Wen-Ke and {Liao}, Mai and {Yuan}, Feng and {Zdziarski}, Andrzej and {Cao}, Xinwu},
        title = "{A Surprising Excess of Radio Emission in Extremely Stable Quasars: A Unique Clue to Jet Launching?}",
      journal = {\apj},
         year = 2024,
        month = aug,
       volume = {971},
       number = {1},
          eid = {60},
        pages = {60},
          doi = {10.3847/1538-4357/ad5a0c},
archivePrefix = {arXiv},
       eprint = {2406.13169},
 primaryClass = {astro-ph.GA},
       adsurl = {https://ui.adsabs.harvard.edu/abs/2024ApJ...971...60K}
}

@ARTICLE{Secunda2020,
       author = {{Secunda}, Amy and {Bellovary}, Jillian and {Mac Low}, Mordecai-Mark and {Ford}, K.~E. Saavik and {McKernan}, Barry and {Leigh}, Nathan W.~C. and {Lyra}, Wladimir and {S{\'a}ndor}, Zsolt and {Adorno}, Jose I.},
        title = "{Orbital Migration of Interacting Stellar Mass Black Holes in Disks around Supermassive Black Holes. II. Spins and Incoming Objects}",
      journal = {\apj},
         year = 2020,
        month = nov,
       volume = {903},
       number = {2},
          eid = {133},
        pages = {133},
          doi = {10.3847/1538-4357/abbc1d},
archivePrefix = {arXiv},
       eprint = {2004.11936},
 primaryClass = {astro-ph.HE},
       adsurl = {https://ui.adsabs.harvard.edu/abs/2020ApJ...903..133S}
}

@ARTICLE{McKernan2024,
       author = {{McKernan}, Barry and {Ford}, K.~E. Saavik and {Cook}, Harrison E. and {Delfavero}, Vera and {McPike}, Emily and {Nathaniel}, Kaila and {Postiglione}, Jake and {Ray}, Shawn and {O'Shaughnessy}, Richard},
        title = "{McFACTS I: Testing the LVK AGN channel with Monte Carlo For AGN Channel Testing \& Simulation (McFACTS)}",
      journal = {arXiv e-prints},
         year = 2024,
        month = oct,
          eid = {arXiv:2410.16515},
        pages = {arXiv:2410.16515},
          doi = {10.48550/arXiv.2410.16515},
archivePrefix = {arXiv},
       eprint = {2410.16515},
 primaryClass = {astro-ph.HE},
       adsurl = {https://ui.adsabs.harvard.edu/abs/2024arXiv241016515M}
}

@ARTICLE{Zu2011,
       author = {{Zu}, Ying and {Kochanek}, C.~S. and {Peterson}, Bradley M.},
        title = "{An Alternative Approach to Measuring Reverberation Lags in Active Galactic Nuclei}",
      journal = {\apj},
         year = 2011,
        month = jul,
       volume = {735},
       number = {2},
          eid = {80},
        pages = {80},
          doi = {10.1088/0004-637X/735/2/80},
archivePrefix = {arXiv},
       eprint = {1008.0641},
 primaryClass = {astro-ph.CO},
       adsurl = {https://ui.adsabs.harvard.edu/abs/2011ApJ...735...80Z}
}

@ARTICLE{Luo2025,
       author = {{Luo}, Jun and {An}, Haipeng and {Bian}, Ligong and {Cai}, Rong-Gen and {Cao}, Zhoujian and {Han}, Wenbiao and {He}, Jianhua and {Hendry}, Martin A. and {Hu}, Bin and {Hu}, Yi-Ming and {Huang}, Fa Peng and {Huang}, Shun-Jia and {Kim}, Sang Pyo and {Li}, En-Kun and {Liu}, Yu-Xiao and {Milyukov}, Vadim and {Pi}, Shi and {Postnov}, Konstantin and {Sasaki}, Misao and {Shao}, Cheng-Gang and {Shao}, Lijing and {Shi}, Changfu and {Sun}, Shuo and {Wang}, Anzhong and {Wang}, Pan-Pan and {Wang}, Sai and {Wang}, Shao-Jiang and {Xianyu}, Zhong-Zhi and {Yang}, Huan and {Yang}, Tao and {Zhang}, Jian-dong and {Zhang}, Xin and {Zhao}, Wen and {Zhu}, Liang-Gui and {Mei}, Jianwei},
        title = "{Fundamental Physics and Cosmology with TianQin}",
      journal = {arXiv e-prints},
         year = 2025,
        month = feb,
          eid = {arXiv:2502.20138},
        pages = {arXiv:2502.20138},
          doi = {10.48550/arXiv.2502.20138},
archivePrefix = {arXiv},
       eprint = {2502.20138},
 primaryClass = {gr-qc},
       adsurl = {https://ui.adsabs.harvard.edu/abs/2025arXiv250220138L}
}

@INPROCEEDINGS{Collin2001,
       author = {{Collin}, S. and {Abrassart}, A. and {Czerny}, B. and {Dumont}, A. -M. and {Mouchet}, M.},
        title = "{Accretion and emission processes in AGN: the UV-X connection}",
    booktitle = {EAS Publications Series},
         year = 2001,
       editor = {{Rocca-Volmerange}, Brigitte and {Sol}, H{\'e}l{\`e}ne},
       series = {EAS Publications Series},
       volume = {1},
        month = jan,
        pages = {35-51},
          doi = {10.48550/arXiv.astro-ph/0003108},
archivePrefix = {arXiv},
       eprint = {astro-ph/0003108},
 primaryClass = {astro-ph},
       adsurl = {https://ui.adsabs.harvard.edu/abs/2001EAS.....1...35C}
}

@ARTICLE{Lewin2024,
       author = {{Lewin}, Collin and {Kara}, Erin and {Barth}, Aaron J. and {Cackett}, Edward M. and {De Rosa}, Gisella and {Homayouni}, Yasaman and {Horne}, Keith and {Kriss}, Gerard A. and {Landt}, Hermine and {Gelbord}, Jonathan and {Montano}, John and {Arav}, Nahum and {Bentz}, Misty C. and {Boizelle}, Benjamin D. and {Dalla Bont{\`a}}, Elena and {Brotherton}, Michael S. and {Dehghanian}, Maryam and {Ferland}, Gary J. and {Fian}, Carina and {Goad}, Michael R. and {Hern{\'a}ndez Santisteban}, Juan V. and {Ili{\'c}}, Dragana and {Kaastra}, Jelle and {Kaspi}, Shai and {Korista}, Kirk T. and {Kosec}, Peter and {Kova{\v{c}}evi{\'c}}, Andjelka and {Mehdipour}, Missagh and {Miller}, Jake A. and {Netzer}, Hagai and {Neustadt}, Jack M.~M. and {Panagiotou}, Christos and {Partington}, Ethan R. and {Popovi{\'c}}, Luka {\v{C}}. and {Sanmartim}, David and {Vestergaard}, Marianne and {Ward}, Martin J. and {Zaidouni}, Fatima},
        title = "{AGN STORM 2. VII. A Frequency-resolved Map of the Accretion Disk in Mrk 817: Simultaneous X-Ray Reverberation and UVOIR Disk Reprocessing Time Lags}",
      journal = {\apj},
         year = 2024,
        month = oct,
       volume = {974},
       number = {2},
          eid = {271},
        pages = {271},
          doi = {10.3847/1538-4357/ad6b08},
archivePrefix = {arXiv},
       eprint = {2409.09115},
 primaryClass = {astro-ph.HE},
       adsurl = {https://ui.adsabs.harvard.edu/abs/2024ApJ...974..271L}
}

@ARTICLE{Grier2017,
       author = {{Grier}, C.~J. and {Trump}, J.~R. and {Shen}, Yue and {Horne}, Keith and {Kinemuchi}, Karen and {McGreer}, Ian D. and {Starkey}, D.~A. and {Brandt}, W.~N. and {Hall}, P.~B. and {Kochanek}, C.~S. and {Chen}, Yuguang and {Denney}, K.~D. and {Greene}, Jenny E. and {Ho}, L.~C. and {Homayouni}, Y. and {I-Hsiu Li}, Jennifer and {Pei}, Liuyi and {Peterson}, B.~M. and {Petitjean}, P. and {Schneider}, D.~P. and {Sun}, Mouyuan and {AlSayyad}, Yusura and {Bizyaev}, Dmitry and {Brinkmann}, Jonathan and {Brownstein}, Joel R. and {Bundy}, Kevin and {Dawson}, K.~S. and {Eftekharzadeh}, Sarah and {Fernandez-Trincado}, J.~G. and {Gao}, Yang and {Hutchinson}, Timothy A. and {Jia}, Siyao and {Jiang}, Linhua and {Oravetz}, Daniel and {Pan}, Kaike and {Paris}, Isabelle and {Ponder}, Kara A. and {Peters}, Christina and {Rogerson}, Jesse and {Simmons}, Audrey and {Smith}, Robyn and {Wang}, Ran},
        title = "{The Sloan Digital Sky Survey Reverberation Mapping Project: H{\ensuremath{\alpha}} and H{\ensuremath{\beta}} Reverberation Measurements from First-year Spectroscopy and Photometry}",
      journal = {\apj},
         year = 2017,
        month = dec,
       volume = {851},
       number = {1},
          eid = {21},
        pages = {21},
          doi = {10.3847/1538-4357/aa98dc},
archivePrefix = {arXiv},
       eprint = {1711.03114},
 primaryClass = {astro-ph.GA},
       adsurl = {https://ui.adsabs.harvard.edu/abs/2017ApJ...851...21G}
}

@ARTICLE{Secunda2024,
       author = {{Secunda}, Amy and {Jiang}, Yan-Fei and {Greene}, Jenny E.},
        title = "{Simulating X-Ray Reverberation in the Ultraviolet-emitting Regions of Active Galactic Nuclei Accretion Disks with Three-dimensional Multifrequency Radiation Magnetohydrodynamic Simulations}",
      journal = {\apjl},
         year = 2024,
        month = apr,
       volume = {965},
       number = {2},
          eid = {L29},
        pages = {L29},
          doi = {10.3847/2041-8213/ad34b0},
archivePrefix = {arXiv},
       eprint = {2311.10820},
 primaryClass = {astro-ph.HE},
       adsurl = {https://ui.adsabs.harvard.edu/abs/2024ApJ...965L..29S}
}

@ARTICLE{Yang2024,
       author = {{Yang}, Sen and {Du}, Pu and {Wang}, Jian-Min},
        title = "{Dependence of Virial Factors on Optical Spectral Properties of Active Galactic Nuclei}",
      journal = {\apjs},
         year = 2024,
        month = oct,
       volume = {274},
       number = {2},
          eid = {24},
        pages = {24},
          doi = {10.3847/1538-4365/ad61e3},
archivePrefix = {arXiv},
       eprint = {2407.04257},
 primaryClass = {astro-ph.GA},
       adsurl = {https://ui.adsabs.harvard.edu/abs/2024ApJS..274...24Y}
}

@ARTICLE{Shen2014,
       author = {{Shen}, Yue and {Ho}, Luis C.},
        title = "{The diversity of quasars unified by accretion and orientation}",
      journal = {\nat},
         year = 2014,
        month = sep,
       volume = {513},
       number = {7517},
        pages = {210-213},
          doi = {10.1038/nature13712},
archivePrefix = {arXiv},
       eprint = {1409.2887},
 primaryClass = {astro-ph.GA},
       adsurl = {https://ui.adsabs.harvard.edu/abs/2014Natur.513..210S}
}

@ARTICLE{Abbott2020,
       author = {{Abbott}, R. and {Abbott}, T.~D. and {Abraham}, S. and {Acernese}, F. and {Ackley}, K. and {Adams}, C. and {Adhikari}, R.~X. and {Adya}, V.~B. and {Affeldt}, C. and {Agathos}, M. and {Agatsuma}, K. and {Aggarwal}, N. and {Aguiar}, O.~D. and {Aich}, A. and {Aiello}, L. and {Ain}, A. and {Ajith}, P. and {Akcay}, S. and {Allen}, G. and {Allocca}, A. and {Altin}, P.~A. and {Amato}, A. and {Anand}, S. and {Ananyeva}, A. and {Anderson}, S.~B. and {Anderson}, W.~G. and {Angelova}, S.~V. and {Ansoldi}, S. and {Antier}, S. and {Appert}, S. and {Arai}, K. and {Araya}, M.~C. and {Areeda}, J.~S. and {Ar{\`e}ne}, M. and {Arnaud}, N. and {Aronson}, S.~M. and {Arun}, K.~G. and {Asali}, Y. and {Ascenzi}, S. and {Ashton}, G. and {Aston}, S.~M. and {Astone}, P. and {Aubin}, F. and {Aufmuth}, P. and {AultONeal}, K. and {Austin}, C. and {Avendano}, V. and {Babak}, S. and {Bacon}, P. and {Badaracco}, F. and {Bader}, M.~K.~M. and {Bae}, S. and {Baer}, A.~M. and {Baird}, J. and {Baldaccini}, F. and {Ballardin}, G. and {Ballmer}, S.~W. and {Bals}, A. and {Balsamo}, A. and {Baltus}, G. and {Banagiri}, S. and {Bankar}, D. and {Bankar}, R.~S. and {Barayoga}, J.~C. and {Barbieri}, C. and {Barish}, B.~C. and {Barker}, D. and {Barkett}, K. and {Barneo}, P. and {Barone}, F. and {Barr}, B. and {Barsotti}, L. and {Barsuglia}, M. and {Barta}, D. and {Bartlett}, J. and {Bartos}, I. and {Bassiri}, R. and {Basti}, A. and {Bawaj}, M. and {Bayley}, J.~C. and {Bazzan}, M. and {B{\'e}csy}, B. and {Bejger}, M. and {Belahcene}, I. and {Bell}, A.~S. and {Beniwal}, D. and {Benjamin}, M.~G. and {Bentley}, J.~D. and {Bergamin}, F. and {Berger}, B.~K. and {Bergmann}, G. and {Bernuzzi}, S. and {Berry}, C.~P.~L. and {Bersanetti}, D. and {Bertolini}, A. and {Betzwieser}, J. and {Bhandare}, R. and {Bhandari}, A.~V. and {Bidler}, J. and {Biggs}, E. and {Bilenko}, I.~A. and {Billingsley}, G. and {Birney}, R. and {Birnholtz}, O. and {Biscans}, S. and {Bischi}, M. and {Biscoveanu}, S. and {Bisht}, A. and {Bissenbayeva}, G. and {Bitossi}, M. and {Bizouard}, M.~A. and {Blackburn}, J.~K. and {Blackman}, J. and {Blair}, C.~D. and {Blair}, D.~G. and {Blair}, R.~M. and {Bobba}, F. and {Bode}, N. and {Boer}, M. and {Boetzel}, Y. and {Bogaert}, G. and {Bondu}, F. and {Bonilla}, E. and {Bonnand}, R. and {Booker}, P. and {Boom}, B.~A. and {Bork}, R. and {Boschi}, V. and {Bose}, S. and {Bossilkov}, V. and {Bosveld}, J. and {Bouffanais}, Y. and {Bozzi}, A. and {Bradaschia}, C. and {Brady}, P.~R. and {Bramley}, A. and {Branchesi}, M. and {Brau}, J.~E. and {Breschi}, M. and {Briant}, T. and {Briggs}, J.~H. and {Brighenti}, F. and {Brillet}, A. and {Brinkmann}, M. and {Brockill}, P. and {Brooks}, A.~F. and {Brooks}, J. and {Brown}, D.~D. and {Brunett}, S. and {Bruno}, G. and {Bruntz}, R. and {Buikema}, A. and {Bulik}, T. and {Bulten}, H.~J. and {Buonanno}, A. and {Buscicchio}, R. and {Buskulic}, D. and {Byer}, R.~L. and {Cabero}, M. and {Cadonati}, L. and {Cagnoli}, G. and {Cahillane}, C. and {Calder{\'o}n Bustillo}, J. and {Callaghan}, J.~D. and {Callister}, T.~A. and {Calloni}, E. and {Camp}, J.~B. and {Canepa}, M. and {Cannon}, K.~C. and {Cao}, H. and {Cao}, J. and {Carapella}, G. and {Carbognani}, F. and {Caride}, S. and {Carney}, M.~F. and {Carullo}, G. and {Casanueva Diaz}, J. and {Casentini}, C. and {Casta{\~n}eda}, J. and {Caudill}, S. and {Cavagli{\`a}}, M. and {Cavalier}, F. and {Cavalieri}, R. and {Cella}, G. and {Cerd{\'a}-Dur{\'a}n}, P. and {Cesarini}, E. and {Chaibi}, O. and {Chakravarti}, K. and {Chan}, C. and {Chan}, M. and {Chandra}, K. and {Chao}, S. and {Charlton}, P. and {Chase}, E.~A. and {Chassande-Mottin}, E. and {Chatterjee}, D. and {Chaturvedi}, M. and {Chatziioannou}, K. and {Chen}, H.~Y. and {Chen}, X.},
        title = "{GW190521: A Binary Black Hole Merger with a Total Mass of $150\,M_{\odot}$}",
      journal = {\prl},
         year = 2020,
        month = sep,
       volume = {125},
       number = {10},
          eid = {101102},
        pages = {101102},
          doi = {10.1103/PhysRevLett.125.101102},
archivePrefix = {arXiv},
       eprint = {2009.01075},
 primaryClass = {gr-qc},
       adsurl = {https://ui.adsabs.harvard.edu/abs/2020PhRvL.125j1102A}
}

@ARTICLE{Kammoun2021,
       author = {{Kammoun}, E.~S. and {Dov{\v{c}}iak}, M. and {Papadakis}, I.~E. and {Caballero-Garc{\'\i}a}, M.~D. and {Karas}, V.},
        title = "{UV/Optical Disk Thermal Reverberation in Active Galactic Nuclei: An In-depth Study with an Analytic Prescription for Time-lag Spectra}",
      journal = {\apj},
         year = 2021,
        month = jan,
       volume = {907},
       number = {1},
          eid = {20},
        pages = {20},
          doi = {10.3847/1538-4357/abcb93},
archivePrefix = {arXiv},
       eprint = {2011.08563},
 primaryClass = {astro-ph.HE},
       adsurl = {https://ui.adsabs.harvard.edu/abs/2021ApJ...907...20K}
}

@ARTICLE{Svensson1994,
       author = {{Svensson}, Roland and {Zdziarski}, Andrzej A.},
        title = "{Black Hole Accretion Disks with Coronae}",
      journal = {\apj},
         year = 1994,
        month = dec,
       volume = {436},
        pages = {599},
          doi = {10.1086/174934},
       adsurl = {https://ui.adsabs.harvard.edu/abs/1994ApJ...436..599S}
}

@ARTICLE{Wang2025,
       author = {{Wang}, Yi-Lin and {Liu}, Jun-Rong and {Wang}, Jian-Min},
        title = "{Continuum reverberation mapping of accretion disks depending on the vertical structures in active galactic nuclei}",
      journal = {\aap},
         year = 2025,
        month = mar,
       volume = {695},
          eid = {A143},
        pages = {A143},
          doi = {10.1051/0004-6361/202452114},
       adsurl = {https://ui.adsabs.harvard.edu/abs/2025A&A...695A.143W}
}

@ARTICLE{Tagawa2023,
       author = {{Tagawa}, Hiromichi and {Kimura}, Shigeo S. and {Haiman}, Zolt{\'a}n and {Perna}, Rosalba and {Bartos}, Imre},
        title = "{Observable Signatures of Stellar-mass Black Holes in Active Galactic Nuclei}",
      journal = {\apjl},
         year = 2023,
        month = mar,
       volume = {946},
       number = {1},
          eid = {L3},
        pages = {L3},
          doi = {10.3847/2041-8213/acc103},
archivePrefix = {arXiv},
       eprint = {2303.02172},
 primaryClass = {astro-ph.HE},
       adsurl = {https://ui.adsabs.harvard.edu/abs/2023ApJ...946L...3T}
}

@ARTICLE{Netzer2022,
       author = {{Netzer}, Hagai},
        title = "{Continuum reverberation mapping and a new lag-luminosity relationship for AGN}",
      journal = {\mnras},
         year = 2022,
        month = jan,
       volume = {509},
       number = {2},
        pages = {2637-2646},
          doi = {10.1093/mnras/stab3133},
archivePrefix = {arXiv},
       eprint = {2110.05512},
 primaryClass = {astro-ph.GA},
       adsurl = {https://ui.adsabs.harvard.edu/abs/2022MNRAS.509.2637N}
}

@ARTICLE{Liu2024,
       author = {{Liu}, Jun-Rong and {Wang}, Yi-Lin and {Wang}, Jian-Min},
        title = "{Accretion-modified Stars in Accretion Disks of Active Galactic Nuclei: Observational Characteristics in Different Regions of the Disks}",
      journal = {\apj},
         year = 2024,
        month = jul,
       volume = {969},
       number = {1},
          eid = {37},
        pages = {37},
          doi = {10.3847/1538-4357/ad463a},
archivePrefix = {arXiv},
       eprint = {2405.02855},
 primaryClass = {astro-ph.HE},
       adsurl = {https://ui.adsabs.harvard.edu/abs/2024ApJ...969...37L}
}

@ARTICLE{Cantiello2021,
       author = {{Cantiello}, Matteo and {Jermyn}, Adam S. and {Lin}, Douglas N.~C.},
        title = "{Stellar Evolution in AGN Disks}",
      journal = {\apj},
         year = 2021,
        month = apr,
       volume = {910},
       number = {2},
          eid = {94},
        pages = {94},
          doi = {10.3847/1538-4357/abdf4f},
archivePrefix = {arXiv},
       eprint = {2009.03936},
 primaryClass = {astro-ph.SR},
       adsurl = {https://ui.adsabs.harvard.edu/abs/2021ApJ...910...94C}
}

@ARTICLE{Samsing2022,
       author = {{Samsing}, J. and {Bartos}, I. and {D'Orazio}, D.~J. and {Haiman}, Z. and {Kocsis}, B. and {Leigh}, N.~W.~C. and {Liu}, B. and {Pessah}, M.~E. and {Tagawa}, H.},
        title = "{AGN as potential factories for eccentric black hole mergers}",
      journal = {\nat},
         year = 2022,
        month = mar,
       volume = {603},
       number = {7900},
        pages = {237-240},
          doi = {10.1038/s41586-021-04333-1},
archivePrefix = {arXiv},
       eprint = {2010.09765},
 primaryClass = {astro-ph.HE},
       adsurl = {https://ui.adsabs.harvard.edu/abs/2022Natur.603..237S}
}

@ARTICLE{Graham2023,
       author = {{Graham}, Matthew J. and {McKernan}, Barry and {Ford}, K.~E. Saavik and {Stern}, Daniel and {Djorgovski}, S.~G. and {Coughlin}, Michael and {Burdge}, Kevin B. and {Bellm}, Eric C. and {Helou}, George and {Mahabal}, Ashish A. and {Masci}, Frank J. and {Purdum}, Josiah and {Rosnet}, Philippe and {Rusholme}, Ben},
        title = "{A Light in the Dark: Searching for Electromagnetic Counterparts to Black Hole-Black Hole Mergers in LIGO/Virgo O3 with the Zwicky Transient Facility}",
      journal = {\apj},
         year = 2023,
        month = jan,
       volume = {942},
       number = {2},
          eid = {99},
        pages = {99},
          doi = {10.3847/1538-4357/aca480},
archivePrefix = {arXiv},
       eprint = {2209.13004},
 primaryClass = {astro-ph.HE},
       adsurl = {https://ui.adsabs.harvard.edu/abs/2023ApJ...942...99G}
}

@ARTICLE{Graham2020,
       author = {{Graham}, M.~J. and {Ford}, K.~E.~S. and {McKernan}, B. and {Ross}, N.~P. and {Stern}, D. and {Burdge}, K. and {Coughlin}, M. and {Djorgovski}, S.~G. and {Drake}, A.~J. and {Duev}, D. and {Kasliwal}, M. and {Mahabal}, A.~A. and {van Velzen}, S. and {Belecki}, J. and {Bellm}, E.~C. and {Burruss}, R. and {Cenko}, S.~B. and {Cunningham}, V. and {Helou}, G. and {Kulkarni}, S.~R. and {Masci}, F.~J. and {Prince}, T. and {Reiley}, D. and {Rodriguez}, H. and {Rusholme}, B. and {Smith}, R.~M. and {Soumagnac}, M.~T.},
        title = "{Candidate Electromagnetic Counterpart to the Binary Black Hole Merger Gravitational-Wave Event S190521g$^{*}$}",
      journal = {\prl},
         year = 2020,
        month = jun,
       volume = {124},
       number = {25},
          eid = {251102},
        pages = {251102},
          doi = {10.1103/PhysRevLett.124.251102},
archivePrefix = {arXiv},
       eprint = {2006.14122},
 primaryClass = {astro-ph.HE},
       adsurl = {https://ui.adsabs.harvard.edu/abs/2020PhRvL.124y1102G}
}

@ARTICLE{Wang2021a,
       author = {{Wang}, Jian-Min and {Liu}, Jun-Rong and {Ho}, Luis C. and {Du}, Pu},
        title = "{Accretion-modified Stars in Accretion Disks of Active Galactic Nuclei: Slowly Transient Appearance}",
      journal = {\apjl},
         year = 2021,
        month = apr,
       volume = {911},
       number = {1},
          eid = {L14},
        pages = {L14},
          doi = {10.3847/2041-8213/abee81},
archivePrefix = {arXiv},
       eprint = {2103.07708},
 primaryClass = {astro-ph.HE},
       adsurl = {https://ui.adsabs.harvard.edu/abs/2021ApJ...911L..14W}
}

@ARTICLE{Wang2021b,
       author = {{Wang}, Jian-Min and {Liu}, Jun-Rong and {Ho}, Luis C. and {Li}, Yan-Rong and {Du}, Pu},
        title = "{Accretion-modified Stars in Accretion Disks of Active Galactic Nuclei: Gravitational-wave Bursts and Electromagnetic Counterparts from Merging Stellar Black Hole Binaries}",
      journal = {\apjl},
         year = 2021,
        month = aug,
       volume = {916},
       number = {2},
          eid = {L17},
        pages = {L17},
          doi = {10.3847/2041-8213/ac0b46},
archivePrefix = {arXiv},
       eprint = {2106.07334},
 primaryClass = {astro-ph.HE},
       adsurl = {https://ui.adsabs.harvard.edu/abs/2021ApJ...916L..17W}
}

@ARTICLE{Cheng1999,
       author = {{Cheng}, K.~S. and {Wang}, Jian-Min},
        title = "{The Formation and Merger of Compact Objects in the Central Engine of Active Galactic Nuclei and Quasars: Gamma-Ray Burst and Gravitational Radiation}",
      journal = {\apj},
         year = 1999,
        month = aug,
       volume = {521},
       number = {2},
        pages = {502-508},
          doi = {10.1086/307572},
archivePrefix = {arXiv},
       eprint = {astro-ph/9908228},
 primaryClass = {astro-ph},
       adsurl = {https://ui.adsabs.harvard.edu/abs/1999ApJ...521..502C}
}

@ARTICLE{Gilbaum2022,
       author = {{Gilbaum}, Shmuel and {Stone}, Nicholas C.},
        title = "{Feedback-dominated Accretion Flows}",
      journal = {\apj},
         year = 2022,
        month = apr,
       volume = {928},
       number = {2},
          eid = {191},
        pages = {191},
          doi = {10.3847/1538-4357/ac4ded},
archivePrefix = {arXiv},
       eprint = {2107.07519},
 primaryClass = {astro-ph.HE},
       adsurl = {https://ui.adsabs.harvard.edu/abs/2022ApJ...928..191G}
}

@ARTICLE{Zhou2024,
       author = {{Zhou}, Shuying and {Sun}, Mouyuan and {Liu}, Tong and {Wang}, Jian-Min and {Wang}, Jun-Xian and {Xue}, Yongquan},
        title = "{Stellar Black Holes Can ``Stretch'' Supermassive Black Hole Accretion Disks}",
      journal = {\apjl},
         year = 2024,
        month = may,
       volume = {966},
       number = {1},
          eid = {L9},
        pages = {L9},
          doi = {10.3847/2041-8213/ad3c3f},
archivePrefix = {arXiv},
       eprint = {2404.07407},
 primaryClass = {astro-ph.HE},
       adsurl = {https://ui.adsabs.harvard.edu/abs/2024ApJ...966L...9Z}
}

@BOOK{Netzer2013,
       author = {{Netzer}, Hagai},
        title = "{The Physics and Evolution of Active Galactic Nuclei}",
         year = 2013,
       adsurl = {https://ui.adsabs.harvard.edu/abs/2013peag.book.....N}
}

@ARTICLE{Bahcall1972,
       author = {{Bahcall}, John N. and {Kozlovsky}, Ben-Zion and {Salpeter}, E.~E.},
        title = "{On the Time Dependence of Emission-Line Strengths from a Photoionized Nebula}",
      journal = {\apj},
         year = 1972,
        month = feb,
       volume = {171},
        pages = {467},
          doi = {10.1086/151300},
       adsurl = {https://ui.adsabs.harvard.edu/abs/1972ApJ...171..467B}
}

@ARTICLE{Blandford1982,
       author = {{Blandford}, R.~D. and {McKee}, C.~F.},
        title = "{Reverberation mapping of the emission line regions of Seyfert galaxies and quasars.}",
      journal = {\apj},
         year = 1982,
        month = apr,
       volume = {255},
        pages = {419-439},
          doi = {10.1086/159843},
       adsurl = {https://ui.adsabs.harvard.edu/abs/1982ApJ...255..419B}
}

@ARTICLE{Peterson1993,
       author = {{Peterson}, Bradley M.},
        title = "{Reverberation Mapping of Active Galactic Nuclei}",
      journal = {\pasp},
         year = 1993,
        month = mar,
       volume = {105},
        pages = {247},
          doi = {10.1086/133140},
       adsurl = {https://ui.adsabs.harvard.edu/abs/1993PASP..105..247P}
}

@ARTICLE{Peterson2014,
       author = {{Peterson}, Bradley M.},
        title = "{Measuring the Masses of Supermassive Black Holes}",
      journal = {\ssr},
         year = 2014,
        month = sep,
       volume = {183},
       number = {1-4},
        pages = {253-275},
          doi = {10.1007/s11214-013-9987-4},
       adsurl = {https://ui.adsabs.harvard.edu/abs/2014SSRv..183..253P}
}

@ARTICLE{Cackett2021,
       author = {{Cackett}, Edward M. and {Bentz}, Misty C. and {Kara}, Erin},
        title = "{Reverberation mapping of active galactic nuclei: from X-ray corona to dusty torus}",
      journal = {iScience},
         year = 2021,
        month = jun,
       volume = {24},
       number = {6},
        pages = {102557},
          doi = {10.1016/j.isci.2021.102557},
archivePrefix = {arXiv},
       eprint = {2105.06926},
 primaryClass = {astro-ph.GA},
       adsurl = {https://ui.adsabs.harvard.edu/abs/2021iSci...24j2557C}
}

@ARTICLE{Kaspi2000,
       author = {{Kaspi}, Shai and {Smith}, Paul S. and {Netzer}, Hagai and {Maoz}, Dan and {Jannuzi}, Buell T. and {Giveon}, Uriel},
        title = "{Reverberation Measurements for 17 Quasars and the Size-Mass-Luminosity Relations in Active Galactic Nuclei}",
      journal = {\apj},
         year = 2000,
        month = apr,
       volume = {533},
       number = {2},
        pages = {631-649},
          doi = {10.1086/308704},
archivePrefix = {arXiv},
       eprint = {astro-ph/9911476},
 primaryClass = {astro-ph},
       adsurl = {https://ui.adsabs.harvard.edu/abs/2000ApJ...533..631K}
}

@ARTICLE{Bentz2013,
       author = {{Bentz}, Misty C. and {Denney}, Kelly D. and {Grier}, Catherine J. and {Barth}, Aaron J. and {Peterson}, Bradley M. and {Vestergaard}, Marianne and {Bennert}, Vardha N. and {Canalizo}, Gabriela and {De Rosa}, Gisella and {Filippenko}, Alexei V. and {Gates}, Elinor L. and {Greene}, Jenny E. and {Li}, Weidong and {Malkan}, Matthew A. and {Pogge}, Richard W. and {Stern}, Daniel and {Treu}, Tommaso and {Woo}, Jong-Hak},
        title = "{The Low-luminosity End of the Radius-Luminosity Relationship for Active Galactic Nuclei}",
      journal = {\apj},
         year = 2013,
        month = apr,
       volume = {767},
       number = {2},
          eid = {149},
        pages = {149},
          doi = {10.1088/0004-637X/767/2/149},
archivePrefix = {arXiv},
       eprint = {1303.1742},
 primaryClass = {astro-ph.CO},
       adsurl = {https://ui.adsabs.harvard.edu/abs/2013ApJ...767..149B}
}

@ARTICLE{Du2016,
       author = {{Du}, Pu and {Wang}, Jian-Min and {Hu}, Chen and {Ho}, Luis C. and {Li}, Yan-Rong and {Bai}, Jin-Ming},
        title = "{The Fundamental Plane of the Broad-line Region in Active Galactic Nuclei}",
      journal = {\apjl},
         year = 2016,
        month = feb,
       volume = {818},
       number = {1},
          eid = {L14},
        pages = {L14},
          doi = {10.3847/2041-8205/818/1/L14},
archivePrefix = {arXiv},
       eprint = {1601.01391},
 primaryClass = {astro-ph.GA},
       adsurl = {https://ui.adsabs.harvard.edu/abs/2016ApJ...818L..14D}
}

@ARTICLE{Du2018,
       author = {{Du}, Pu and {Zhang}, Zhi-Xiang and {Wang}, Kai and {Huang}, Ying-Ke and {Zhang}, Yue and {Lu}, Kai-Xing and {Hu}, Chen and {Li}, Yan-Rong and {Bai}, Jin-Ming and {Bian}, Wei-Hao and {Yuan}, Ye-Fei and {Ho}, Luis C. and {Wang}, Jian-Min and {SEAMBH Collaboration}},
        title = "{Supermassive Black Holes with High Accretion Rates in Active Galactic Nuclei. IX. 10 New Observations of Reverberation Mapping and Shortened H{\ensuremath{\beta}} Lags}",
      journal = {\apj},
         year = 2018,
        month = mar,
       volume = {856},
       number = {1},
          eid = {6},
        pages = {6},
          doi = {10.3847/1538-4357/aaae6b},
archivePrefix = {arXiv},
       eprint = {1802.03022},
 primaryClass = {astro-ph.GA},
       adsurl = {https://ui.adsabs.harvard.edu/abs/2018ApJ...856....6D}
}

@ARTICLE{Du2019,
       author = {{Du}, Pu and {Wang}, Jian-Min},
        title = "{The Radius-Luminosity Relationship Depends on Optical Spectra in Active Galactic Nuclei}",
      journal = {\apj},
         year = 2019,
        month = nov,
       volume = {886},
       number = {1},
          eid = {42},
        pages = {42},
          doi = {10.3847/1538-4357/ab4908},
archivePrefix = {arXiv},
       eprint = {1909.06735},
 primaryClass = {astro-ph.GA},
       adsurl = {https://ui.adsabs.harvard.edu/abs/2019ApJ...886...42D}
}

@ARTICLE{Wang1999,
       author = {{Wang}, Jian-Min and {Zhou}, You-Yuan},
        title = "{Self-similar Solution of Optically Thick Advection-dominated Flows}",
      journal = {\apj},
         year = 1999,
        month = may,
       volume = {516},
       number = {1},
        pages = {420-424},
          doi = {10.1086/307080},
       adsurl = {https://ui.adsabs.harvard.edu/abs/1999ApJ...516..420W}
}

@ARTICLE{Gravity2018,
       author = {{Gravity Collaboration} and {Sturm}, E. and {Dexter}, J. and {Pfuhl}, O. and {Stock}, M.~R. and {Davies}, R.~I. and {Lutz}, D. and {Cl{\'e}net}, Y. and {Eckart}, A. and {Eisenhauer}, F. and {Genzel}, R. and {Gratadour}, D. and {H{\"o}nig}, S.~F. and {Kishimoto}, M. and {Lacour}, S. and {Millour}, F. and {Netzer}, H. and {Perrin}, G. and {Peterson}, B.~M. and {Petrucci}, P.~O. and {Rouan}, D. and {Waisberg}, I. and {Woillez}, J. and {Amorim}, A. and {Brandner}, W. and {F{\"o}rster Schreiber}, N.~M. and {Garcia}, P.~J.~V. and {Gillessen}, S. and {Ott}, T. and {Paumard}, T. and {Perraut}, K. and {Scheithauer}, S. and {Straubmeier}, C. and {Tacconi}, L.~J. and {Widmann}, F.},
        title = "{Spatially resolved rotation of the broad-line region of a quasar at sub-parsec scale}",
      journal = {\nat},
         year = 2018,
        month = nov,
       volume = {563},
       number = {7733},
        pages = {657-660},
          doi = {10.1038/s41586-018-0731-9},
archivePrefix = {arXiv},
       eprint = {1811.11195},
 primaryClass = {astro-ph.GA},
       adsurl = {https://ui.adsabs.harvard.edu/abs/2018Natur.563..657G}
}

@ARTICLE{Shen2012,
       author = {{Shen}, Yue and {Kelly}, Brandon C.},
        title = "{The Demographics of Broad-line Quasars in the Mass-Luminosity Plane. I. Testing FWHM-based Virial Black Hole Masses}",
      journal = {\apj},
         year = 2012,
        month = feb,
       volume = {746},
       number = {2},
          eid = {169},
        pages = {169},
          doi = {10.1088/0004-637X/746/2/169},
archivePrefix = {arXiv},
       eprint = {1107.4372},
 primaryClass = {astro-ph.CO},
       adsurl = {https://ui.adsabs.harvard.edu/abs/2012ApJ...746..169S}
}

@ARTICLE{Vestergaard2009,
       author = {{Vestergaard}, M. and {Osmer}, Patrick S.},
        title = "{Mass Functions of the Active Black Holes in Distant Quasars from the Large Bright Quasar Survey, the Bright Quasar Survey, and the Color-selected Sample of the SDSS Fall Equatorial Stripe}",
      journal = {\apj},
         year = 2009,
        month = jul,
       volume = {699},
       number = {1},
        pages = {800-816},
          doi = {10.1088/0004-637X/699/1/800},
archivePrefix = {arXiv},
       eprint = {0904.3348},
 primaryClass = {astro-ph.CO},
       adsurl = {https://ui.adsabs.harvard.edu/abs/2009ApJ...699..800V}
}

@ARTICLE{Abramowicz1988,
       author = {{Abramowicz}, M.~A. and {Czerny}, B. and {Lasota}, J.~P. and {Szuszkiewicz}, E.},
        title = "{Slim Accretion Disks}",
      journal = {\apj},
         year = 1988,
        month = sep,
       volume = {332},
        pages = {646},
          doi = {10.1086/166683},
       adsurl = {https://ui.adsabs.harvard.edu/abs/1988ApJ...332..646A}
}

@ARTICLE{Shakura1973,
       author = {{Shakura}, N.~I. and {Sunyaev}, R.~A.},
        title = "{Black holes in binary systems. Observational appearance.}",
      journal = {\aap},
         year = 1973,
        month = jan,
       volume = {24},
        pages = {337-355},
       adsurl = {https://ui.adsabs.harvard.edu/abs/1973A&A....24..337S}
}

\clearpage

\begin{figure*}
\centering
\includegraphics[angle=0,width=0.9\textwidth]{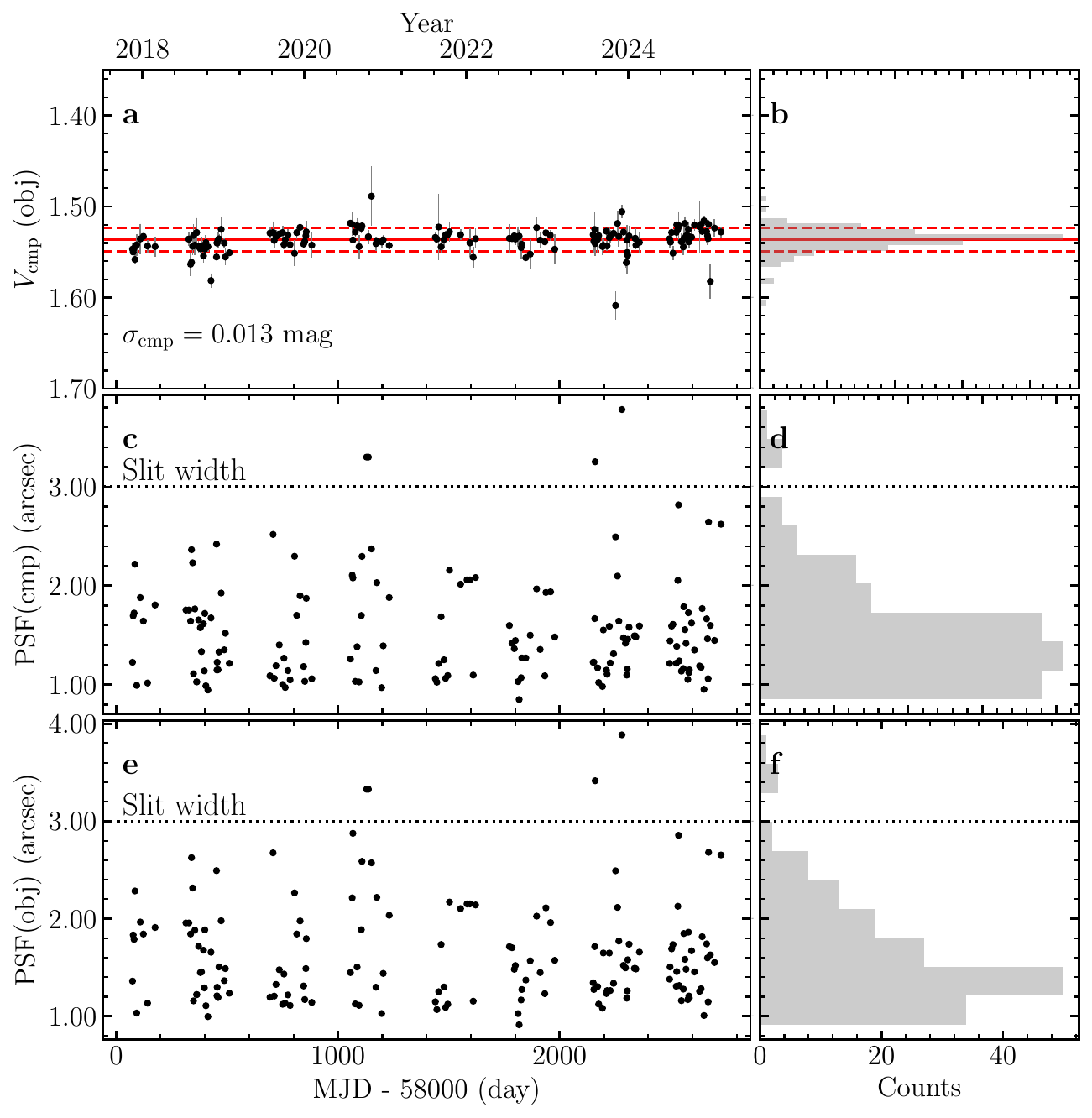}
\caption{\footnotesize
Panel {\textbf{a}}: $V$-band LC of the comparison star indicates that the star keeps constant with a very small scatter. 
Panel {\textbf{c}} and {\textbf{e}}: image PSF FWHMs of the comparison star and the target.
We find ${\rm FWHM(cmp)\approx FWHM(obj)}$, showing that the host of PHL\,1092 weakly contaminates the nuclear emissions.
Panel {\textbf{b,\,d,\,f}} are distributions of the comparison star and seeing.
}
\label{fig:seeing_cmp_lc}
\end{figure*}

\begin{figure*}
\centering
\includegraphics[angle=0,width=0.45\textwidth]{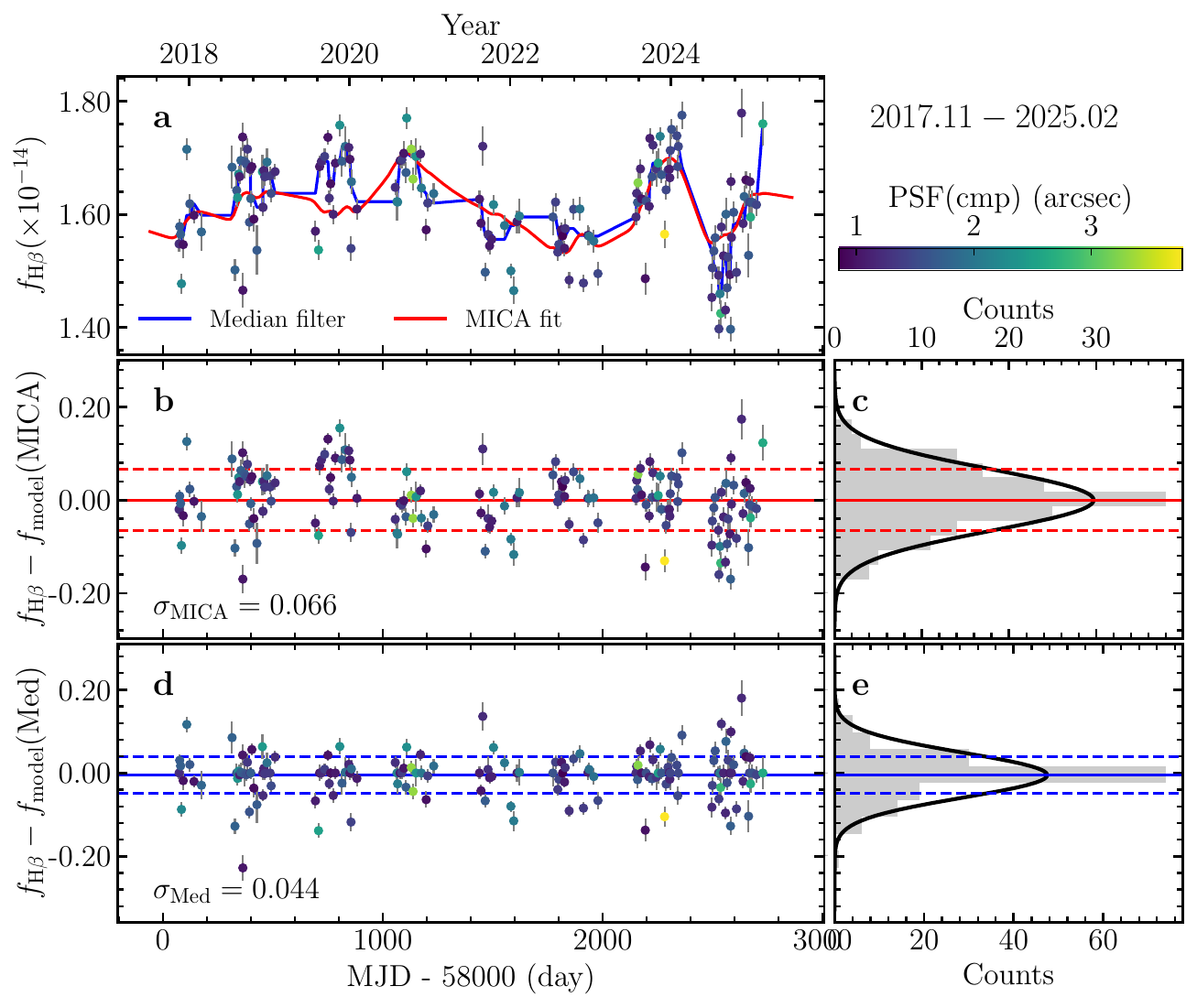}
\includegraphics[angle=0,width=0.45\textwidth]{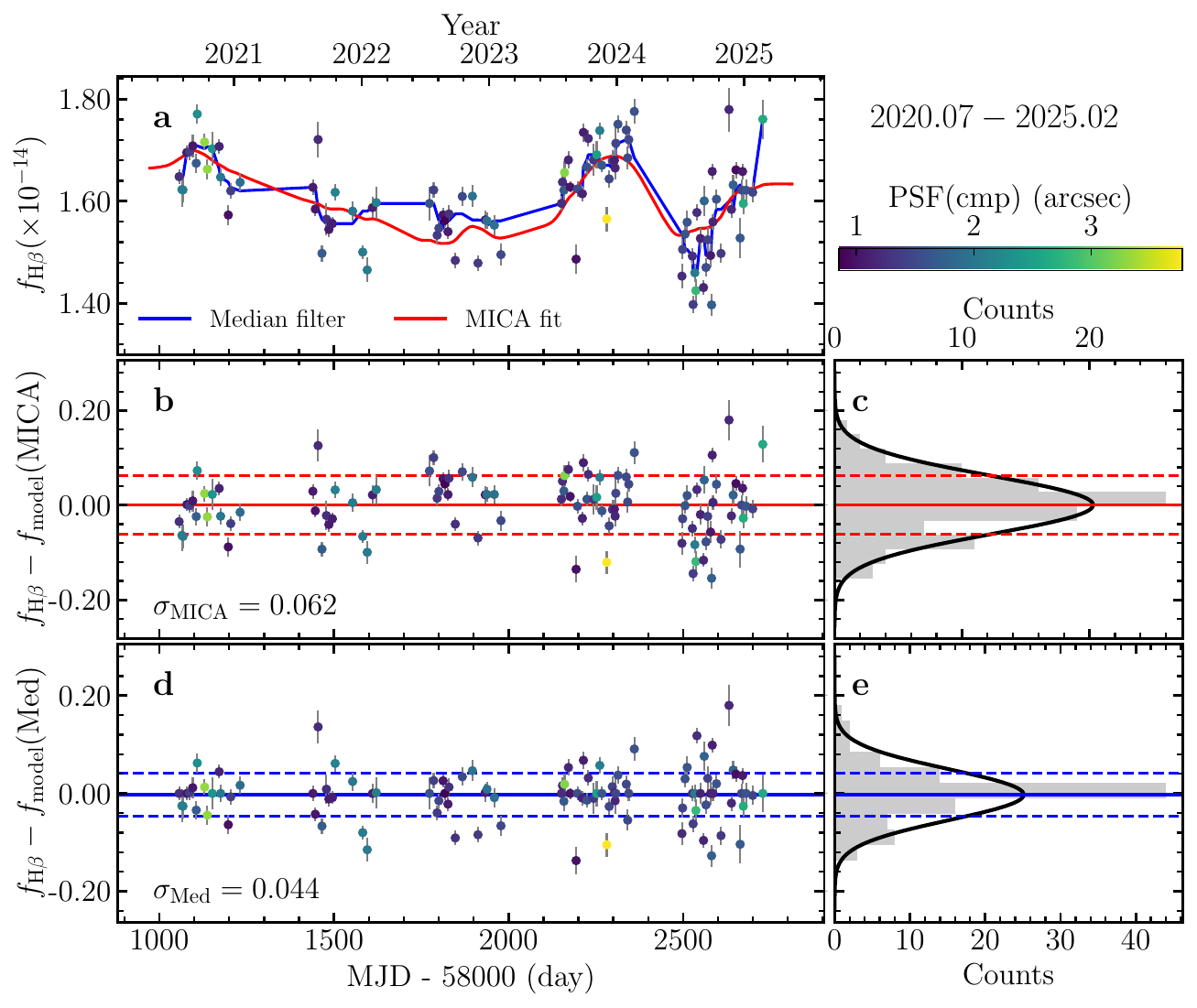}
\includegraphics[angle=0,width=0.45\textwidth]{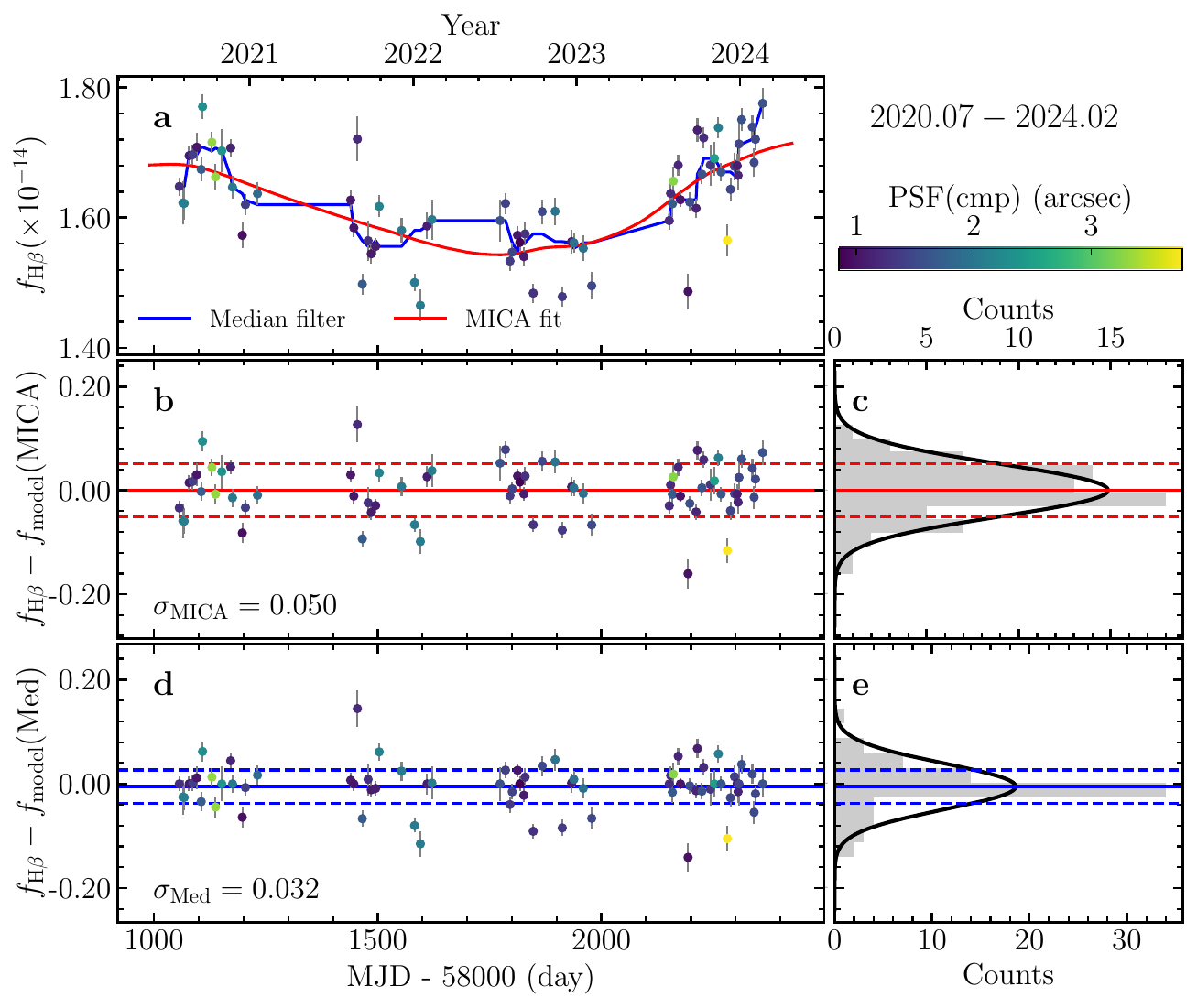}
\includegraphics[angle=0,width=0.45\textwidth]{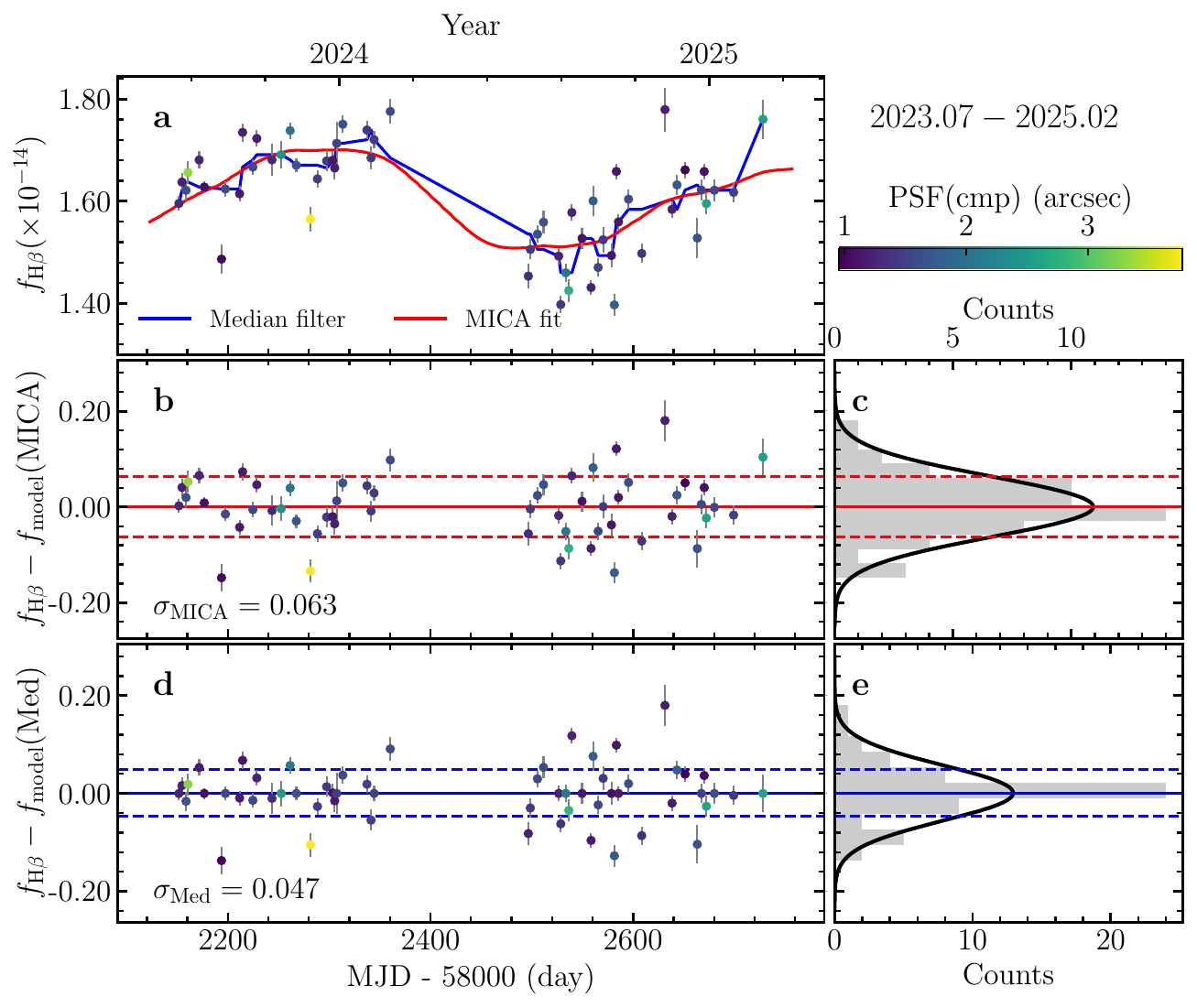}
\caption{\footnotesize 
Panel \textbf{a} shows H$\beta$ LC with median-filter (blue) and \MICA\ (red) model fits curves, colour-coded by the FWHM of the target’s image, which serves as a proxy for change of atmospheric seeing. 
Panels \textbf{d} and \textbf{c} present the residuals from the \MICA\ and median-filter, respectively, with their corresponding residual distributions shown in panels {\textbf{d}} and {\textbf{e}}. Red/Blue solid and dashed lines in panels \textbf{b}, \textbf{d}/\textbf{c}, \textbf{e} mark the mean and standard deviations of residuals, respectively. In panels \textbf{b} and \textbf{c}, no correlation between seeing variations and residuals, indicating that the long-term variability is not caused by seeing changes. Solid black lines in panels \textbf{d} and \textbf{e} indicate Gaussian fits to the residual distributions, demonstrating the robustness of our results against aliasing effects. }
\label{fig:Gauss}
\end{figure*}

\begin{figure*}
\centering
\includegraphics[width=0.9\textwidth]{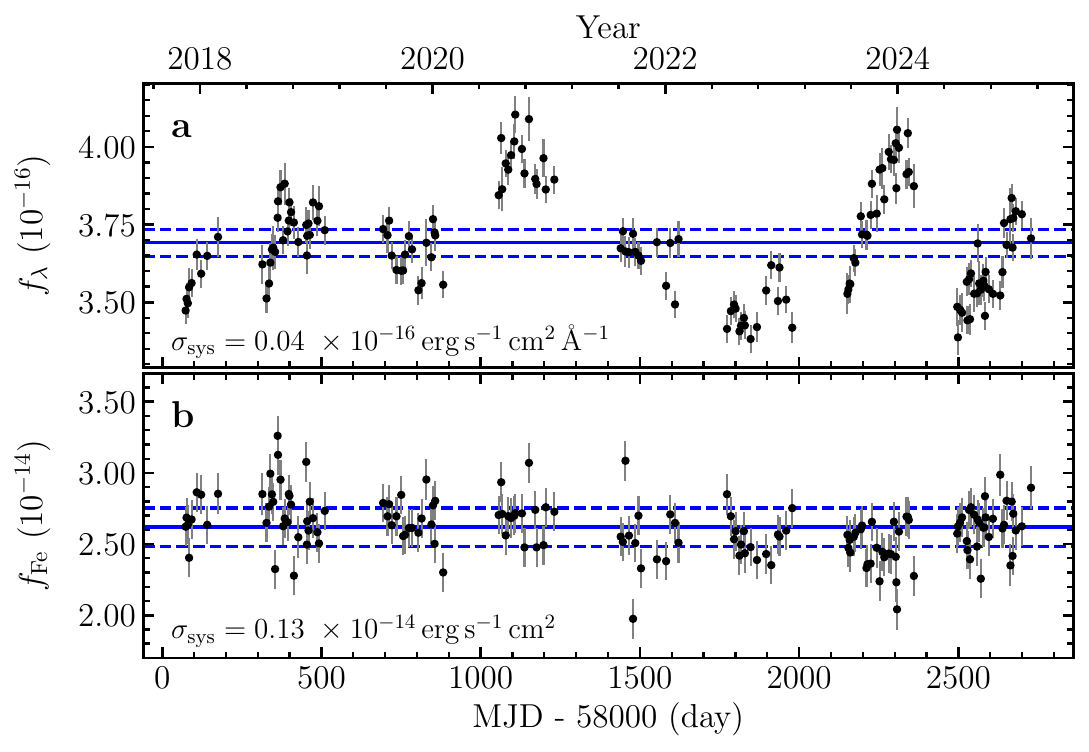}
\caption{\footnotesize LCs of 5100\,\AA\ continuum (panel {\textbf{a}}) and integrated \feii\, line (panel {\textbf{b}}), with the corresponding systematic uncertainties ($\sigma_{\rm sys}$) included. Blue solid lines are the mean fluxes of 5100\,\AA\ continuum and \feii\ line, while the dashed lines are the corresponding scatters, namely systematic uncertainties. No significant variation of the \feii\ line is detected within $1\,\sigma_{\rm sys}$.}
\label{fig:LC_Fe}
\end{figure*}

\begin{figure*}
\centering
\includegraphics[width=0.7\textwidth]{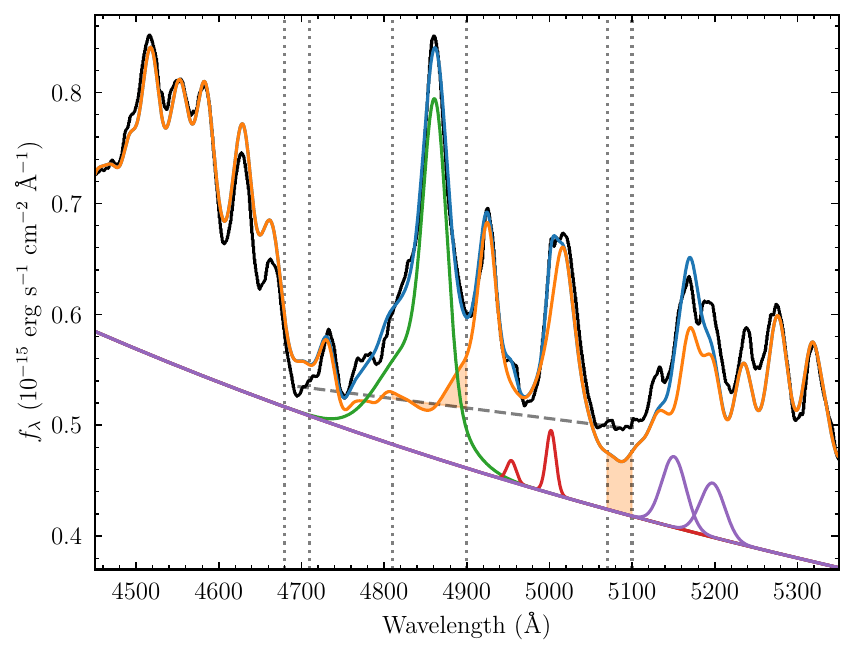}
\caption{\footnotesize
Possible Fe {\sc ii} residuals in the continuum and H$\beta$ windows. The black line represents the mean spectrum from 2017 in the rest frame, while the blue line indicates the best-fit result. 
The orange line corresponds to the Fe {\sc ii} template from Ref.\cite{Boroson1992}. 
The green line depicts the H$\beta$ component, the red lines represent the [O {\sc iii}] emission lines, and the purple lines correspond to the coronal lines. 
Vertical dotted lines mark the positions of the H$\beta$ line and the two continuum windows. 
The dashed lines indicate the continuum background determined by the integration method. 
The orange shaded area represents the possible Fe {\sc ii} contribution that may not be accounted for by the integration method within the measurement windows.
}
\label{fig:Fe_fit}
\end{figure*}

\clearpage

\begin{figure*}
\centering
\includegraphics[width=0.49\textwidth]{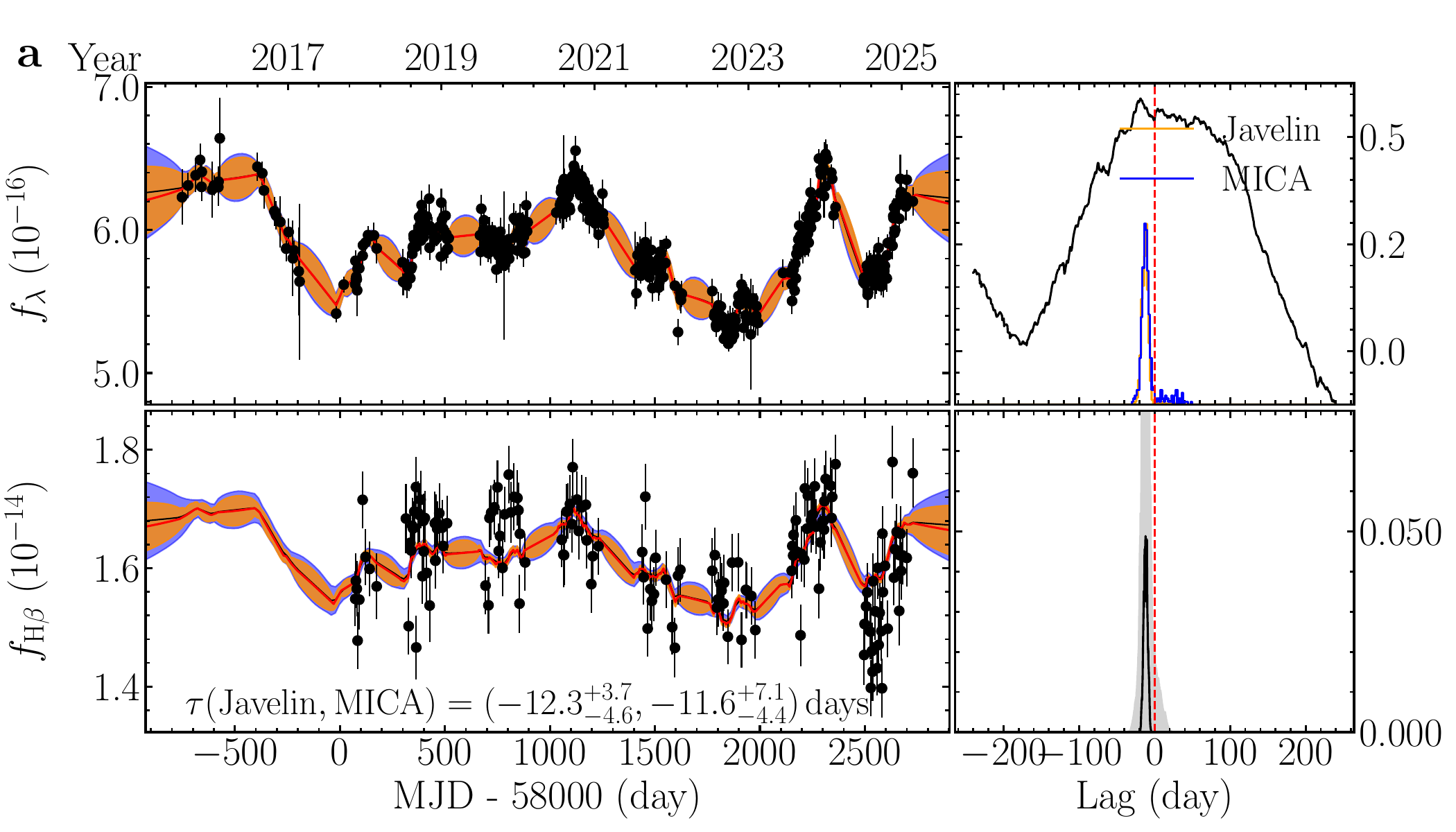}
\includegraphics[width=0.49\textwidth]{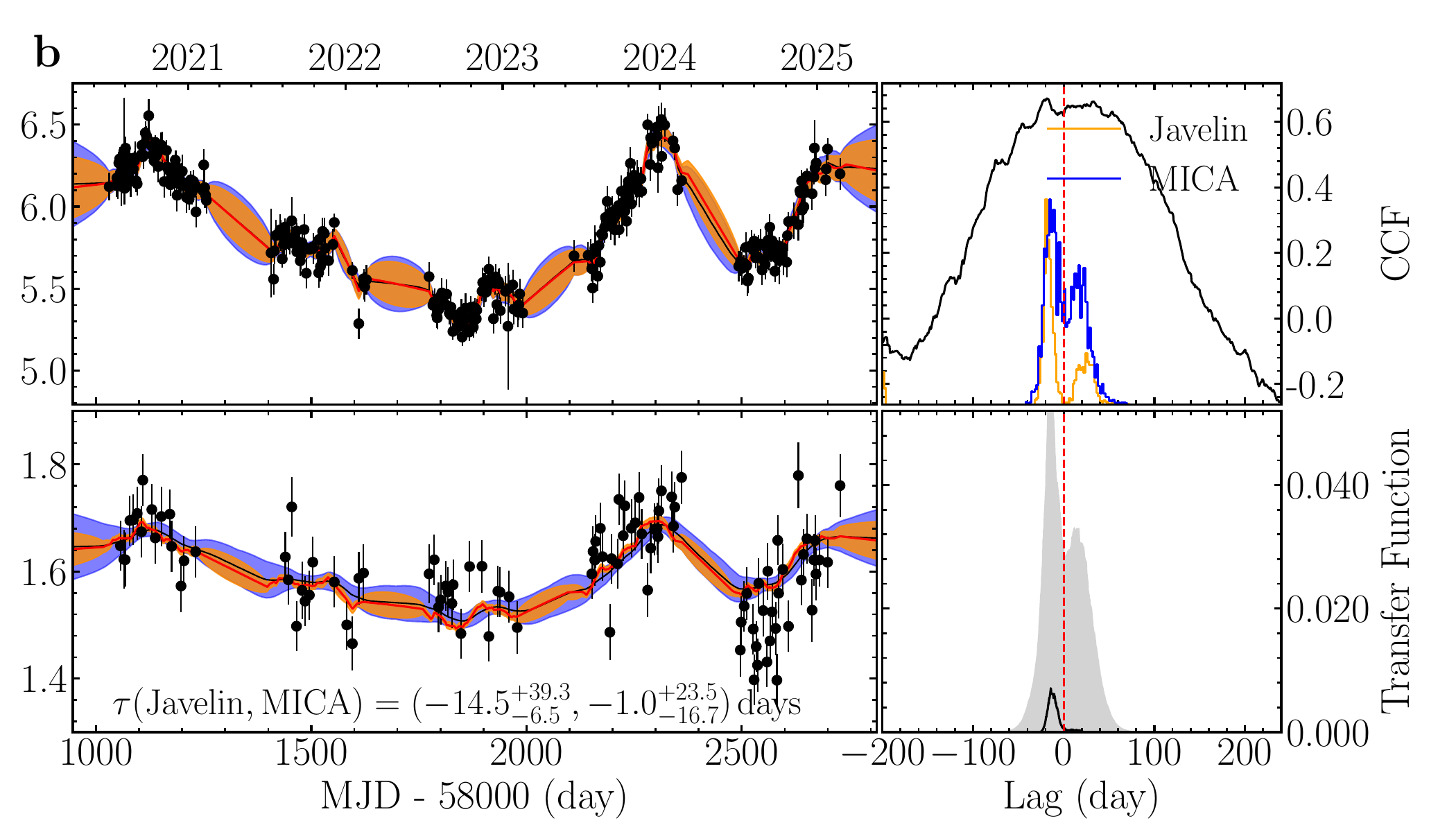}
\includegraphics[width=0.49\textwidth]{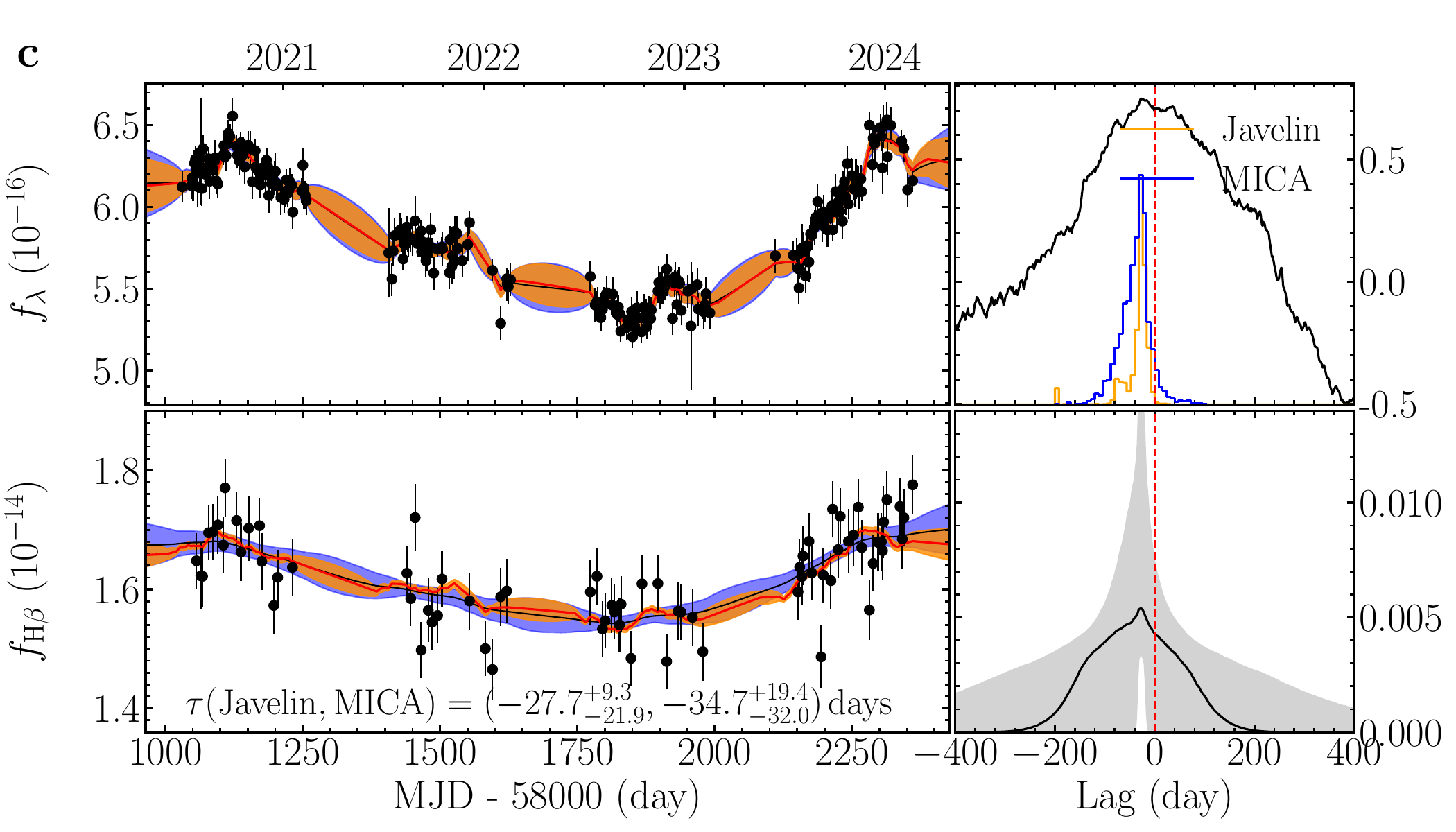}
\includegraphics[width=0.49\textwidth]{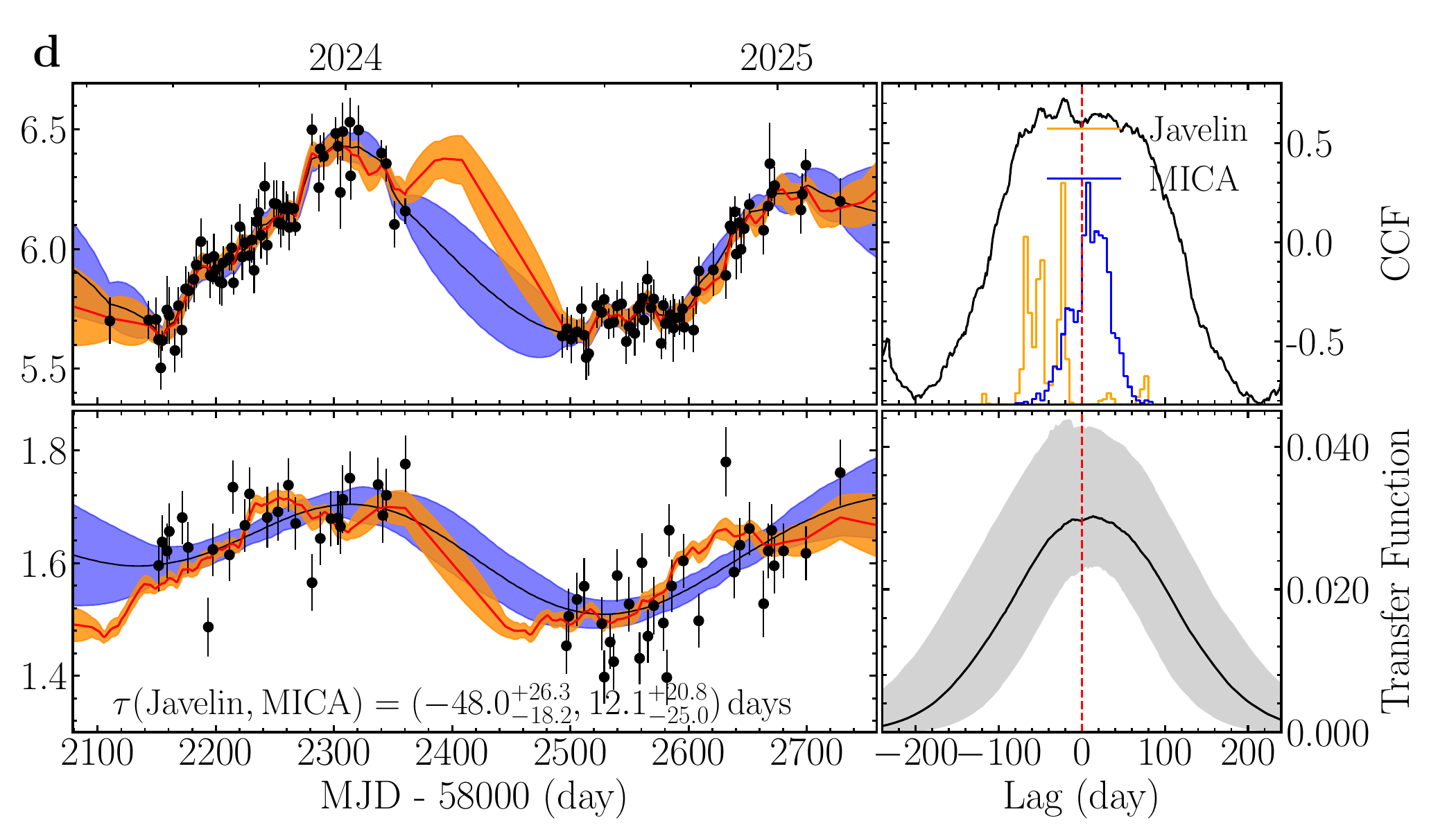}
\caption{\footnotesize H$\beta$ lags with respect to broad photometric LCs are determined for the four different periods.
The results are consistent with those in Fig.\,\ref{fig:MICA-lags}.
The photometric data are obtained with observations from CAHA-$V$, ZTF-$r$, and ATLAS-$c$.
After performing intercalibration of the dataset, the LC is binned into two-day intervals.
}
\label{fig:lags_photometry}
\end{figure*}

\clearpage

\begin{figure*}
    \centering
    \includegraphics[width=\linewidth]{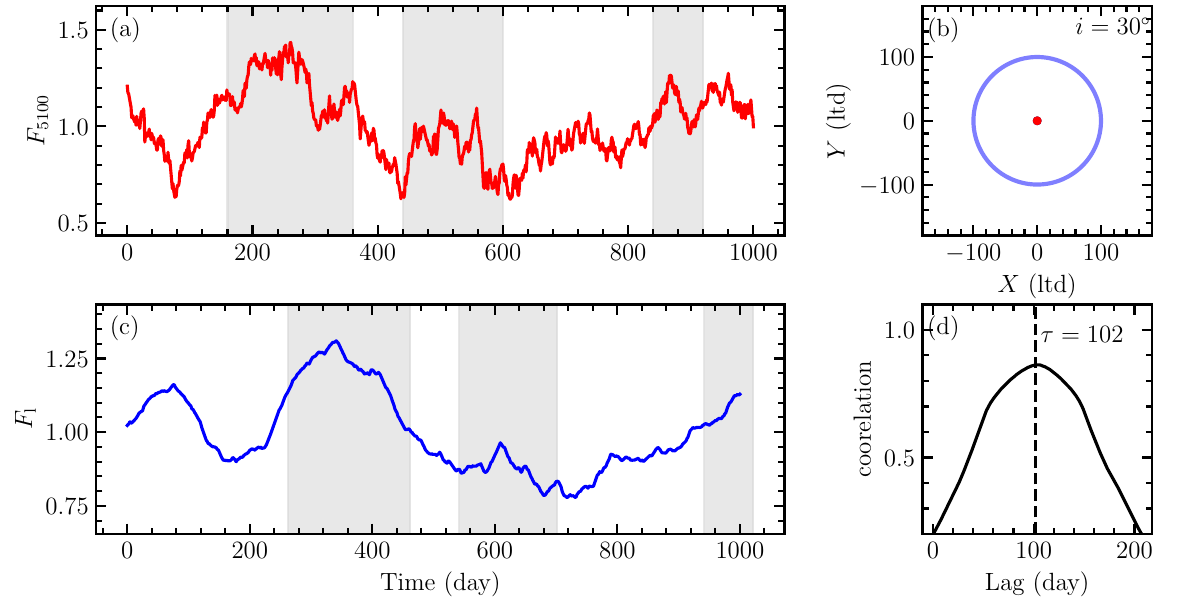}
    \caption{\footnotesize The 5100\,\AA\, and line LCs in the case of point source approximation (PSA). In such a case, X-ray, the ionizing photons and the 5100\,\AA\ share the same LCs. We take $R_{5100}=R_{\rm ion}=R_{\rm X}=0$ and $R_{\rm BLR}=R_{\rm H\beta}=100\,$ltd in this simulation.
    (a) The 5100\,\AA\ LC; (b) The position of ionizing sources (red dots) and BLR clouds (blue ring). (c) The line LC; (d) The cross correlation between continuum at 5100\,\AA\, and line LC. The most features of the H$\beta$ LC is that the extended BLRs smear variabilities with shorter timescales (than $R_{\rm H\beta}/c$), but keep the large structures of the 5100\,{\AA} variations.}
    \label{fig:lc0}
\end{figure*}

\clearpage

\begin{figure*}
\centering
\includegraphics[width=0.49\linewidth]{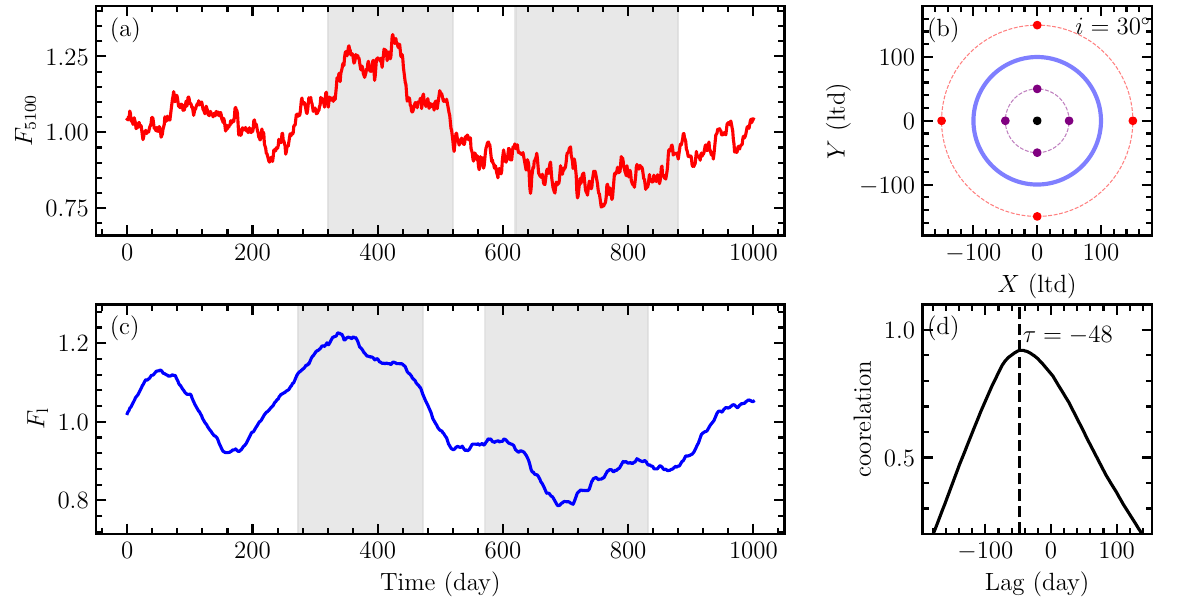}
\includegraphics[width=0.49\linewidth]{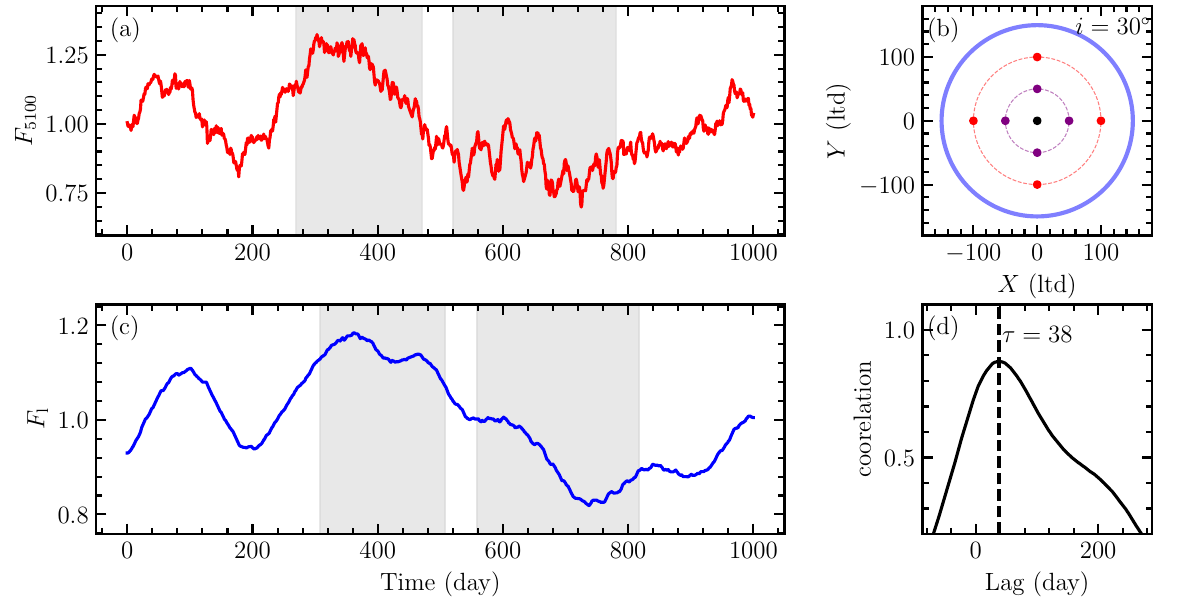}
\caption{\footnotesize Left Panel: The 5100\,\AA\, and line LCs ionized by $4$ uniformly distributed sources. The BLR is inside the 5100\,\AA\, sources. We take $(R_{\rm ion},R_{5100},R_{\rm H\beta})=(50,150,100)\,$ltd in the left panel.
(a) the 5100\,\AA\, LC; (b) the position of X-ray source (black dots), ionizing sources (purple dots), BLR clouds (blue ring), and 5100\,\AA\, sources (red dots); 
(c) the line LC; (d) the cross correlation between the 5100\,\AA\, and line LCs. The LDRs appear in this case because of $R_{5100}>R_{\rm H\beta}$.
Right panel: we take $(R_{\rm ion},R_{5100},R_{\rm H\beta})=(50,100,150)\,$ltd, namely, the BLR is outside of the 5100\,\AA\ region ($R_{5100}<R_{\rm H\beta}$), showing positive delays. The shadowed regions in two panel show the non-concordant reverberations with the 5100\,\AA. This is caused by the non-point source effects.}
\label{fig:lc1-2}
\end{figure*}

\begin{figure*}
    \centering
    \includegraphics[width=0.49\linewidth]{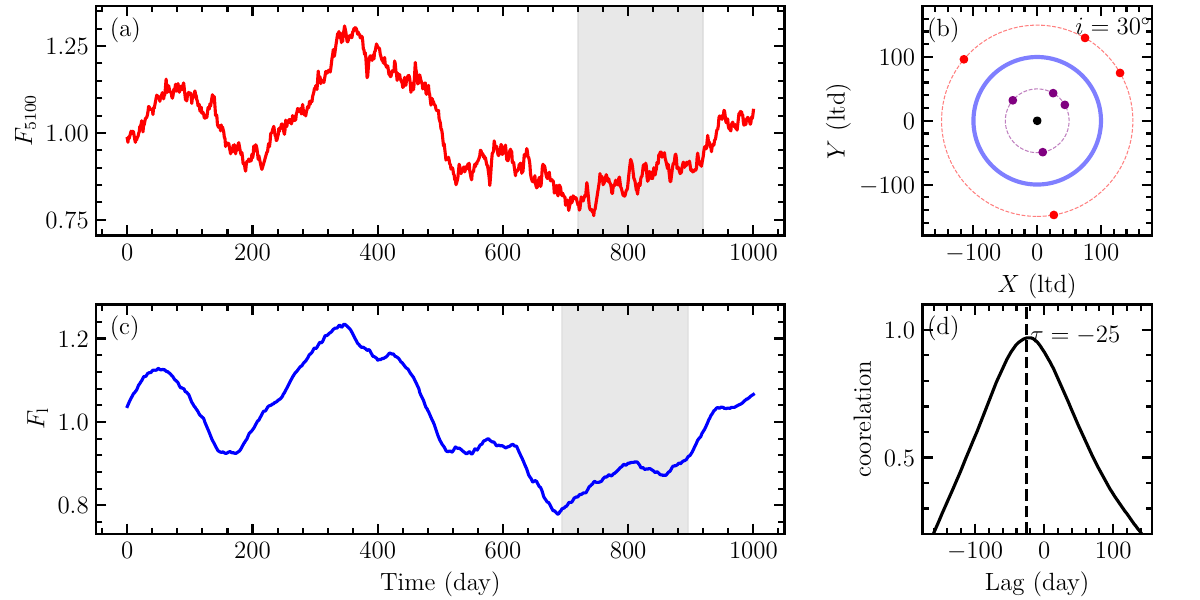}
    \includegraphics[width=0.49\linewidth]{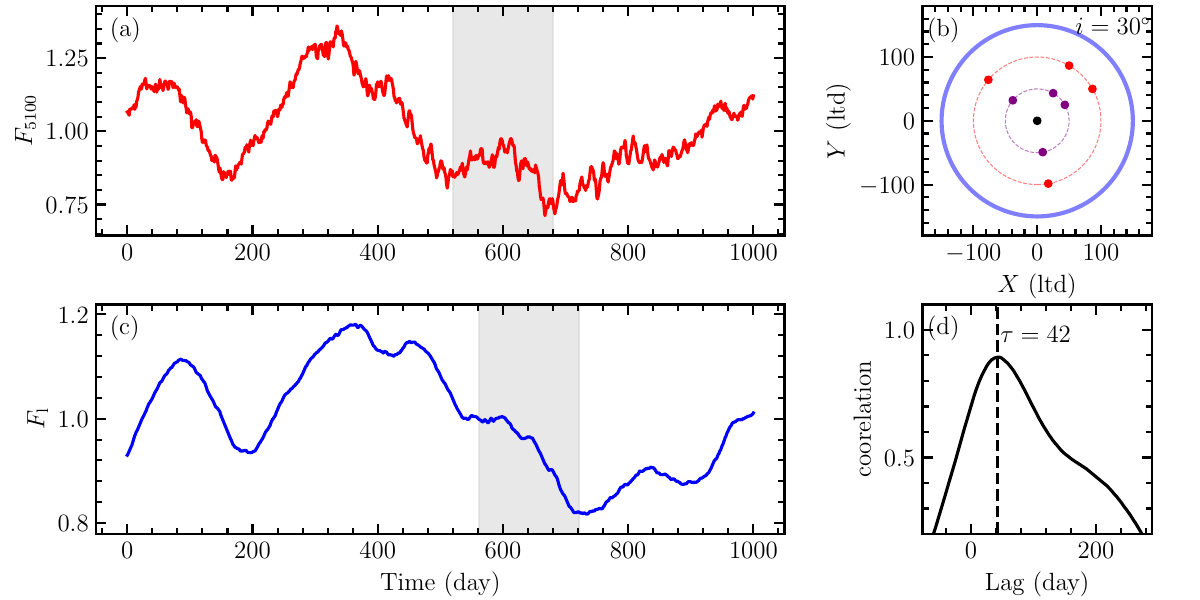}
    \caption{\footnotesize Left panel: The continuum and line LCs of a BLR ionized by $4$ randomly distributed sources. The BLR is inside the 5100 region showing LRDs. We take 
    $(R_{\rm ion},R_{5100},R_{\rm H\beta})=(50,150,100)\,$ltd in the left panel. Right panel: we take $(R_{\rm ion},R_{5100},R_{\rm H\beta})=(50,100,150)\,$ltd. The shadowed regions show non-concordant reverberations.}
    \label{fig:lc3-4}
\end{figure*}

\clearpage 

\begin{figure*}
    \centering
    \includegraphics[width=0.49\linewidth]{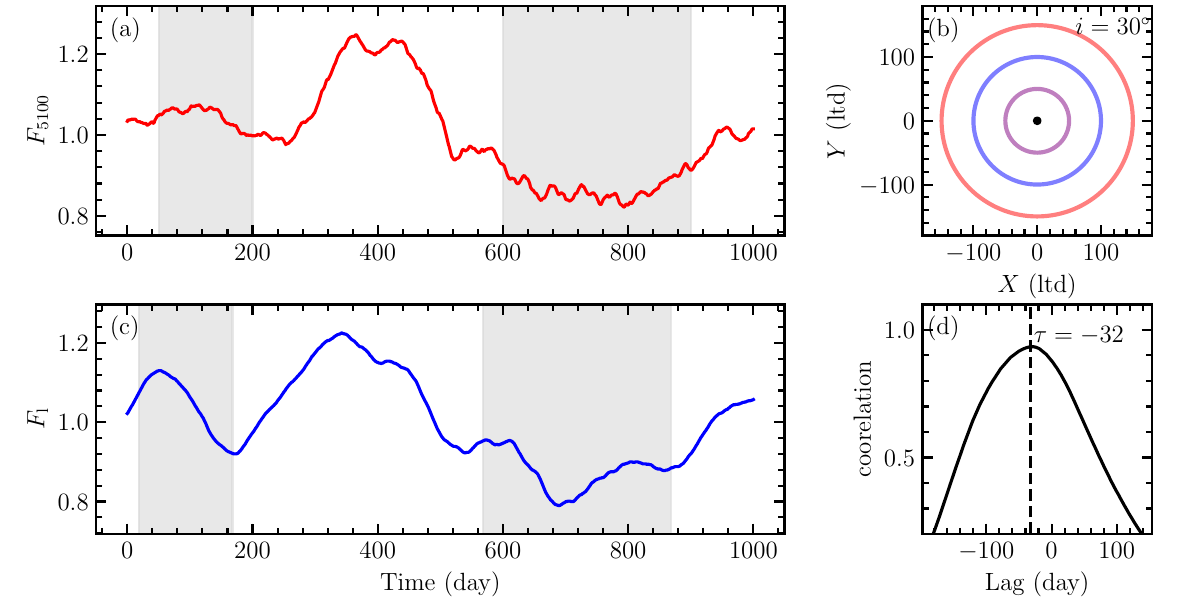}
    \includegraphics[width=0.49\linewidth]{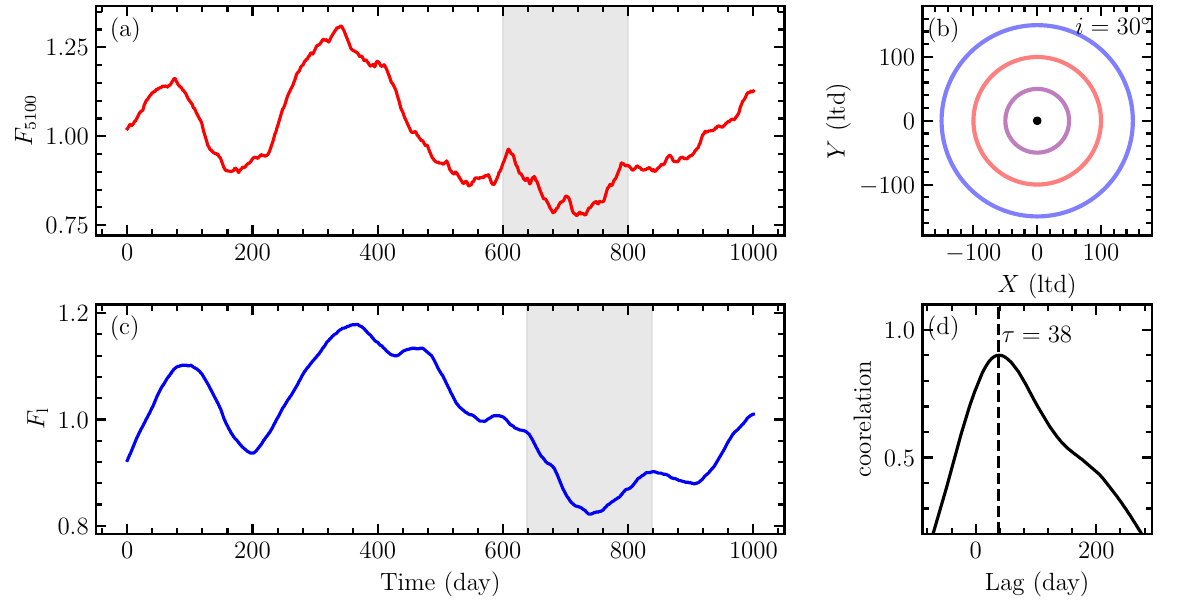}
    \caption{\footnotesize Left panel: The continuum and line LCs of a BLR ionized by a uniform ring of ionizing sources. The BLR is inside the 5100\,\AA\ region exhibiting the LDRs. We take $(R_{\rm ion},R_{5100},R_{\rm H\beta})=(50,150,100)\,$ltd in the left panel. Right panel: $(R_{\rm ion},R_{5100},R_{\rm H\beta})=(50,100,150)\,$ltd. The NPS effects are clear.}
    \label{fig:lc5-6}
\end{figure*}

\begin{figure*}
\centering
\includegraphics[width=\linewidth]{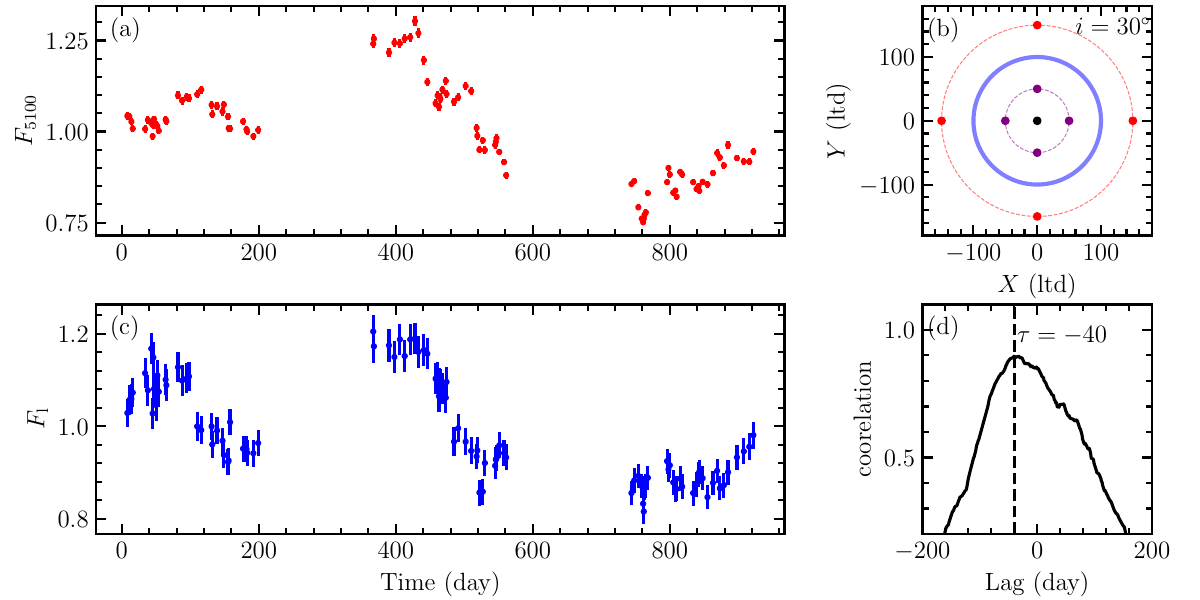}
\caption{\footnotesize The same as Fig.\,\ref{fig:lc0}, but the LC is sampled with a non-uniform cadence of 6\,days and a season gap of $160$ days. The relative uncertainties of continuum and line are 1\% and 3\%, respectively. The leading delays are consistent with that in Fig.\,\ref{fig:lc0}, demonstrating the robustness of the present results from our eight-year campaign of PHL\,1092.}
\label{fig:gap}
\end{figure*}

\begin{figure*}
\centering
\includegraphics[angle=0,width=0.48\textwidth]{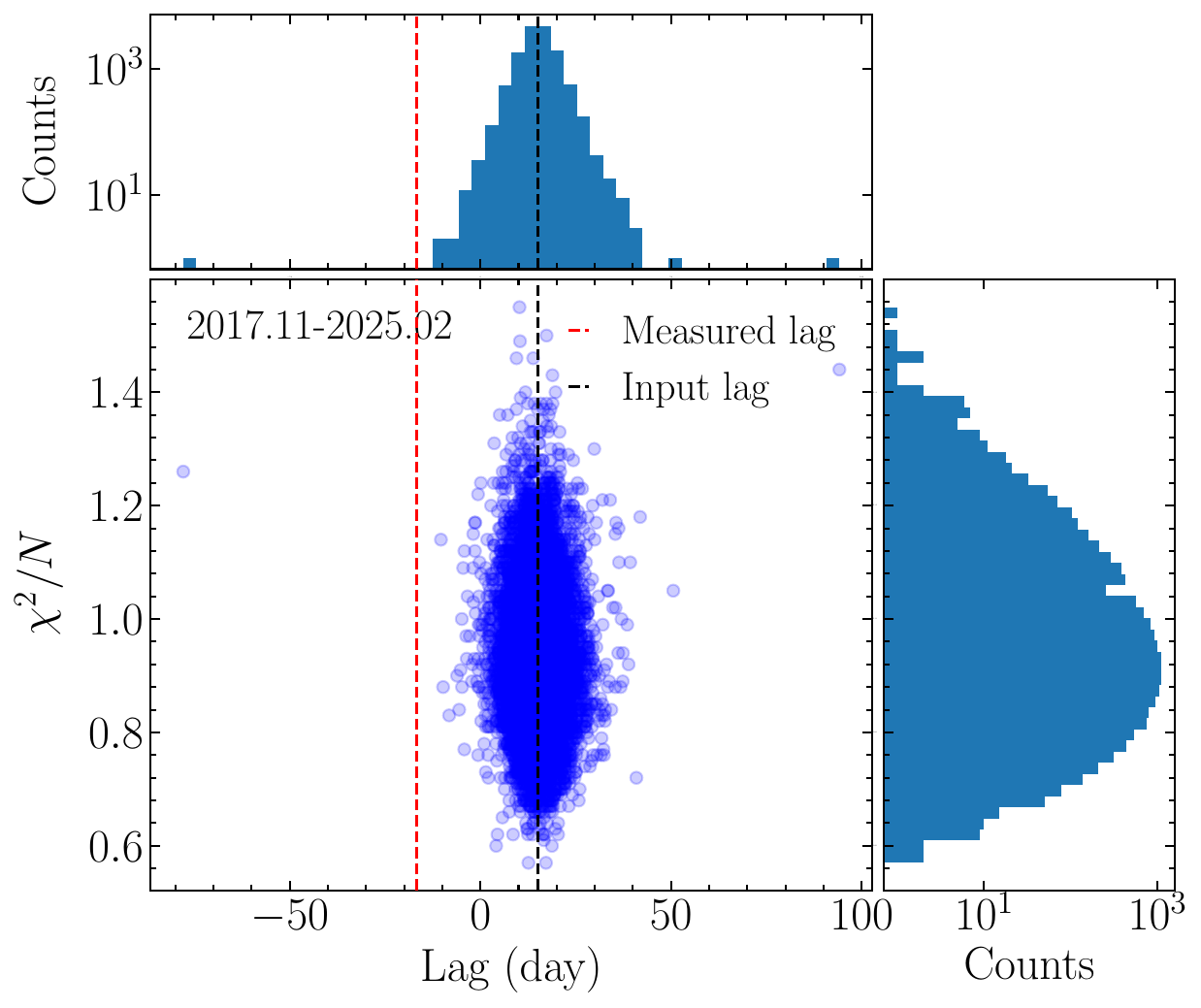}
\includegraphics[angle=0,width=0.48\textwidth]{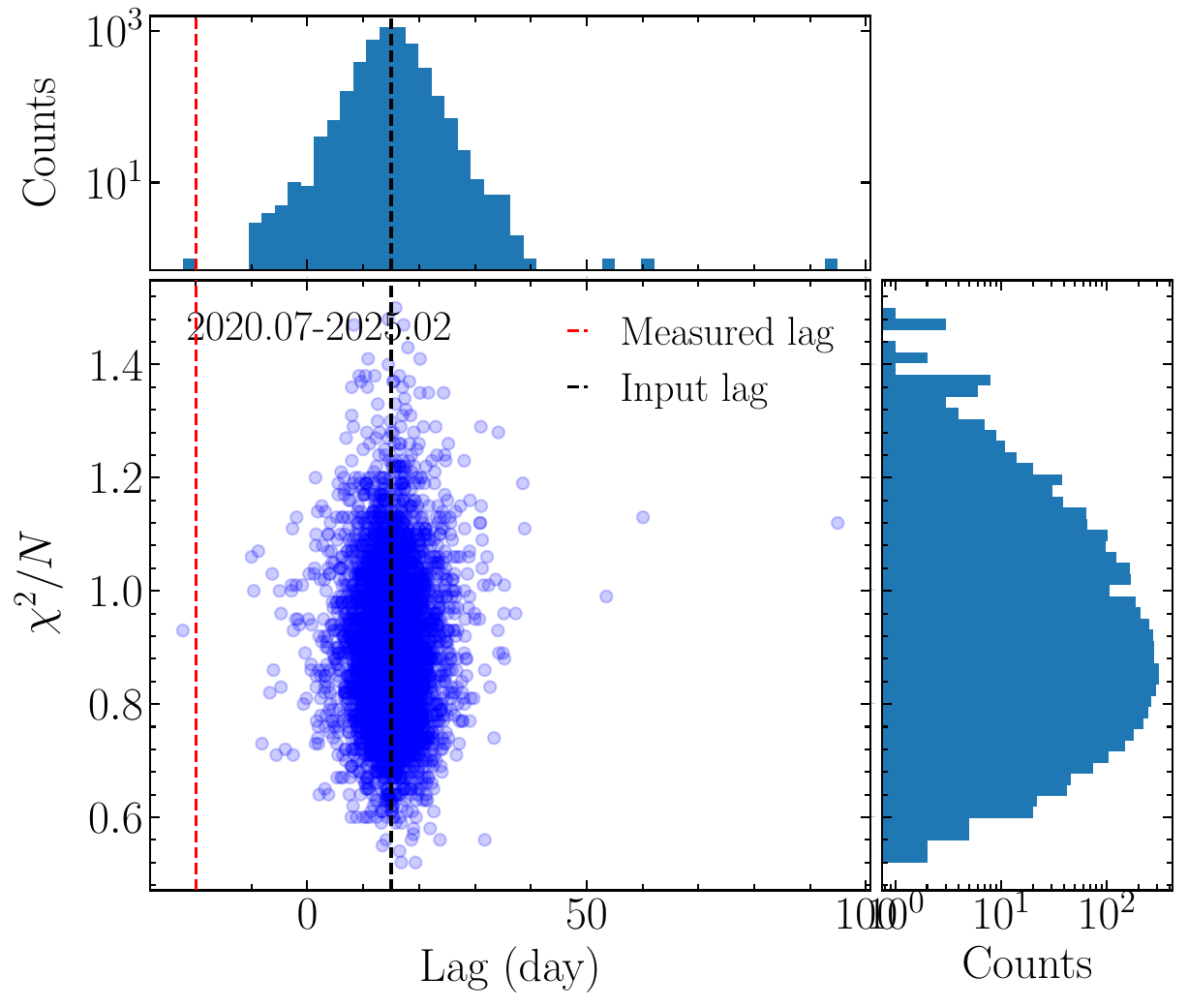}
\includegraphics[angle=0,width=0.48\textwidth]{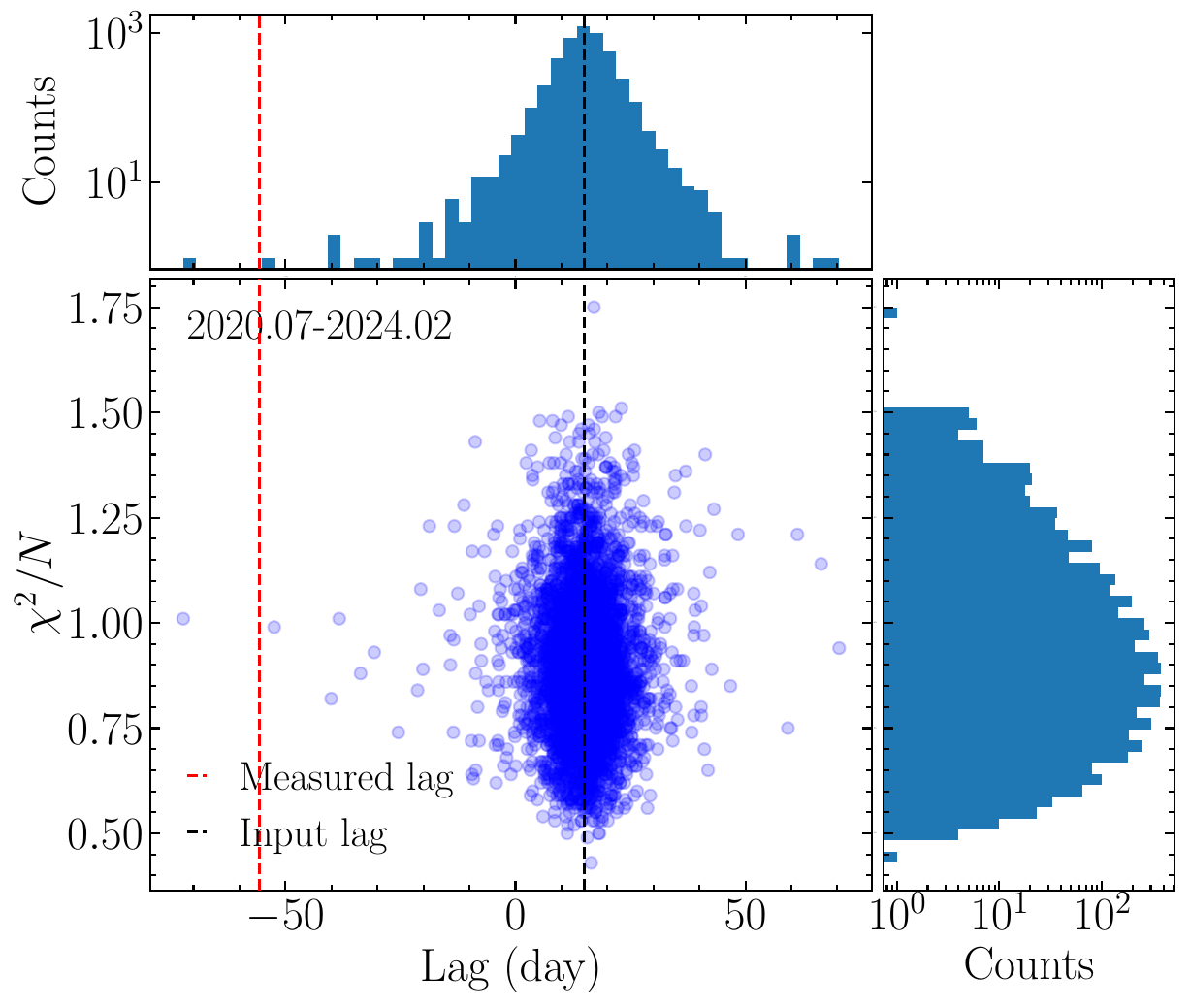}
\includegraphics[angle=0,width=0.48\textwidth]{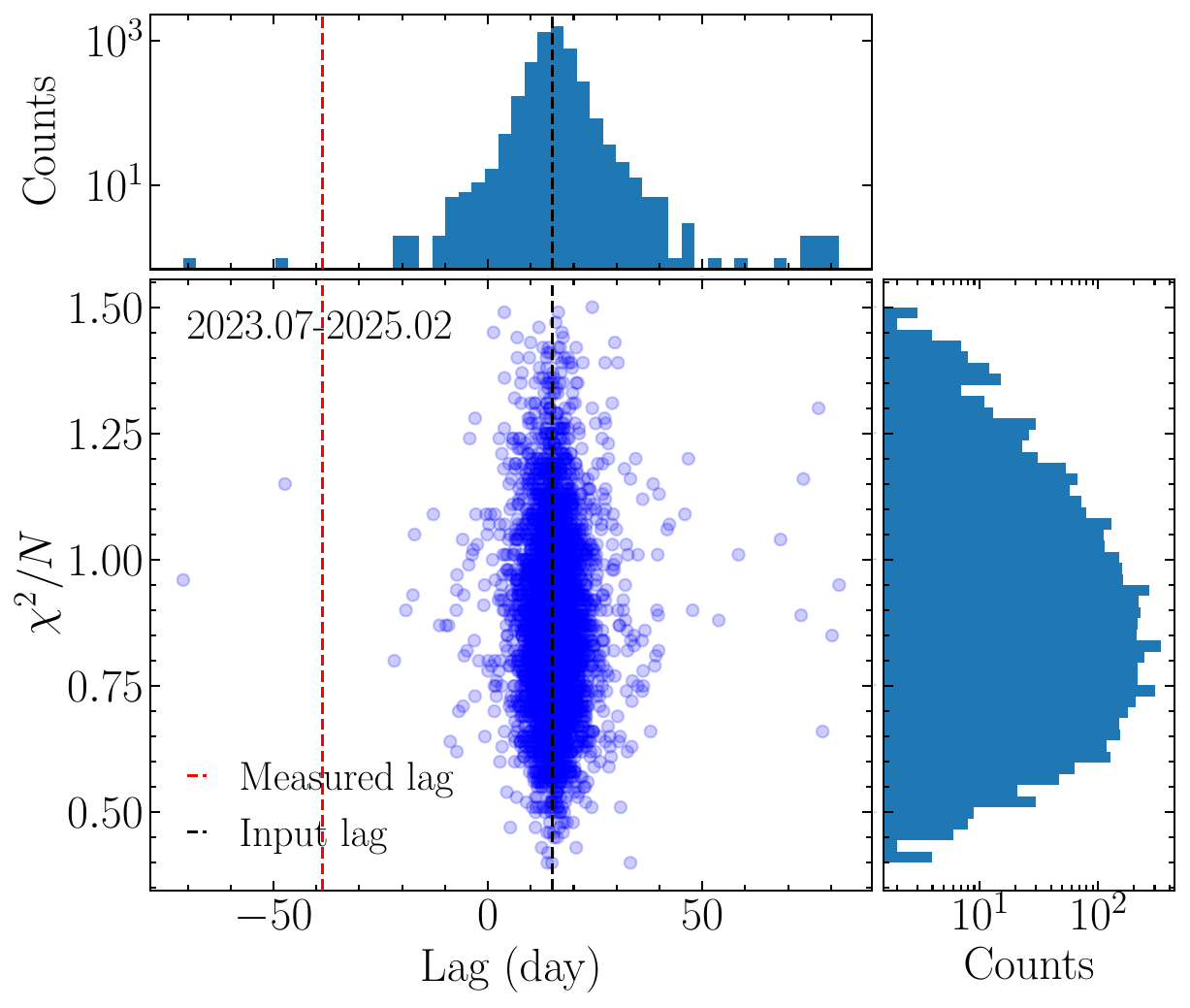}
\caption{\footnotesize Distribution of H$\beta$ lags (\MICA\ measured) from Monte Carlo simulations.
The black dashed line indicates the input time lags of $\tauhb^0=16\,$days used to generate the mock LCs of \hb\ line, while the red dashed line shows the lag measured from the real data for different epochs.
}
\label{fig:simlag_distribution}
\end{figure*}

\end{methods}

\end{document}